\documentclass[11pt]{article}
\usepackage{amssymb,amsmath}
\usepackage{bm} 
\usepackage{booktabs} 
\usepackage{array}
\usepackage{latexsym}
\usepackage{graphicx}
\usepackage{color}
\usepackage{datetime}
\usepackage[nosort]{cite}
\usepackage{verbatim}
\usepackage{chngpage} 

\usepackage{psfrag}

\usepackage{mciteplus}

\usepackage[colorlinks=true,      linkcolor=blue,      urlcolor=blue,      
            filecolor=blue,      citecolor=blue,       pdfstartview=FitH,     
						pdfpagemode=UseNone,      bookmarksopen=true]{hyperref}  
\usepackage[all]{hypcap}     

\def\eq#1{(\ref{#1})}

\newcommand{\comm}[1]{} 

\newcommand{\ket}[1]{|#1 \rangle}

\def\({\left(}
\def\){\right)}
\def\[{\left[}
\def\]{\right]}

\def\half{\frac12}
\def\coeff#1#2{{\textstyle \frac{#1}{#2}}}

\def\One{{\hbox{ 1\kern-.8mm l}}}

\def\barray{\begin{array}}
\def\earray{\end{array}}
\def\be{\begin{equation}}
\def\ee{\end{equation}}
\def\bea{\begin{eqnarray}}
\def\eea{\end{eqnarray}}
\def\bal{\begin{align}}
\def\eal{\end{align}}
\def\nn{\nonumber}

\def\mm{{\bf m}}
\def\mmb{{\bf\bar m}}
\def\Abar{\dot A}
\def\Bbar{\dot B}
\def\alphabar{\dot\alpha}
\def\chibar{\bar\chi}
\def\Xclock{{\bf X}}
\def\chiclock{{\bm \chi}}
\def\chiclockbar{{\bar\chiclock}}
\def\xmode{{\bf x}}
\def\chimode{{\mathcal X}}
\def\xcover{{\widehat \Xclock}}
\def\chicover{{\widehat\chiclock}}
\def\chibarcover{{\widehat{\overline\chiclock}}}
\def\u{w} 
\def\v{\hat{v}} 
\def\R{R_y} 

\def\hn{\hat{n}}
\def\hm{\hat{m}}
\def\hp{\hat{p}}

\def\kppo{k\hp\!+\!1}

\def\knpo{k\hn\!+\!1}

\def\kmpo{k\hm\!+\!1}
\def\kmmo{k\hm\!-\!1}

\def\ntwop{n_{2,\hat p}}
\def\nthreep{n_{3,\hat p}}
\def\nfourp{n_{4,\hat p}}
\def\barntwop{{\bar n}_{2,\hat p}}
\def\barnthreep{{\bar n}_{3,\hat p}}
\def\barnfourp{{\bar n}_{4,\hat p}}

\def\nthreen{n_{3,\hat n}}
\def\nfourm{n_{4,\hat m}}

\def\rplus{\rho_{+}}
\def\rminus{\rho_{-}}
\def\rpm{\rho_{\pm}}

\def\rhoone{\varrho_{1}}
\def\rhotwo{\varrho_{2}}
\def\rhoA{\varrho_{A}}
\def\rhofour{\varrho_{4}}
\def\rhohat{\hat\varrho}

\def\mR{\mathbb{R}}
\def\mZ{\mathbb{Z}}

\numberwithin{equation}{section} 

\makeatletter
\g@addto@macro\bfseries{\boldmath}
\makeatother

\definecolor{cardinal}{rgb}{0.6,0,0}
\definecolor{darkgreen}{rgb}{0,0.4,0}
\definecolor{golden}{rgb}{0.92, 0.7, 0}
\definecolor{midnight}{rgb}{0, 0, 0.5}
\definecolor{darkblue}{rgb}{0, 0, 0.7}

\def\Neql#1{{\cal N}\!=\!{#1}}

\def\IR{\mathbb{R}}
\def\ZZ{\mathbb{Z}}

\def\cA{{\cal A}}
\def\cB{{\cal B}}
\def\cC{{\cal C}}

\def\cF{{\cal F}}

\def\cL{{\cal L}}
\def\cM{{\cal M}}
\def\cN{{\cal N}}
\def\cP{{\cal P}}

\def\cR{{\cal R}}

\def\nBPS#1{$\frac{1}{#1}$-BPS}

\def\cO{{\cal O}}

\def\bbR{\mathbb{R}}
\def\bbS{\mathbb{S}}
\def\bbT{\mathbb{T}}
\def\bbZ{\mathbb{Z}}

\def\ket#1{|{#1}\rangle}
\def\half{\frac{1}{2}}

\begin{document}


\begin{flushright}
IPHT-T16/004\\
\end{flushright}

\vspace{8mm}

\begin{center}

\bigskip

{\huge \textsc{Momentum Fractionation on Superstrata}}

\vspace{14mm}

{\large
\textsc{Iosif Bena$^1$,~ Emil Martinec$^{2}$,~ David Turton$^{1}$,~ Nicholas P. Warner$^{3}$}}

\vspace{12mm}

$^1$Institut de Physique Th\'eorique, \\
Universit\'e Paris Saclay,\\
CEA, CNRS, F-91191 Gif sur Yvette, France \\
\medskip
\centerline{$^2$Enrico Fermi Inst. and Dept. of Physics, }
\centerline{University of Chicago,  5640 S. Ellis Ave., }
\centerline{Chicago, IL 60637-1433, USA}
\medskip
\centerline{$^3$Department of Physics and Astronomy}
\centerline{and Department of Mathematics,}
\centerline{University of Southern California,} 
\centerline{Los Angeles, CA 90089, USA}

\vspace{4mm} 
{\small\upshape\ttfamily iosif.bena @ cea.fr, ejmartin @ uchicago.edu,} \\ 
 {\small\upshape\ttfamily  david.turton @ cea.fr, warner @ usc.edu} \\

\vspace{13mm}
 
\textsc{Abstract}

\end{center}

\begin{adjustwidth}{13.5mm}{13.5mm} 
 
\vspace{1mm}
\noindent
Superstrata are bound states in string theory that carry D1, D5, and momentum charges, and whose supergravity descriptions are parameterized by arbitrary functions of (at least) two variables. 
In the D1-D5 CFT, typical three-charge states reside in high-degree twisted sectors, and their momentum charge is carried by modes that individually have fractional momentum. 
Understanding this momentum fractionation holographically is crucial for understanding typical black-hole microstates in this system.
We use solution-generating techniques to add momentum to a multi-wound supertube and thereby construct the first examples of asymptotically-flat superstrata. The resulting supergravity solutions are horizonless and smooth up to well-understood orbifold singularities.
Upon taking the AdS$_3$ decoupling limit, our solutions are dual to CFT states with momentum fractionation. We give a precise proposal for these dual CFT states.
Our construction establishes the very nontrivial fact that large classes of CFT states with momentum fractionation can be realized in the bulk as smooth horizonless supergravity solutions.

\end{adjustwidth}

\thispagestyle{empty}
\newpage


\baselineskip=14.5pt
\parskip=3pt

\tableofcontents

\baselineskip=15pt
\parskip=3pt

\section{Introduction}
\label{Sect:introduction}

String theory has been successful in counting the microstates of black holes in the regime of parameters where stringy effects overwhelm gravitational effects at the horizon scale. When supersymmetry is present, this counting carries over to the regime of parameters where gravitational effects are dominant at the horizon scale, and the entropy of these microstates reproduces the Bekenstein-Hawking entropy of the black hole \cite{Strominger:1996sh, Sen:1995in}. However, the exploration of the implications of this achievement for resolving the information paradox~\cite{Hawking:1976ra} and for understanding the physics of an infalling observer~\cite{Almheiri:2012rt,*Almheiri:2013hfa,Mathur:2012jk,*Mathur:2013gua} is still in its infancy. Indeed, very little is known about the fate of the individual stringy microstates, counted in the zero-gravity regime, as one increases the gravitational coupling and goes to the regime in which the configuration corresponds to a classical black hole with a large event horizon. 

There are several possibilities as to what this fate might be.  One is that, as gravity becomes stronger, all these microstates develop a horizon and end up looking identical to the black hole~\cite{Horowitz:1996nw,Damour:1999aw,Papadodimas:2012aq}.  Another is that some of the microstates that one constructs at zero gravitational coupling will develop a horizon, and others will remain horizonless.  A third possibility is that none of these microstates develop a horizon, and they all grow into horizon-sized bound states that have the same mass, charges and angular momentum as the black hole, but have no horizon~\cite{Lunin:2001jy,Mathur:2005zp,Bena:2007kg,Skenderis:2008qn, Balasubramanian:2008da, Chowdhury:2010ct,Mathur:2012zp,Bena:2013dka}.  There are then a range of ``sub-possibilities'':  At one extreme, typical black-hole microstates would not be describable in supergravity, but will be intrinsically quantum or non-geometrical; at the other extreme, in the sector dual to the typical microstates, one could find a {\it basis} of Hilbert space vectors that correspond to coherent states that have a supergravity description, or at least a stringy limit thereof.

In the context of the AdS-CFT correspondence~\cite{Maldacena:1997re}, one can similarly ask whether a typical CFT microstate corresponds to a classical black hole with an event horizon, or to some horizonless configuration.
The latter might either be impossible to describe in supergravity because of large quantum fluctuations or stringy corrections, or might be described using a Hilbert state basis given by smooth low-curvature solutions, or might correspond to some hybrid configuration (such as an intrinsically quantum configuration lying in a smooth, horizonless supergravity solution).

There exist pieces of evidence that can be taken as bringing support to any of these possible outcomes, some founded on calculations, and some based more on intuition and conjecture.  
Perhaps the strongest evidence that at least some microstates become smooth horizonless supergravity solutions at strong gravitational coupling comes from the explicit construction of numerous families of smooth horizonless solutions that have the same charges as black holes \cite{Bena:2006kb}. The largest family of solutions are parametrized by arbitrary continuous functions of two variables \cite{Bena:2015bea}, and come from the back-reaction of certain families of {\it superstrata}~\cite{Bena:2011uw}.  Superstrata are string theory bound states whose counting has been argued to reproduce a finite fraction of the entropy of three-charge supersymmetric black holes \cite{Bena:2014qxa}. 

However, even if the existence of these large families of solutions rules out the possibility that all the microstates one counts at zero gravity develop a horizon, it does not prove that all microstates remain horizonless, nor does it establish whether typical horizonless configurations are smooth and describable in supergravity, or are instead non-geometric or strongly-curved. 

For example, it has been argued \cite{Chen:2014loa} that for the two-charge D1-D5 black hole, the typical states of the dual symmetric product orbifold CFT~\cite{Vafa:1995bm,Douglas:1995bn,Vafa:1995zh,Bershadsky:1995qy,Maldacena:1997re} are not well-described by the microstate geometries of \cite{Lunin:2001jy,Lunin:2002iz,Kanitscheider:2007wq} when the average harmonic of the two-charge profile function becomes larger than $\sqrt{N_1 N_5}$. 
The harmonics of the profile function correspond to the winding of strands in the D1-D5 orbifold CFT; since typical three-charge microstates come from adding momentum to CFT strands whose length is of order $N_1 N_5$, one might naively conclude that all typical three-charge microstate geometries would be strongly-curved and hence not describable by supergravity. 

There are also arguments that the bulk configurations dual to typical CFT states will be non-geometrical. One such argument comes from an analysis of the possible supertube transitions that can occur in three-charge systems, which indicate that the configurations resulting from these transitions will be generically non-geometric \cite{deBoer:2010ud,*deBoer:2012ma}. 
It has also been suggested that the states that carry fractionated momentum modes, which are the typical states that contribute in the entropy counting, will involve multi-valued wavefunctions~\cite{Bena:2015bea}.

Furthermore, there are also conjectures that when tracking microstates of the D1-D5-P system to the regime of parameters where gravity becomes important, only very few states will give rise to horizonless geometries, while most states will correspond to a black hole with a horizon \cite{Sen:2009bm}. According to this perspective, the more typical the state, the larger the likelihood that its bulk dual will not be a horizonless solution, but will be a solution with a horizon.

The purpose of this paper is to provide evidence that these alternative scenarios are not realized, by showing that highly-nontrivial CFT states whose momentum is carried by fractionated carriers are dual to smooth horizonless supergravity solutions (with localized orbifold singularities). We construct  these solutions using a combination of two solution-generating techniques: 
{\it Spectral interchange} (also known as {\it spectral inversion}) and adding charge density oscillations to a supertube. Spectral interchange is a transformation of the D1-D5(-P) BPS solutions that interchanges the null coordinate along the D1 and D5 branes, $v=t+y$, with the Gibbons-Hawking fiber of the transverse space \cite{Bena:2008wt,Niehoff:2013kia}.  Modifying the charge density distribution along the supertube source profile has been studied, for example, in~\cite{Lunin:2002iz,Bena:2008dw,Bena:2010gg}. 

In this paper we show that by combining these techniques one can add $y$-momentum to a seed solution with D1 and D5 charges, as follows: First perform a spectral inversion, then use a charge density oscillation to introduce $\psi$-dependence and associated angular momentum, then spectrally invert back to the original frame to obtain a new solution carrying $v$-dependence and momentum. The $\psi$-dependent solutions in the spectrally inverted frame can be generated by integrating the Green functions against the  modified charge and angular momentum densities along the supertube.

For our explicit construction, we apply this combination of techniques to a simple seed solution -- a multiwound circular D1-D5 supertube. The multiwinding of the seed solution is what will allow us to study the physics of momentum fractionation.
While in principle the Green function/spectral interchange method can be used to construct new general classes of superstrata, a particular class of examples is amenable to a direct analysis of the equations governing all supersymmetric solutions of six-dimensional supergravity \cite{Gutowski:2003rg,Cariglia:2004kk,Bena:2011dd,Giusto:2013rxa}.
These equations determine the various potentials that enter in the supergravity solution, and are arranged in stages or layers, where the potentials to be solved for one layer satisfy linear equations sourced by the potentials determined in previous layers \cite{Bena:2011dd}. 

Our solutions are regular up to the usual orbifold singularity at the location of the multiwound supertube.
We arrange the regularity of our supergravity solutions by imposing constraints on Fourier modes and coefficients; this procedure is known as {\it coiffuring} \cite{Mathur:2013nja,Bena:2013ora,Bena:2014rea}.
We find two classes of regular solutions, corresponding to two ``Styles'' of coiffuring. We analyze the conserved charges and other properties of the solutions.

Our construction also yields the first examples of asymptotically-flat superstrata; in a particular limit our solutions contain the generalization to asymptotically-flat space of one class of the asymptotically-AdS superstrata constructed in~\cite{Bena:2015bea}.

Upon taking the decoupling limit, we obtain solutions that are asymptotically AdS$_3\times$S$^3\times \cM$ (where $\cM$ is either  $\bbT^4$ or $K3$) and we investigate the corresponding dual CFT description. 
We do this by assembling a variety of clues. We observe that the relation between the $y$-momentum and the angular momenta of the solutions suggest that the dual CFT states involve repeated applications of fractionally moded $SU(2)$ $\cR$-symmetry generators, and also that they can be generated by {\it fractional spectral flow}~\cite{Martinec:2001cf,*Martinec:2002xq} applied to a {\it subset} of strands of certain two-charge seed states.

We then study the vevs of \nBPS{4} operators and find that they have the right properties to reproduce the vevs of the supergravity fields at the linearized level, using the technology of~\cite{Skenderis:2006ah,*Kanitscheider:2006zf,Kanitscheider:2007wq,Taylor:2007hs}. We find however that the supergravity regularity constraints are not visible at this order.  Finally, by  analyzing the possible two-charge seed solutions, we determine the precise proposal for the CFT states dual to both styles of coiffuring in supergravity. 

Prior to the present work, there were only two classes of supergravity solutions, one BPS and one non-BPS~\cite{Jejjala:2005yu,Bena:2005va,*Berglund:2005vb}, which had been shown to be dual to CFT states involving momentum fractionation \cite{Giusto:2012yz, Chakrabarty:2015foa}.\footnote{There is a sense in which states obtained by the action of integer-moded generators acting on multi-wound strands can be argued to involve momentum fractionation, however this fractionation is somewhat trivial and does not correspond to degrees of freedom deep inside a throat~\cite{Mathur:2011gz,Mathur:2012tj,Lunin:2012gp}. Thus, by ``CFT states involving momentum fractionation'' we mean states which cannot be written in terms of integer-moded generators acting on R-R ground states. }
These states came from fractional spectral flow applied to {\it all} strands of certain two-charge states, and hence are very special. One way to see this is that the $AdS$ region of their dual bulk solutions can be obtained from global $AdS_3 \times S^3$ by a coordinate transformation.\footnote{The same is true of the three-charge solutions obtained by integer spectral flow~\cite{Giusto:2004id,Giusto:2004ip,Lunin:2004uu}.} In contrast, our technology produces supergravity solutions that are much more general, and cannot be written in this way. 

The remainder of this paper is structured as follows.
In Section~\ref{Sect:sixD}, we review the class of five- and six-dimensional supergravity solutions of interest, the BPS equations they satisfy, and the multiwound circular D1-D5 supertube.
In Section~\ref{Sect:SpecInter}, we apply the sequence of solution-generating techniques to add momentum to the seed solution.
We perform a direct analysis of the BPS equations in Section~\ref{Sect:MomST}, and find two classes of regular solutions via coiffuring.
In Section~\ref{Sect:CFT}, we first review the \nBPS{4} states in the CFT, the \nBPS{4} operators that are dual to linearized supergravity field modes, and spectral flow. We then develop the precise proposal for the CFT states dual to our  supergravity solutions.
Section~\ref{Sect:Discussion} summarizes our results and discusses open questions.


\section{BPS solutions in supergravity}
\label{Sect:sixD}

We work in type IIB string theory on $\mathbb{R}^{4,1}\times \bbS^1\times \cM$ where $\cM$ is  $\bbT^4$ or $K3$. We take the size of $\cM$ to be microscopic and the $\bbS^1$ to be macroscopic. The $\bbS^1$ is parameterized by the coordinate $y$ which we take to have radius $R_y$,
\begin{equation}
y ~\sim~ y \,+\, 2 \pi R_y  \,. 
\label{yperiod}
\end{equation}
We reduce on $\cM$ and work in the supergravity limit. The six-dimensional truncation of interest is an $\Neql1$ supergravity  coupled to two (anti-self-dual) tensor multiplets. This is the theory in which the first superstrata were constructed \cite{Bena:2015bea}; the theory contains all the fields expected from D1-D5-P string emission calculations \cite{Giusto:2011fy}. The BPS system of equations describing all 1/8-BPS D1-D5-P solutions of this theory has been found in  \cite{Giusto:2013rxa}, and is a generalization of the system discussed in \cite{Gutowski:2003rg,Cariglia:2004kk} and greatly simplified in \cite{Bena:2011dd}.

\subsection{The BPS equations in six dimensions}
\label{ss:BPSeqns}

To exploit the structure of the six-dimensional BPS equations, we work with null coordinates $u$ and $v$, defined by:
\begin{equation}
u ~\equiv~   \frac{1}{\sqrt{2}}\, (t-y) \,,  \qquad v ~\equiv~    \frac{1}{\sqrt{2}}\, (t+y)  \,. \label{uvdefn}
\end{equation}
The periodicity of the $y$ circle induces an identification on $u$ and $v$. It will be convenient to parameterize this as follows:
\begin{equation}
(u,v) ~\sim~ (u,v) + (-4\pi R, 4\pi R)\,, \qquad   R ~\equiv~ \frac{R_y}{2 \sqrt{2}} \,.
\label{uviden}
\end{equation}

For supersymmetric solutions, the metric is required to have the local form:
\begin{equation}
ds_6^2 ~=~    -\frac{2}{\sqrt{\cP}} \, (dv+\beta) \big(du +  \omega + \tfrac{1}{2}\, \mathcal{F} \, (dv+\beta)\big) 
~+~  \sqrt{\cP} \, ds_4^2(\cB)\,,  \label{sixmet}
\end{equation}
Note that we can always shift $\mathcal{F}$ by a constant, $c$, by sending $u \to u- \frac{1}{2} c v$ and $\omega \to \omega - \frac{1}{2} c \beta$. Given our choice of $t$ and $y$ coordinates in \eq{uvdefn}, to obtain our desired asymptotics we require that $\mathcal{F}$ vanishes at infinity throughout this paper.  

Introducing the quantities $Z_3$ and $\mathbf{k}$ via\footnote{Note that in our conventions $ \mathcal{F}$ is always negative.}
\bea \label{eq:Z3k}
Z_3 ~=~ 1- \frac{\cF}{2} \,, \qquad  \mathbf{k} ~=~ \frac{\omega+\beta}{\sqrt{2}} \,, 
\eea
one can write the metric in the form
\begin{equation}
ds_6^2 ~=  -\frac{1}{Z_3\sqrt{\cP} } \, (dt +  \mathbf{k})^2 \,+\, 
\frac{Z_3}{\sqrt{\cP}}\, \left[dy  +\left(1- Z_3^{-1}\right)  (dt +  \mathbf{k}) +\frac{\beta-\omega}{\sqrt{2}} \right]^2   +  \sqrt{\cP} \, ds_4^2(\cB)\,.
\label{sixmet-sqty}
\end{equation}

This form of the metric is useful in the analysis of closed time-like curves (CTC's).  In particular, if there are closed curves whose length in the metric $ds_4^2(\cB)$ vanishes, then it is essential that the remaining part of the metric does not make these curves time-like.  The relevant condition is manifest from  (\ref{sixmet-sqty}): The danger arises if one chooses a curve along which $dy$ is related to the other angles such that the second square vanishes.\footnote{To see this, let us suppose that such curves are timelike, and let $\cC_1$ be such a curve. $\cC_1$ itself is not necessarily closed; denote the $y$ values at the start and end of the curve by $y_1$ and $y_2$. If $y_2$ is not equal to $y_1$ (modulo $2\pi R_y$), consider $y_2$ as the starting point of a new curve $\cC_2$, similarly defined so that $dy$ is related to the other angles such that the second square vanishes. By iterating, one obtains a sequence of timelike-related points along the $y$ direction, with fixed values of the other coordinates. Since $y$ is periodic, by iterating this procedure one either obtains a CTC or comes arbitrarily close to obtaining a CTC, meaning that the spacetime has `almost-closed' timelike curves and so fails to be `strongly causal' as defined in~\cite{Hawking:1973uf}.}  We thus require that for any such curve, in the limit where the length of the curve in $ds_4^2(\cB)$ tends to zero, the one-form $\mathbf{k}$ acting on the tangent vector to the curve must also tend to zero (appropriately quickly). 

The four-dimensional base, $\cB$, has a metric, $ds_4^2$, and is required to be an ``almost hyper-K\"ahler'' manifold \cite{Gutowski:2003rg}. However we are going to simplify things by assuming that the base has a Gibbons-Hawking metric:
\begin{equation}
ds_4^2 ~=~ V^{-1} \, \big( d\psi + A)^2  ~+~ V\, d \vec y \cdot d
\vec y \,, \label{GHmetric}
\end{equation}
where the periodicity of $\psi$ will be given below in \eq{eq:psi-phi-periods} and where, on the flat $\IR^3$ defined by the coordinates $\vec y$, one has:
\begin{equation}
\nabla^2 V ~=~ 0\,, \qquad \vec \nabla \times \vec A ~=~ \vec \nabla V\,.
\label{AVreln}
\end{equation}
We take $V$ to have the form
\begin{equation}
 V ~=~ h ~+~ \sum_{j=1}^N \, \frac{q_j}{|\vec y - \vec y^{(j)}| } \,, 
\label{Vform}
\end{equation}
for some fixed points, $\vec y^{(j)} \in \IR^3$, some charges, $q_j \in \ZZ$, and some constant $h$.

We will also require that the one-form, $\beta$, is $v$-independent and then the BPS equations require that  $\beta$ has self-dual field strength:
 \begin{equation}\label{eqbeta}
\Theta_3 ~\equiv~ d \beta = *_4 d\beta\,,
 \end{equation}
where $*_4$ denotes the four-dimensional Hodge dual in the Gibbons-Hawking metric.  We will also assume that $\beta$ is  $\psi$-independent and solve the self-duality by taking
\begin{equation}
\beta ~=~ \frac{K^3}{V} \, (d\psi +A) ~+~ \vec \sigma^{(3)} \cdot \vec{dy} \,, 
\label{betaform1}
\end{equation} 
where $K^3$ is harmonic on $\IR^3$ and
\begin{equation}
 \vec \nabla \times \vec \sigma^{(3)}~=~ -\vec \nabla K^3 \,.
\label{Asigform}
\end{equation} 

The supergravity theory has three tensor gauge fields (one is in the graviton multiplet) and two scalars  (one in each tensor multiplet). The scalars may be thought of as the dilaton, $\Phi$, and axion, $C_0$, of the IIB theory.  The tensor fields of BPS solutions may be described in terms of three potential functions,  $Z_1$, $Z_2$, $Z_4$ and three sets of two-forms, $\Theta_1$, $\Theta_2$, $\Theta_4$, on the base $\cB$.

The BPS condition then requires a suitable generalization of the ``floating brane Ansatz'' \cite{Bena:2009fi} in which the metric warp factor and scalars are expressed in terms of the potentials:
\begin{equation}
\cP   ~=~     Z_1 \, Z_2  -  Z_4^2 \,, \qquad  e^{2\Phi}~=~ \frac{Z_1^2}{\cP}\,,  \qquad     C_0~=~\frac{Z_4}{Z_1}   \,.
\label{Psimp}
\end{equation}
Since we are allowing the scalars and tensor gauge fields (but not $\beta$ or $ds_4^2$) to depend upon $v$, the BPS equations impose the following linear differential equations on the potentials and  the two-forms $(Z_I, \Theta_I)$:\footnote{We define the $d$-dimensional Hodge star $*_d$ acting on a $p$-form to be
$$
 *_d\, (dx^{m_1}\wedge\cdots\wedge dx^{m_p})
 ~=~  \frac{1}{(d-p)!} \, dx^{n_1}\wedge\cdots\wedge dx^{n_{d-p}}\,  \epsilon_{n_1\dots n_{d-p}}{}^{m_1\dots m_p} \,,
$$
where we use the orientation $\epsilon^{+-1234} \equiv \epsilon^{vu1234} = \epsilon^{1234} = 1$. These are the conventions used in~\cite{Gutowski:2003rg} and note that they differ from the typical conventions for the Hodge dual.}
 \begin{equation}\label{BPSlayer1a}
 *_4 \mathcal{D} \dot{Z}_1 =  \mathcal{D} \Theta_2\,,\quad \mathcal{D}*_4\mathcal{D}Z_1 = -\Theta_2\wedge d\beta\,,\quad \Theta_2=*_4 \Theta_2\,,
 \end{equation}
  \begin{equation}\label{BPSlayer1b}
 *_4 \mathcal{D} \dot{Z}_2 =  \mathcal{D} \Theta_1\,,\quad \mathcal{D}*_4\mathcal{D}Z_2 = -\Theta_1\wedge d\beta\,,\quad \Theta_1=*_4 \Theta_1\,,
  \end{equation}
  \begin{equation}\label{BPSlayer1c}
 *_4 \mathcal{D} \dot{Z}_4 =  \mathcal{D}  \Theta_4\,,\quad \mathcal{D}*_4\mathcal{D}Z_4 = -\Theta_4\wedge d\beta\,,\quad \Theta_4=*_4 \Theta_4\,.
\end{equation}
where the dot denotes $ \frac{\partial}{\partial v}$,  $\mathcal{D}$ is defined by
\begin{equation}
\mathcal{D} \equiv \tilde d - \beta\wedge \frac{\partial}{\partial v}\,,
\end{equation}
and $\tilde d$ denotes the exterior differential on the spatial base $\cB$.

In (\ref{BPSlayer1a})--(\ref{BPSlayer1c}),  the first equation in each set involves four component equations, while the second  equation in each set is essentially an integrability condition for the first equation.  The self-duality condition reduces each $ \Theta_I$ to three independent components and adding in the corresponding $Z_J$ yields four independent functional components upon which there are four constraints. 

If we separate the $Z_I$ into their $v$-independent (zero-mode) and $v$-dependent parts, $Z_I = Z_I^{(0)}+ Z_I^{(v)}$, then the $v$-dependent parts $Z_I^{(v)}$ satisfy simpler equations, as follows. It is convenient to define two-forms $\xi_I$ via:
\begin{equation}
\Theta_I ~\equiv~ \partial_v \xi_I \,, \qquad I =1,2,4\,.
\label{ThetaIdefn}
\end{equation} 
Then for the $v$-dependent parts,
one can simplify the BPS equations (\ref{BPSlayer1a})--(\ref{BPSlayer1c}) by integrating, as follows:
\begin{equation}
*_4 \,  \mathcal{D} Z_1^{(v)} ~=~   \mathcal{D} \xi_2 \,, \qquad *_4 \,  \mathcal{D} Z_2^{(v)}  ~=~   \mathcal{D} \xi_1 \,, \qquad*_4 \,  \mathcal{D} Z_4^{(v)}  ~=~   \mathcal{D} \xi_4    \,.  \label{xieqn1}
\end{equation}

The final set of BPS equations are linear differential equations for $\omega$ and $\mathcal{F}$:
 \begin{equation}
\label{BPSlayer2a}
\mathcal{D} \omega + *_4  \mathcal{D}\omega  = Z_1 \Theta_1+ Z_2 \Theta_2  - \mathcal{F}\Theta_3 -2\,Z_4 \Theta_4 \,,
\end{equation}
and a second-order constraint that follows from the $vv$ component of Einstein's equations\footnote{This simplified form is equivalent to (2.9b) of~\cite{Giusto:2013bda}.},
\begin{equation}\label{BPSlayer2b}
\begin{aligned}
 *_4\mathcal{D} *_4\!\big(\dot{\omega} + \coeff{1}{2}\,\mathcal{D} \mathcal{F} \big)&~=~\dot{Z}_1\dot{Z}_2+Z_1 \ddot{Z}_2 + Z_2 \ddot{Z}_1 -(\dot{Z}_4)^2 -2 Z_4 \ddot{Z}_4-\coeff{1}{2} *_4\!\big(\Theta_1\wedge \Theta_2 - \Theta_4 \wedge \Theta_4\big) \\
 &~=~\partial_v^2 (Z_1 Z_2 - {Z}_4^2)  -(\dot{Z}_1\dot{Z}_2  -(\dot{Z}_4)^2 )-\coeff{1}{2} *_4\!\big(\Theta_1\wedge \Theta_2 - \Theta_4 \wedge \Theta_4\big)\,.
\end{aligned}
\end{equation}
%

\subsection{BPS solutions in five dimensions}
\label{ss:5dim}

We now recall how $v$-independent solutions reduce to five-dimensions and our discussion will closely follow that of \cite{Bena:2010gg}.  We will  assume that the magnetic $2$-forms, $\Theta^{(I)}$, are independent of the GH fiber coordinate, $\psi$.  This means that one may use the same class of solutions as in (\ref{betaform1}) by introducing more harmonic functions, $K^{I}$, on $\IR^3$ and taking
\begin{equation}
\Theta^{(I)} ~=~  d B^{(I)} \,, \label{GHdipoleforms}
\end{equation}
with 
\begin{equation}
 B^{(I)}=V^{-1} K^{I} \, (d \psi + A) ~+~ \vec{\sigma}^{(I)}\cdot d\vec{y} \,,
 \qquad \vec \nabla \times \vec \sigma^{(I)}  ~\equiv~ - \vec \nabla K^I \,.
\label{vecpotdefns}
\end{equation}

The sources in BPS equations for $Z_I$  ($I=1,2,3,4$) are independent of  $v$ and $\psi$ and so the inhomogeneous solutions for the functions $Z_I$  follow the standard form:
\begin{equation}
Z_I  ~=~ \coeff{1}{2}\, C_{IJK}  V^{-1} K^{J}K^{K}   ~+~ L_I  \,, 
\label{ZIform}
\end{equation}
where $C_{IJK}$ are the usual  (completely symmetric) structure constants for supergravity coupled to vector multiplets.  The particular theory that we use can be written in this form if one sends $Z_4 \to -Z_4$ and takes
\begin{equation}
C_{123}  ~=~  1 \,, \qquad  C_{344}  ~=~  -2 \,,
\label{RecC}
\end{equation}
with other (non-cyclically related) components equal to zero.

The functions $L_I$ in (\ref{ZIform}) are required to be harmonic on the GH base, $\cB$,  and can be allowed to depend upon all the coordinates, including $\psi$.
Thus we have 
\begin{equation}
\nabla^2_{(4)} L_I  ~=~0  \,.
\label{Lharmcond}
\end{equation}

One can then make a simple Ansatz for the angular momentum, one-form $\omega$:  
\begin{equation}
\omega ~=~ \mu (d\psi+ A) + \vec{\varpi}\cdot d\vec{y} \,.
\label{omansatz}
\end{equation}
If one introduces the covariant derivative
\begin{equation}
\vec { D} ~\equiv~ \vec \nabla ~-~   \vec A \,\partial_\psi   \,,
\label{covD}
\end{equation}
then the last BPS equation can be written as:
\begin{equation}
( \mu \vec {D} V - V\vec { D}  \mu  ) ~+~      \vec {D}  \times \vec \varpi ~+~    V \partial_\psi  \vec \varpi  ~=~ -  V\, \sum_{I=1}^3 \, Z_I \, \vec \nabla \big(V^{-1} K^I  \big) \,.
\label{covomeqn}
\end{equation}

The  BPS equations have a gauge invariance:  $\omega \to \omega + df$ and this reduces to:
\begin{equation}
\mu \to \mu ~+~ \partial_\psi f \,, \qquad  \vec \varpi \to  \vec \varpi ~+~ \vec {D} f \,,
\label{fgaugetrf}
\end{equation}
The Lorentz gauge-fixing condition, $d\star_4\omega =0$, reduces to
\begin{equation}
V^2 \, \partial_\psi \mu   ~+~\vec {D}  \cdot \vec \varpi  ~=~ 0 \,,
\label{Lorgauge}
\end{equation}
and we will impose this gauge choice.

Taking the covariant divergence, using $\vec { D}$,  of (\ref{covomeqn}) and using the Lorentz gauge choice,  one obtains:
\begin{equation}
V^{^2}\, \nabla^2_{(4)}  \mu  ~=~   \vec { D} \cdot \Big( V\, \sum_{I=1}^3 \, Z_I \, \vec { D} \big(V^{-1} K^I  \big) \Big) \,.
\label{mueqn}
\end{equation}
It is  useful to note that the  four-dimensional Laplacian may be written as:
\begin{equation}
\nabla^2_{(4)}  F ~=~ V^{-1} \big[ V^2 \, \partial_\psi^2  F   ~+~ \vec { D}  \cdot  \vec {D}    F \big] \,.
\label{Lapl}
\end{equation}

The equation for $\mu$ is  solved by taking:
\begin{equation}
\mu ~=~ \coeff{1}{6}\,  V^{-2}C_{IJK}  K^{I}K^{J}K^{K} ~+~ \coeff{1}{2}\, V^{-1} K^{I}L_{I}  ~+~  M\,,
\label{muform}
\end{equation}
where, once again, $M$ is a harmonic function on $\cB$.

Finally, we can use this solution back in (\ref{covomeqn}) to simplify the right-hand side and obtain:
\begin{equation}
\vec {D} \times \vec \varpi ~+~ V \partial_\psi \vec \varpi ~=~ V \vec {D} M - M\vec {D} V
+\frac{1}{2} \big( K^{I} \vec {D} L_{I} - L_{I}   \vec { D} K^{I} \big).
\label{omegaeqn}
\end{equation}
Once again one sees the emergence of the familiar symplectic form on the right-hand side of this equation.  One can also verify that the covariant divergence (using $\vec { D}$) generates an identity that is trivially satisfied as a consequence of  $ \vec\nabla V =  \vec \nabla \times \vec A$, (\ref{Lorgauge}),  (\ref{muform})  and
\begin{equation}
\nabla^2_{(4)} L_I ~=~\nabla^2_{(4)}   M ~=~ 0 \,.
\label{harmonicLM}
\end{equation}
An explicit, closed form for all the components of $\vec \varpi$ was not given in  \cite{Bena:2010gg}, but for our solutions we will be able to construct them.

\subsection{A round supertube in flat space}
\label{ss:rndstube}

The simplest supertube Ansatz is to take the base, $\cB$, to be flat $\IR^4$ and set $\Theta_3$ and $\beta$ to be that of a simple magnetic monopole.
There are two convenient ways to formulate this.  First, one can  take $\beta$ given by (\ref{betaform1})  and write  $\IR^4$ in  Gibbons-Hawking form using spherical polar coordinates $(\rho_{-},\vartheta_{-},\phi)$: 
\begin{equation}
ds_4^2 ~=~ V^{-1} \, (d\psi +A)^2 ~+~ V\,(d\rho_{-}^2 + \rho_{-}^2 \, d\vartheta_{-}^2 + \rho_{-}^2 \, \sin^2 \vartheta_{-}\, d\phi^2 ) \,,
\label{GHmet}
\end{equation} 
where in terms of the three-dimensional Cartesian coordinates $y_1, y_2, y_3$ we have
\begin{equation}
V  ~=~  \frac{1}{\rminus}  \,, \qquad K^3  ~=~  \frac{k R}{\rplus}  \,,  \qquad \rpm ~\equiv~ \sqrt{y_1^2  + y_2^2 +({y_3 \mp \coeff{1}{2}\ell})^2} \,, 
\label{GHVdefn}
\end{equation} 
where the dipole moment $k$ is an integer.  One then has: 
\begin{equation}
A  ~=~\frac{(y_3 +  \coeff{1}{2}\ell)} { \rminus}   \, d\phi   \,,  \qquad  \sigma ~=~- k R\, \frac{(y_3 -  \coeff{1}{2}\ell)} {\rplus}   \, d\phi \,.
\label{Asigres1}
\end{equation} 
The periodicity identifications on $\psi$ and $\phi$ are as usual
\begin{equation}
\psi ~\sim~ \psi + 4\pi \,,
 \qquad (\psi,\phi) ~\sim~ (\psi,\phi) + (2\pi,-2\pi) \,.
\label{eq:psi-phi-periods}
\end{equation}
One can then follow through with the construction outlined in Section \ref{ss:5dim}.   However, we  subsequently want to make heavy use of the results and formalism employed in \cite{Bena:2015bea} and so we will use this as an opportunity to introduce the geometry and flux components that make up the second convenient description of supertubes.  

One starts by describing the base manifold in terms of spherical bipolar coordinates, defined by\footnote{Our spherical bipolar angles $\varphi_1$ and $\varphi_2$ are related to those of~\cite{Bena:2015bea} by $\varphi_1^{\rm here} = \phi^{\rm there}$, $\varphi_2^{\rm here} = \psi^{\rm there}$.}
\begin{align}
4\, \rplus    &~=~ \Sigma ~\equiv~(r^2 + a^2 \cos^2 \theta)  \,, \qquad\quad~ 4\,  \rminus   =~\Lambda ~\equiv~(r^2 + a^2 \sin^2 \theta) \,, \\
\cos\frac{\vartheta_-}{2} &~=~  \frac{(r^2 + a^2)^{1/2}}{ \Lambda^{1/2}}  \sin \theta \,, \qquad\qquad
\sin\frac{\vartheta_-}{2} ~=~  \frac{r \cos \theta}{ \Lambda^{1/2}} \,, \\
\psi   &~=~ \varphi_1 +  \varphi_2  \,, \qquad~~ \phi   ~=~ \varphi_1 -  \varphi_2 \,,  \qquad~~ \ell ~\equiv~ \coeff{1}{4}\, a^2  \,.
\label{polarchng}
\end{align} 
The metric becomes:
\begin{equation}
ds_4^2 ~=~ \Sigma\, \bigg( \frac{dr^2}{(r^2 + a^2)} ~+~ d \theta^2 \bigg)  ~+~ (r^2 + a^2) \sin^2 \theta \, d\varphi_1^2 ~+~ r^2  \cos^2 \theta \, d\varphi_2^2  \,, 
\label{bipolmet}
\end{equation} 
and we choose the natural system of frames 
\begin{equation}
e_1 ~=~ \frac{\Sigma^{1/2} }{(r^2 + a^2)^{1/2}} \, dr\,, \quad   e_2 ~=~\Sigma^{1/2}   \, d\theta\,, \quad   e_3 ~=~(r^2 + a^2)^{1/2}  \sin \theta  \, d\varphi_1\,, \quad   e_4 ~=~ r  \cos  \theta \, d\varphi_2 \,.
\label{frames1}
\end{equation} 
Following \cite{Bena:2015bea}, it is convenient to introduce the self-dual two-forms $\Omega^{(1)}$, $\Omega^{(2)}$ and $\Omega^{(3)}$:
\begin{equation}\label{selfdualbasis}
\begin{aligned}
\Omega^{(1)} &\equiv \frac{dr\wedge d\theta}{(r^2+a^2)\cos\theta} + \frac{r\sin\theta}{\Sigma} d\varphi_1\wedge d\varphi_2 ~=~  \frac{1}{\Sigma \, (r^2+a^2)^\frac{1}{2}  \cos\theta} \,(e_1 \wedge e_2 +  e_3 \wedge e_4)\,,\\
\Omega^{(2)} &\equiv  \frac{r}{r^2+a^2} dr\wedge d\varphi_2 + \tan\theta\, d\theta\wedge d\varphi_1  ~=~  \frac{1}{\Sigma^\frac{1}{2}\, (r^2+a^2)^\frac{1}{2} \cos\theta} \,(e_1 \wedge e_4 +  e_2 \wedge e_3)   \,,\\
 \Omega^{(3)} &\equiv \frac{dr\wedge d\varphi_1}{r} - \cot\theta\, d\theta\wedge d\varphi_2~=~  \frac{1}{\Sigma^\frac{1}{2}\, r \sin\theta} \,(e_1 \wedge e_3 -  e_2 \wedge e_4)  \,,
\end{aligned}
\end{equation}
and note that
\begin{equation}
\begin{aligned}
  *_4(\Omega^{(1)}\wedge \Omega^{(1)})
 &={2\over (r^2+a^2)\Sigma^2\cos^2\theta},\quad&
 *_4(\Omega^{(2)}\wedge \Omega^{(2)})
 &={2\over (r^2+a^2)\Sigma\cos^2\theta},\\
 *_4(\Omega^{(3)}\wedge \Omega^{(3)})
 &={2\over r^2\Sigma\sin^2\theta},&
\Omega^{(i)}\wedge \Omega^{(j)} &=0,\quad i\neq j.
\end{aligned}
\end{equation}

The vector field $\beta$ corresponding to the harmonic functions in (\ref{GHVdefn})  is
\begin{equation} 
\hat\beta ~=~\frac{2\,k R a^2 }{\Sigma} \, ( \sin^2 \theta \, d\varphi_1-   \cos^2 \theta \, d\varphi_2 )+ k R\,( d\varphi_1+ d\varphi_2) \,.
\label{betaform2}
\end{equation} 
To obtain flat asymptotics, we see from \eq{sixmet-sqty} that $\beta$ and $\omega$ must vanish at infinity. We thus make a coordinate transformation to gauge away the constant part of $\hat\beta$, obtaining
\begin{equation} 
\beta ~\equiv~\beta_1 \, d\varphi_1+ \beta_2\,  d\varphi_2 ~=~\frac{2\,k R a^2 }{\Sigma} \, ( \sin^2 \theta \, d\varphi_1-   \cos^2 \theta \, d\varphi_2 ) \,.
\label{betaform3}
\end{equation} 
The two-form $\Theta_3=d\beta$ is given by
\begin{equation} 
\Theta_3 ~=~ d\beta ~=~\frac{4\,k R a^2 }{\Sigma^2} \, ( (r^2 + a^2)\cos^2 \theta \,  \Omega^{(2)}   - r^2 \sin^2 \theta \,  \Omega^{(3)} ) \,.
\label{Theta3form1}
\end{equation} 

The basic, round $v$-independent asymptotically-flat supertube solution is then given by:
\begin{align}
Z_1 &~=~ 1 ~+~ \frac{Q_1}{\Sigma} \,, \qquad Z_2 ~=~ 1 ~+~ \frac{Q_2}{\Sigma} \,, \qquad \mathcal{F} ~=~ 0   \,, \qquad Z_4 ~=~ 0 \,,\nonumber  \\
\Theta_3 &~=~ d\beta \,,   \qquad  \Theta_I ~=~ 0\,, \ \  I=1,2,4  \nonumber \\
\omega &~\equiv~\omega_1 \, d\varphi_1+ \omega_2\,  d\varphi_2 ~=~ \Big(c_1 + \frac{J\,(r^2 + a^2) }{a^2 \, \Sigma} \Big)\, d \varphi_1 ~+~ \Big(c_2 - \frac{J\,r^2}{a^2 \, \Sigma} \Big)\, d \varphi_2   \,,
\label{STsol1}
\end{align} 
where $c_1$ and $c_2$ are  constants to be determined via regularity and asymptotics. 
The constants $Q_1$, $Q_2$ and $J$ are harmonic sources that encode charges and angular momentum.

At the center of space ($r=0, \theta =0$) the size of the $\varphi_1$-circle and of the $\varphi_2$-circle collapse to zero size as measured in the spatial base metric, $ds_4^2$, in  (\ref{sixmet}).
Moreover,  ${\cal P}$  goes to a constant at the center of space. 
It is evident from this and the discussion around (\ref{sixmet-sqty}) that to avoid closed time-like curves at the center of space one must have $\omega + \beta =0$ at $r=0, \theta =0$.  This implies:
\begin{equation} 
c_1 ~=~  - \frac{J}{a^2} \,, \qquad  c_2 ~=~ 2 k R \,.
\label{cres1}
\end{equation} 
In addition, $\omega$ must also fall off when $r \to \infty$ and hence we require 
\begin{equation} 
J  ~=~2 \, k R \, a^2 \,.
\label{STreg1}
\end{equation} 
Thus $\omega$ is given by
\begin{equation} 
\omega ~=~\frac{2\,k R a^2 }{\Sigma} \, ( \sin^2 \theta \, d\varphi_1 +   \cos^2 \theta \, d\varphi_2 ) \,.
\label{omegaring}
\end{equation} 

Finally there is the regularity of the metric near the supertube, which means that as one approaches $\Sigma =0$, or  $r=0, \theta =\frac{\pi}{2}$, the metric must remain smooth.  One can easily check that the only potentially singular parts of the metric are the $d\varphi_1^2$ terms and these are proportional to:
\begin{equation} 
- \frac{2}{\sqrt{\cP}}\,\beta_1 \, \omega_1 ~+~  \sqrt{\cP} \,a^2  \, d\varphi_1^2 \, 
\label{JSTdiv1}
\end{equation} 
The vanishing of the  singularity at $\Sigma=0$  requires
\begin{equation} 
J ~=~ \frac{Q_1Q_2}{4kR} \, \qquad \Rightarrow \qquad a^2 ~=~ \frac{Q_1Q_2}{k^2R_y^2} \,.
\label{STreg0}
\end{equation} 
Thus  supertube regularity determines the radius, $a$, and the angular momentum, $J$, in terms of the charges $Q_1$, $Q_2$ and the dipole charge $k$.
We thus recover the supertube metric of\cite{Balasubramanian:2000rt,*Maldacena:2000dr} and its Gibbons-Hawking description \cite{Giusto:2004kj}. 
 
Having made these choices, the $\psi$-fiber limits to a fixed size as one approaches the supertube while the remaining part of the spatial metric limits to (in spherical polar coordinates ($\rplus,\vartheta_+,\phi$) centered around the supertube):
\begin{equation} 
\widetilde{ds}_4^2 ~=~  \frac{ \sqrt{Q_1Q_2}}{4\,\ell} \,  \bigg[  \frac{16 \,  \ell}{Q_1Q_2} \,  \rplus  \big(dy+\coeff{1}{\sqrt{2}}\, ( \sigma-\varpi)\big)^2~+~ \frac{1}{ \rplus} \,(d\rplus^2 + \rplus^2 \, d\vartheta_+^2 + \rplus^2 \, \sin^2 \vartheta_+\, d\phi^2 ) \bigg]\,.
\label{nearST1}
\end{equation} 
Setting $\rplus = \coeff{1}{4} r_+^2$ and using (\ref{polarchng}) and  (\ref{STreg0}) one obtains:
\begin{equation} 
\widetilde{ds}_4^2 ~=~  \frac{ \sqrt{Q_1Q_2}}{4\,\ell} \,  \Big[ dr_+^2 +\coeff{1}{4}\, r_+^2  \, \big(d\vartheta_+^2 +   \sin^2 \vartheta_+\, d\phi^2  +  \coeff{1}{k^2}\, \big[ \coeff{1}{\sqrt{2} \,R} \,  \big(dy+\coeff{1}{\sqrt{2}}\, ( \sigma-\varpi)\big)\big]^2  \big)\Big]\,.
\label{nearST2}
\end{equation} 
Since $R_y = 2 \sqrt{2} R$, one has  $y \sim y + 4 \pi  \sqrt{2} R$ and so  the  coordinate $\coeff{y}{\sqrt{2} \,R} $ has period $4 \pi$, which means that the metric in (\ref{nearST2}) represents 
 the standard $\ZZ_k$ orbifold of $\IR^4$.

\section{Supertubes with momentum via spectral interchange}
\label{Sect:SpecInter}

The original D1-D5 supertube solution \cite{Lunin:2001jy,Lunin:2002iz} was defined in terms of an arbitrary profile function, $\vec F(\v)$,  in $\IR^4$.  While this manifestly describes the shape of the supertube, the supertube solution is {\it not} invariant under reparameterizations of $\v$, indeed, reparameterizations encode the choice of the charge-density functions.  Put differently, the supertube can be given two charge densities, $\rhoone$ and $\rhotwo$, and an angular momentum density, $\rhohat$.  However, supertube regularity and the absence of closed time-like curves (CTC's) places two functional constraints (local analogues of (\ref{STreg0})) on these densities leaving a free choice of one function.  This function encodes the degrees of freedom represented by the choice of reparameterization in the original formulation.  

Spectral interchange can then be combined with the addition of such a charge-density fluctuation so as to generate a third (momentum) charge.

\subsection{Spectral interchange in general}
\label{ss:SIflip}

The idea behind spectral interchange is extremely simple.  When the base space, $\cB$, has a Gibbons-Hawking form then the entire solution can be written as a torus fibration over a flat $\IR^3$.  The torus is, of course, described by $(v,\psi)$ and one can  act on this torus with elements of $GL(2,\ZZ)$\footnote{Technically, one should restrict to the global diffeomorphisms, $SL(2,\ZZ)$, but if one allows orbifolds it is sometimes convenient to use $GL(2,\ZZ)$.}.  Since these transformations are generated by simple changes of coordinate, they must map BPS solutions to BPS solutions.    Some elements of this transformation group generate what are known as gauge transformations \cite{Bena:2005ni} and generalized spectral flows \cite{Bena:2008wt}, that mix $K^3$ and $V$.   Of relevance later will be the gauge transformations:
 \begin{eqnarray}
K^I &\to& K^I + \alpha^I V \cr
L_I &\to&  L_I ~-~   C_{IJK}\,\alpha^J K^K ~-~  \coeff{1}{2} \, C_{IJK}\, \alpha^J \alpha^K V \cr
M &\to& M ~-~  \coeff{1}{2} \,\alpha^I L_I ~+~  \coeff{1}{12} \, C_{IJK}\,
\left( V \alpha^I \alpha^J \alpha^K +3 \alpha^I \alpha^J K^K \right) \,. \label{BPSgaugetrf}
\end{eqnarray}
Such transformations are pure gauge in that, while they reshuffle the potentials, while leaving the physical properties of the solution invariant.

Spectral interchange is a subset of the generalized spectral flow transformations \cite{Bena:2008wt}, and is simply the modular inversion that interchanges $v$ and $\psi$ on the torus  \cite{Niehoff:2013kia}. It corresponds to a global diffeomorphism on the fibers:
\begin{equation}
v \to  - \psi \,, \quad \psi  \to  - v  \,;  \qquad \Leftrightarrow    \qquad  V ~\leftrightarrow~ K_3\,,    \quad   A \to  - \xi \,, \quad \xi  \to- A \,.
\label{SpecInv}
\end{equation}
This mapping also interchanges all the harmonic functions that make up the BPS solutions outlined in the previous section, as we now describe. 

To make the mapping more precise, we must normalize the torus angles that we interchange. 
The periodicity of the $y$ circle \eq{yperiod} induces an identification on $u$ and $v$. As described in \eq{uviden} above, we parameterize this as
\begin{equation}
(u,v) ~\sim~ (u,v) + (-4\pi R, 4\pi R) \,.
\label{uviden-2}
\end{equation}
Recalling the periodicity identifications on $\psi$ and $\phi$ given in \eq{eq:psi-phi-periods},
we see that the relevant lengths are $4\pi R$ for $v$ and $4\pi$ for $\psi$.
Thus the spectral interchange is more precisely written as:
\begin{eqnarray}
\frac{v}{R} &\to& -\psi \,, \qquad \psi \,~\to~\, -\frac{v}{R} \,.
\label{eq:specinvdef-2}
\end{eqnarray}

Setting $Z_4=0$ and $\Theta^{(4)}=0$, spectral interchange implies that the following must also give a BPS solution:
\begin{eqnarray}
\widetilde V &=& \frac{K^3}{R} \,, \quad \widetilde K^3 ~=~ R\, V \,, \quad
\widetilde K^1 ~=~ \frac{L_2}{R} \,, \quad \widetilde K^2 ~=~ \frac{L_1}{R} \,, \cr
\widetilde L_1 &=& R K_2 \,, \quad \widetilde L_2 ~=~ R\,  K_1   \,, \quad
\widetilde L_3 ~=~ -\frac{2\, M}{R} \,, \quad \widetilde M ~=~ -\coeff{1}{2}\, R\,  L_3 
\label{eq:specinv5d}
\end{eqnarray}
where any $\psi$-dependence is converted to $v$-dependence in accordance with (\ref{eq:specinvdef-2}). Observe, in particular, that if the $L_I$ have some non-trivial $\psi$-dependence, then $\widetilde K^1$, $\widetilde K^2$ and $\widetilde L_3$ and hence $\widetilde{\mathcal{F}}$ inherit a non-trivial $v$-dependence.  Thus the new solution involves a momentum wave and carries a momentum charge.  We now implement this general idea in a specific explicit construction.

\subsection{Spectral interchange: An example}
\label{ss:SIexample}

Our goal it to obtain a supertube with a magnetic dipole, $k$, and generic momentum densities and we will do this via spectral interchange.

Performing spectral interchange on the round $k$-wound supertube \eq{STsol1}, combined with a gauge transformation with parameters
\begin{equation}
\alpha^1 \,=\, -\frac{\bar{Q}_2}{k R} \,, \qquad   \alpha^2 \,=\, -\frac{\bar{Q}_1}{k R} \,, \qquad   \alpha^3 ~=~ 0\,,
\qquad\quad \bar{Q}_i ~\equiv~ \frac{Q_i}{4} \,, \quad i=1,2 \,,
\label{eq:gauge-trans}
\end{equation}
results in a solution specified by the harmonic functions
\begin{align}
V &~=~ \frac{k}{\rplus} \,, \qquad 
K^1 ~=~ K^2 ~=~  \frac{1}{R}\,, \qquad
K^3  ~=~  \frac{R}{\rminus}  \,,\label{ConstSeedSol1} \\
L_1 &~=~ \frac{\bar{Q}_1}{k} \frac{1}{\rminus} \,, \qquad
L_2 ~=~  \frac{\bar{Q}_2}{k} \frac{1}{\rminus} \,,  \qquad
L_3 ~=~  \left( k + \frac{\bar{Q}_1+\bar{Q}_2}{k R^2} \right)  \,,\label{ConstSeedSol2} \\
M &~=~ - \frac{1}{2} R
~+~  \frac12 \frac{\bar{Q}_1\bar{Q}_2}{k^2R} \frac{1}{\rminus} \,.
\label{ConstSeedSol3}
\end{align}
This solution describes a supertube that is singly-wound, in a base space which is $\mR^4/\mZ_k$.
The spectral interchange has thus had the effect of exchanging the original $\mZ_k$ orbifold at the location of the supertube for a  $\mZ_k$ orbifold at the center of space, and the original smooth center of space has become the location of a singly-wound supertube.

On this supertube in the spectrally-inverted frame, we introduce charge densities as studied in~\cite{Bena:2010gg},
\begin{align}
V &~=~ \frac{k}{\rplus} \,, \qquad 
K^1 ~=~ K^2 ~=~  \frac{1}{R}\,, \qquad
K^3  ~=~  \frac{R}{\rminus}  \,,\label{SeedSol1} \\
L_1 &~=~ \frac{\bar{Q}_1}{k}\, \lambda_1(\psi,\vec{y})  \,, \qquad
L_2 ~=~  \frac{\bar{Q}_2}{k} \, \lambda_2(\psi,\vec{y}) \,,  \qquad
L_3 ~=~  \left( k + \frac{\bar{Q}_1+\bar{Q}_2}{k R^2} \right)  \,,\label{SeedSol2} \\
M &~=~ - \frac{1}{2} R
~+~  \frac12 \frac{\bar{Q}_1\bar{Q}_2}{k^2R} \,  j(\psi,\vec{y}) \,,
\label{SeedSol3}
\end{align}
where the $\lambda_A$ and $j$ are harmonic functions on $\IR^4$ written as a Gibbons-Hawking space, and are sourced by normalized densities $\rhoone$, $\rhotwo$, and   $\rhohat$ localized at the supertube location $\vec{y}=\vec{y}_{-}$, that is $\rho_-=0$ or $(y_1=0, y_2=0, y_3=-\frac{\ell}{2})$:
\begin{eqnarray}
\lambda_A(\psi,\vec{y})  ~=~  4\pi \int\!d^3 y' \int_0^{4\pi} \!d \psi' \; 
\widehat G(\psi,\vec{y}; \,  \psi',\vec{y}\;\!')\,  \rhoA (\psi' - k \phi') \delta^3(\vec{y}\,'-\vec{y}_{-})  \,, \nonumber \\
  j(\psi,\vec{y})  ~=~  4\pi  \int\!d^3 y' \int_0^{4\pi} \!d \psi' \; 
\widehat G(\psi,\vec{y}; \,  \psi',\vec{y}\;\!')\,  \rhohat (\psi' - k \phi') \delta^3(\vec{y}\,'-\vec{y}_{-})   \,.
\label{wigglybits-2}
\end{eqnarray}
The dependence of the densities on the combination of angles $\psi - k \phi$ will become clear when we use the Green function on $\IR^4/\mZ_k$ in the next subsection to construct explicit solutions. For now, we keep the discussion general to explain our overall strategy.

We now transform back to the original supertube frame, first performing the inverse gauge transformation to \eq{eq:gauge-trans} and then performing spectral inversion.
This results in the new BPS solution:
\begin{align}
V &~=~  \frac{1}{\rminus} \,, \qquad
K_1 ~=~ \frac{\bar{Q}_2}{k R}\,  \Big( \lambda_2(-\tfrac{v}{R},\vec{y}) ~-~ \frac{1}{\rminus} \Big) \,, \qquad
K_2 ~=~ \frac{\bar{Q}_1}{k R}\, \Big( \lambda_1(-\tfrac{v}{R},\vec{y}) ~-~  \frac{1}{\rminus}\Big)  \,,  \label{SIresult1}\\
K_3 &~=~ \frac{ k R}{\rplus}  \,,  \qquad
L_1  ~=~ 1~+~ \frac{\bar{Q}_1}{\rplus} \,, \qquad
L_2 ~=~  1~+~ \frac{\bar{Q}_2}{\rplus} \,, \label{SIresult2}\\
L_3 & ~=~  1 ~-~ \frac{\bar{Q}_1\bar{Q}_2}{(k R)^2} \, 
\left( j(-\tfrac{v}{R},\vec{y}) ~-~ \lambda_1(-\tfrac{v}{R},\vec{y}) - \lambda_2(-\tfrac{v}{R},\vec{y}) +  \frac{1}{\rminus}\right)  \,, \label{SIresult3}\\
M &~=~  -\frac{k R}{2} ~+~ \frac{1}{2} \frac{\bar{Q}_1\bar{Q}_2}{k R}\, \frac{1}{\rplus} \,.
\label{SIresult4}
\end{align}

The form of $V$ means that the base, $\cB$, has returned to flat $\IR^4$.  There is a supertube with a dipole charge $k$ (corresponding to a pole in $K^3$), and charges $\bar{Q}_A$ located at $\rplus =0$.  
In addition, the harmonic functions $\lambda_A$ and $j$ describe a momentum wave along the $v$ direction that is sourced at $\rminus  =0$.  We have therefore succeeded in adding momentum to a standard two charge supertube solution.  
 
Spectral interchange is simply a global diffeomorphism and so regularity conditions can be imposed on the supertube in the spectral-inverted frame.  Before we do this, one should note that in the original seed solution  (\ref{SeedSol1})--(\ref{SeedSol3}), the parameters, $\bar{Q}_A$, could  be absorbed into the normalization of the charge densities, $\rhoA$ and $\rhohat$.  We are therefore free to adjust them in some convenient manner and we choose to impose the constraint:
\begin{equation}
\ell ~=~ \frac{\bar{Q}_1\bar{Q}_2}{(k R)^2} \,.
\label{basicreg}
\end{equation}
As we will see, this choice will mean that one of the supertube regularity conditions is automatically satisfied for $\rhoA = \rhohat =1$.
 
Supertube regularity with varying charge density was extensively studied  in \cite{Bena:2010gg} (following~\cite{Bena:2008dw}) where it was shown that the supertube (\ref{SeedSol1})--(\ref{SeedSol3}) is regular if one imposes the following functional constraints at each point of the GH fiber:
\begin{equation}
\lim_{\rminus \to 0}\,  \rminus \,  \left[ V  \mu ~-~ Z_3  K^3 \right] ~=~ 0  \,, \label{regcondb}
\end{equation}
\begin{equation}
\lim_{\rminus \to 0}\,  \rminus^2 \, \left[  V Z_1 Z_2  ~-~ Z_3 (K^3)^2  \right] ~=~ 0  \,.
\label{regcondc}
\end{equation}

Using (\ref{basicreg}), the  first equation can be reduced to
\begin{equation}
{k R} \left( \rhohat - 1 \right)  ~+~ \frac{1}{k R} \left[ \bar Q_1 \left( \rhoone - 1 \right) + \bar Q_2 \left( \rhotwo - 1 \right) \right] ~=~ 0 \,,
 \label{regcondd}
\end{equation}
where $\rhoA$ and $\rhohat$ are defined in (\ref{wigglybits-2}).  The second regularity condition  (\ref{regcondc}), when combined with (\ref{regcondd}) reduces to a simple, local constraint on the charge densities~\cite{Bena:2010gg},
\begin{equation}
 \label{regconde}
\rhohat ~=~ \rhoone \;\! \rhotwo \,.
\end{equation}

The regularity conditions (\ref{regcondd}) and (\ref{regconde}) can be thought of as ``coiffuring'' the charge densities so as to achieve regularity.  One should note that while one can certainly satisfy (\ref{regcondd}) using finite sets of Fourier modes, the charge density condition, (\ref{regconde}), generically requires one of the Fourier series to be infinite.  As we will see below, coiffuring and the holographic interpretation of the modes is somewhat simpler if one switches on $(Z_4, \Theta_4)$. One could repeat the foregoing analysis by introducing an additional charge density $\rhofour$, however for ease of presentation we will continue without introducing $\rhofour$ explicitly, and introduce $(Z_4, \Theta_4)$ in Section \ref{Sect:MomST}.

\subsection{The Green function and mode expansions on an  $\mR^4/\mZ_k$  base}
\label{ss:GreenModes}

To construct explicit solutions of the form (\ref{SeedSol1})--(\ref{SeedSol3}), we need the scalar Green function for a GH base space with $V = \frac{k}{\rplus}$ and with a source located at $\rminus =0$.
It is straightforward to obtain this via a coordinate transformation of the standard flat  $\IR^4$ Green function, or one can use  the general result of Page \cite{Page:1979ga}.  
One finds that the Green function for the response at the point $(\psi, \vec{y})$ caused by a source at the point $(\psi',\vec{y}\,'=\vec{y}_{-})$ defined by $\rminus =0$ is:
\begin{equation}
\widehat G(\psi,\vec{y};\psi',\vec{y}\;\!')~=~{1\over 16\pi^2 \rminus} {\sinh\Bigl[ {k \over 2} \log
{ \rplus +\ell + \rminus \over  \rplus +\ell - \rminus}\Bigr]\over 
\cosh \Bigl[{k \over 2} \log
{ \rplus +\ell + \rminus \over  \rplus +\ell - \rminus}\Bigr] 
-\cos {\left[ \tfrac12(\psi-\psi') - \tfrac{k}{2} (\phi-\phi') \right]}}\,.
\label{PageGF}
\end{equation}
Note that this function depends upon the combination of angular coordinates:
\begin{equation}
\psi - k \phi\,.
\end{equation}
This should not be surprising because the GH fiber is defined by $(d \psi +A)$ and, at $\rminus =0$, this becomes $(d \psi - k d \phi)$.  Thus the charge density functions and solutions will depend upon precisely this mixture of angles, explaining the form of Eq.\;\eq{wigglybits-2}.
If one expands the charge densities into Fourier modes,
\begin{equation}
\rhoA(\psi-k \phi) ~=~  \sum_n {b_{A,n} \,  e^{i \frac{n}{2}(\psi-k\phi) }} \,, 
\end{equation}
then the solutions are elementary to obtain from the Green function using contour integration (see for example \cite{Bena:2013ora}):
\begin{equation}
\lambda_A(\psi, \vec{y}) ~=~ \sum_n {\frac{b_{A,n}}{\rminus}
\left[
\left( \frac{\rplus - \rminus +\ell}{\rplus + \rminus +\ell}
\right)^{\frac{k}{2}}
 e^{\frac{i}{2}(\psi-k \phi) }
\right]^{n} }
~\equiv~ \sum_n \, {\frac{b_{A,n}}{\rminus} \, \hat{F}^n} \,
\label{eq:lambda-sol}
\end{equation}
where $\hat{F}^n$ is defined through the above equation.
Similarly, for $j$ we have  
\begin{equation}
\rhohat(\psi-k\phi) ~=~ \sum_n \,  {\hat{b}_{n} \, e^{i \frac{n}{2}(\psi-k\phi) }} \,, \qquad j(\psi, \vec{\rminus}) ~=~ \sum_n \, {\frac{\hat{b}_{n}}{\rminus} \, \hat{F}^n} \,.
\label{eq:jsol}
\end{equation}
Note that in the limit $\rminus \to 0$ these reduce to the following simple forms: 
\begin{equation}
\lambda_A(\psi, \vec{y}) ~\to~\sum_n {\frac{b_{A,n}}{\rminus} \, 
 e^{i\frac{n}{2}(\psi-k\phi) }}
~=~ \frac{\rhoA(\psi-k\phi)}{\rminus} \,, \qquad
j(\psi, \vec{y}) ~\to~ \frac{\rhohat(\psi-k\phi)}{\rminus} \,. \label{eq:r0limit}
\end{equation}
Introducing spherical polar coordinates $(\rho,\vartheta,\phi)$ centered at the origin (halfway between the supertube and the GH center), we observe that for $\rho \gg \ell$,
\begin{equation}
\rpm ~\simeq ~ \rho \left(1 ~\mp~ \frac{\ell}{2\, \rho}\cos\vartheta \right) \,.
\end{equation}
This means that $\hat F^n$ falls off as $\rho^{-{\frac{k n}{2}}}$ at large $\rho$:
\begin{equation}
\hat F^n ~\sim~  \left( \frac{\ell(1-\cos\vartheta)}{2\, \rho} \right)^{\frac{k n}{2}} e^{\frac{in}{2}(\psi-k\phi)} 
\end{equation}
and so higher orbifolds lead to more rapid fall-off at infinity.

When we come to imposing regularity constraints, we will find it useful to introduce non-zero $(Z_4, \Theta_4)$. In principle one could repeat the above analysis with an additional density profile function $\rhofour$ and analyze the modified supertube regularity conditions in the spectral inverted frame.
Rather than pursue this route, we will find it more convenient to perform a direct analysis of the BPS equations 
using the techniques of~\cite{Bena:2015bea} to construct our explicit solutions.
This will lead to the complete solution in a manner that is well-adapted to coiffuring and holography.

\section{Adding momentum to the supertube}
\label{Sect:MomST}

As we have seen, adding momentum to a supertube naturally leads us to consider $v$-dependent fluctuations.  We now do this by generalizing the circular supertube seed solution described in Section  \ref{ss:rndstube}.  In this way we will also obtain the complete solution including all components of the angular-momentum vector.

A natural way to construct $v$-dependent solutions is to introduce fluctuating charge-density sources along the $v$-fiber above the center of space, $r=0, \theta =0$ or $\rminus =0$, as described in \cite{Niehoff:2013kia}.  Indeed, the $\psi$-fiber pinches off at the center of space while the $v$-fiber remains finite: 
\begin{equation} 
 ( dv + \beta) ~\to~  ( dv - 2 k R \, d\varphi_2)  \,.
\label{vfiber1}
\end{equation} 
This means that a single-valued source introduced along the $v$-fiber must have a Fourier expansion with the following dependence:
\begin{equation} 
e^{-i p(\frac{v}{2 R} - k \varphi_2)}\,, \qquad p \in \ZZ \,.
\label{Fourier1}
\end{equation} 
We will therefore seek solutions based upon these Fourier modes. Thus we define the phase:
\begin{equation} 
\zeta ~=~ \frac{v}{2 R} - k \varphi_2 \,.
\label{zetadefn}
\end{equation} 
%

\subsection{The first layer of equations}
\label{ss:Layer1}

Based upon the form of Eqs.\;\eq{SIresult4}, \eq{eq:lambda-sol} and \eq{eq:jsol} and the results of \cite{Niehoff:2013kia,Shigemori:2013lta,Giusto:2013bda,Bena:2015bea} it is not hard to infer a solution to the first layer of BPS equations. Define 
\begin{equation} 
\Delta ~\equiv~  \frac{a \, \cos  \theta}{(r^2 + a^2)^\frac{1}{2}} \,,
\label{Deltadefn}
\end{equation} 
then a somewhat lengthy computation shows that the following fields satisfy the first layer of equations (\ref{BPSlayer1a})--(\ref{BPSlayer1c}) for some complex Fourier coefficients, $b_1$ and $b_2$:
\begin{align} 
Z_A &~=~ 1 ~+~ \frac{Q_A}{\Sigma} \, \big(1 + \Delta^{k n} \,( b_A \, e^{-i n \zeta}    + \bar b_A \, e^{i n \zeta}) \big) \,, \quad  A= 1,2  \,, \label{Z12osc1} \\
\Theta_1 &~=~ -\frac{n\, Q_2 }{2 \, R} \,  \Delta^{k n} \,\Big[  \, b_2\, e^{-i n \zeta} \,(\Omega^{(2)} + i r \sin \theta \, \Omega^{(1)}) +  \bar b_2\, e^{i n \zeta} \,(\Omega^{(2)} - i r \sin \theta\,  \Omega^{(1)}) \Big] \,,\label{T1osc1}\\
\Theta_2 &~=~ -\frac{n\, Q_1 }{2 \, R} \,  \Delta^{k n} \,\Big[  \, b_1\, e^{-i n \zeta} \,(\Omega^{(2)} + i r \sin \theta \, \Omega^{(1)}) +  \bar b_1\, e^{i n \zeta} \,(\Omega^{(2)} - i r \sin \theta\,  \Omega^{(1)}) \Big] \,,\label{T2osc1}
\end{align} 
To these fields one can add a completely independent, purely oscillating set of modes for $(Z_4, \Theta_4)$:
\begin{align} 
Z_4 &~= \frac{\Delta^{k p}}{\Sigma} \,  ( b_4\, e^{-i p \zeta}    + \bar b_4 \, e^{i p \zeta})  \,,  \label{Z4osc1} \\
\Theta_4 &~=~ -\frac{p}{2 \, R} \,  \Delta^{k p} \,\Big[  \, b_4\, e^{-i p \zeta} \,(\Omega^{(2)} + i r \sin \theta \, \Omega^{(1)}) +  \bar b_4\, e^{i p \zeta} \,(\Omega^{(2)} - i r \sin \theta\,  \Omega^{(1)}) \Big] \,.  \label{T4osc1} 
\end{align} 
Regularity of the metric and dilaton factors mean that one should have $Z_A >0$ for $A=1,2$.  This means that the terms in the parentheses in  (\ref{Z12osc1}) must be strictly positive and since $|\Delta| < 1$ away from the source, one can certainly guarantee $Z_A >0$ by taking:
\begin{equation} 
|b_A|  ~\le~\frac{1}{2}\,, \qquad A=1,2  \,.  \label{babound} 
\end{equation} 
One may be able to improve this bound slightly, but the important point is that $|b_A|$ will always be bounded by a number of order $1$.
  
\subsection{The second layer of equations}
\label{ss:Layer2}

Consider a single mode of $\omega$ and $\mathcal{F}$: 
\begin{equation}
\omega ~=~ e^{- i q \zeta}(\hat \omega_r dr + \hat \omega_\theta d \theta + \hat \omega_1 d \varphi_1 +\hat \omega_2 d \varphi_2)  \,, \qquad \mathcal{F} =~ - W \;\! e^{- i q \zeta}
\end{equation}
then the differential operators in (\ref{BPSlayer2a}) and  (\ref{BPSlayer2b}) may be written as:
\begin{align}
&  \mathcal{D}\omega   +*_4 \mathcal{D}\omega  +  {\mathcal{F}}\, \Theta_3  \nonumber \\ 
& \qquad \qquad~\equiv~ e^{- i q \zeta}  \Big[ (r^2 + a^2)  \cos \theta \,\Omega^{(1)}\,\mathcal{L}^{(q)}_1 +r \sin \theta \,\Omega^{(3)}\,\mathcal{L}^{(q)}_3   + \frac{(r^2 + a^2)}{r}  \cos \theta \,  \Omega^{(2)}\,\mathcal{L}^{(q)}_2 \Big] \,,   \\
&*_4\mathcal{D} *_4  \Big(\mathcal{D} \mathcal{F}    - 2\,\dot{\omega} \Big)     ~\equiv~     e^{- i q \zeta}  \Big[ \,   \widehat \cL^{(q)} \, W - \frac{i \, q}{R}\,\mathcal{L}^{(q)}_0 \Big] \,, 
\end{align}
where we define
\begin{align}
 \mathcal{L}^{(q)}_0 & \equiv  \frac{1}{\Sigma}\,\Big[\frac{1}{r}\,\partial_r (r \,(r^2+a^2)\, \hat \omega_r) + \frac{1}{\sin\theta\cos\theta}\,\partial_\theta(\sin\theta\cos\theta\,\hat \omega_\theta)+ \frac{i \, k q \, a^2 }{(r^2 + a^2) }\, \hat \omega_1 +  \frac{i \, k q  }{\cos^2\theta}\, \hat \omega_2\Big] \,, \\
\mathcal{L}^{(q)}_1 &\equiv (\partial_r\hat \omega_\theta-\partial_\theta\hat \omega_r)-
\frac{i\, k q}{r(r^2+a^2) \sin\theta\, \cos\theta} 
\,(r^2 \hat \omega_1 - a^2 \sin^2\theta  \,\hat \omega_2)  \,,\\
\mathcal{L}^{(q)}_2 &\equiv \frac{1}{\cos \theta}\, \,\partial_r\hat \omega_2  +   \frac{r}{(r^2 +a^2) \sin\theta}  \,\partial_\theta\hat \omega_1 - \frac{i \, k q \, r^2 }{\Sigma   \cos \theta}\,\hat \omega_r   - \frac{i \, k q  a^2 r\, \sin \theta }{\Sigma \,(r^2 +a^2)}\,\hat \omega_\theta 
- \frac{4 k R a^2 r \cos \theta}{\Sigma^2} \, W \,,\\
\mathcal{L}^{(q)}_3 &\equiv \frac{1}{\sin \theta}\,\partial_r\hat \omega_1-\frac{1}{r \cos \theta}\,\partial_\theta\hat \omega_2 - \frac{i\, k q}{\Sigma  \cos\theta\, } \, (a^2 \sin\theta \cos\theta\,   \hat \omega_r    - r\,\hat \omega_\theta) 
+ \frac{4 k R a^2 r \sin \theta}{\Sigma^2} \, W  \,,\\
   \widehat \cL^{(q)} W  & \equiv  \frac{1}{\Sigma}\,\Big[\frac{1}{r}\,\partial_r (r \,(r^2+a^2)\, \partial_r W) + \frac{1}{\sin\theta\cos\theta}\,\partial_\theta(\sin\theta\cos\theta\,  \partial_\theta W ) -  \frac{k^2 q^2 (r^2 + a^2 \sin^2 \theta)}{(r^2 + a^2)\cos^2 \theta }\, W\Big] \,.
\end{align}

Using the solutions in (\ref{Z12osc1})--(\ref{T4osc1}), the source terms in  (\ref{BPSlayer2a}) and  (\ref{BPSlayer2b})   give rise, {\it a priori},  to four non-trivial kinds of source terms: 
\begin{itemize}
\setlength{\itemsep}{0.1\baselineskip}
\item[1(a).] Terms arising from products of modes with same phase.  These depend upon $e^{\pm 2 i n \zeta}$ and have singularities involving $\Sigma^{-2}$.
\item[1(b).] Terms arising from the product of a mode and a $\frac{Q_A}{\Sigma}$ term.  These depend upon $e^{\pm  i n \zeta}$ and have singularities involving $\Sigma^{-2}$.
\item[2.] Terms arising from the product of a mode and the constant ($1$) in $Z_A$.  These depend upon $e^{\pm  i n \zeta}$ and have singularities involving $\Sigma^{-1}$.
\item[3.] Terms arising from product of modes with the opposite  phase.  These are independent of $\zeta$ and have singularities involving $\Sigma^{-2}$.
\end{itemize}
However, the sources of types 1(a) and 1(b) are not really distinct in that the solution  is the same but simply with a different mode number.   We therefore break down the sources into types 1,2 and 3 and write the particular equations that need to be solved and find the particular solutions. 

These systems of equations are:

\bigskip
\leftline{\underline{\bf Source 1:}}
\begin{equation}
\begin{aligned}
\mathcal{L}^{(q)}_1 ~=~& - \frac{i\, q}{2\,R}\,\frac{\Delta^{k q} \,  r \sin\theta}{\Sigma (r^2 + a^2) \cos\theta} \,, \qquad 
 {
\mathcal{L}^{(q)}_3 ~=~0 \,, 
} \\
 {
\mathcal{L}^{(q)}_2 ~=~ } & - \frac{q}{2\,R}\,\frac{\Delta^{k q} \,  r }{\Sigma (r^2 + a^2) \cos\theta} \,,
\qquad 
\widehat \cL^{(q)} \, W - \frac{i \, q}{R}\,\mathcal{L}^{(q)}_0  ~=~  
\frac{q^2}{2\,R^2 }\,\frac{\Delta^{k q}}{\Sigma^2} \,,
\label{source1}
\end{aligned}
\end{equation}
\bigskip

\leftline{\underline{\bf Source 2:}}
\begin{equation}
\begin{aligned}
\mathcal{L}^{(q)}_1 ~=~& - \frac{i\, q}{2\,R}\,\frac{\Delta^{k q} \,  r \sin\theta}{ (r^2 + a^2) \cos\theta} \,, \qquad 
 {
\mathcal{L}^{(q)}_3 ~=~0 \,,  } \\
 {
\mathcal{L}^{(q)}_2  ~=~ } & 
- \frac{q}{2\,R}\,\frac{\Delta^{k q} \,  r }{ (r^2 + a^2) \cos\theta} \,, \qquad 
\widehat \cL^{(q)} \, W - \frac{i \, q}{R}\,\mathcal{L}^{(q)}_0  ~=~  
\frac{q^2}{2\,R^2 }\,\frac{\Delta^{k q}}{\Sigma} \,, 
\label{source2}
\end{aligned}
\end{equation}
\bigskip
\leftline{\underline{\bf Source 3:}}  
\begin{equation}
\begin{aligned}
\mathcal{L}^{(q=0)}_1 ~=~& 0  \,, \qquad\qquad\qquad\qquad\qquad\quad   
 {
\mathcal{L}^{(q=0)}_3 ~=~0 \,, } \\
 {
\mathcal{L}^{(q=0)}_2 ~=~ } & 
- \frac{m}{R}\,\frac{\Delta^{2mk} \,r}{\Sigma (r^2 + a^2) \cos\theta} \,, \qquad 
\widehat \cL^{(q=0)} \, W   ~=~  
 {
\frac{m^2}{R^2 }\,\frac{\Delta^{2km}}{\Sigma (r^2 + a^2) \cos^2\theta} \,, 
}
\label{source3}
\end{aligned}
\end{equation}

These equations have a gauge invariance  associated with changing the $u$-coordinate:
\begin{equation}\label{gaugeu}
u\to u + f(x^i,v)\,,\quad \omega \to \omega - df + \beta \, \partial_v{f} \,,\quad {\mathcal{F}} \to {\mathcal{F}} - 2\, \partial_v{f}\,.
\end{equation}
In terms of the $q^{\rm th}$ mode this becomes 
\begin{equation}
 (\hat \omega_r, \hat \omega_\theta, \hat \omega_1, \hat \omega_2 ; W) \to  (\hat \omega_r, \hat \omega_\theta, \hat \omega_1, \hat \omega_2 ; W)  +  (\partial_r h, \partial_\theta h, \frac{i\,k q\, a^2 \sin^2 \theta}{\Sigma}\, h , \frac{i\,k q\, r^2 }{\Sigma} \, h ; \frac{i\, q\,}{R} \, h) \,,
 \label{gaugeh}
\end{equation}
for an arbitrary function $h(x^i)$ on the base, $\cB$.  In particular, for $q\ne 0$ one can choose a gauge with $W=0$.

It is relatively easy to find the explicit solutions for each of these sources:

\bigskip
\leftline{\underline{\bf Solution for Source 1:}}
\begin{equation}
 (\hat \omega_r, \hat \omega_\theta, \hat \omega_1, \hat \omega_2 ; W)  ~=~ \frac{\Delta^{k q}}{4 \,k R}\, \Big(- \frac{i }{r (r^2 + a^2) }, 0,  \frac{\sin^2 \theta}{\Sigma}  ,\frac{\cos^2 \theta}{\Sigma} ; 0 \Big)  \,, 
\label{solution1}
\end{equation}
\bigskip
\leftline{\underline{\bf Solution for Source 2:}}
\begin{equation}
 (\hat \omega_r, \hat \omega_\theta, \hat \omega_1, \hat \omega_2 ; W)  ~=~ \frac{\Delta^{k q}}{4 \,k R}\, \Big(- \frac{i }{r}, i \tan \theta,  0 ,1 ; 0 \Big)  \,, 
\label{solution2}
\end{equation}
\bigskip
\leftline{\underline{\bf Solution for Source 3:}}
\begin{align}
 \hat \omega_r ~=~ & \hat \omega_\theta  ~=~ 0 \,, \qquad W~=~ - \frac{1}{4 \,k^2 R^2}\,\frac{1}{(r^2 + a^2 \sin^2 \theta)} \, \big(1 -  \Delta^{2 k m}\big)\,, \label{solution3a} \\
\hat \omega_1 ~=~ &\frac{1}{2 \,k  R }\,\frac{(r^2 + a^2)}{\Sigma} \, \bigg( \frac{\big( \Delta^{2 k m}-1\big)\sin^2 \theta}{(r^2 + a^2 \sin^2 \theta)} ~+~ \frac{1}{a^2} \bigg) + \frac{\widehat J}{a^2  }\,\frac{(r^2 + a^2)}{\Sigma}  + \hat c_1 \,, \label{solution3b} \\
\hat \omega_2 ~=~ &\frac{1}{2 \,k  R }\,\frac{r^2}{\Sigma} \, \bigg( \frac{\big( \Delta^{2 k m}-1\big)\cos^2 \theta}{(r^2 + a^2 \sin^2 \theta)} ~-~ \frac{1}{a^2} \bigg) - \frac{\widehat J}{a^2  }\,\frac{r^2}{\Sigma}  + \hat c_2 \,, \label{solution3c}
\end{align}
where $\widehat J$, $\hat c_1$ and $\hat c_2$ are constants to be determined.

\subsection{The complete angular momentum vector}
\label{ss:compomega}

Writing the components of $\omega$ so as to include all the phases:
\begin{equation}
\omega ~=~ ( \omega_r dr +  \omega_\theta d \theta +  \omega_1 d \varphi_1 + \omega_2 d \varphi_2) \,,
\end{equation}
putting together all the source terms and, for the moment setting $b_4=0$, we find:
\begin{align}
 \omega_r ~=~ & - \frac{i \, Q_1 Q_2}{4 \,k R}\,  \frac{\Delta^{2 k n}}{r (r^2 + a^2) }  \, \big(  b_1 b_2 \, e^{- 2 i n \zeta} -  \bar b_1 \bar b_2 \,  e^{2 i n \zeta} \big) \nonumber \\ 
  & - \frac{i}{4 \,k R}\,  \frac{\Delta^{ k n}}{r (r^2 + a^2) }  \, \big[ \big((b_1 + b_2) Q_1 Q_2 + (r^2 + a^2)(b_1 Q_1+ b_2 Q_2)  \big) e^{-  i n \zeta} \nonumber\\
&-   \big((\bar b_1 + \bar b_2) Q_1 Q_2 + (r^2 + a^2)(\bar b_1 Q_1+ \bar b_2 Q_2)  \big) e^{  i n \zeta}   \big]  \qquad \quad \label{omrsol1} \\
 \omega_\theta  ~=~ &  \frac{i \, \Delta^{ k n} }{4 \,k R}\,\tan \theta \, \big( (b_1 Q_1+  b_2 Q_2)   e^{-  i n \zeta}  -  (\bar b_1 Q_1+ \bar b_2 Q_2)   e^{  i n \zeta} \big) \,, \label{omthsol1} \\
\omega_1 ~=~ & \frac{ Q_1 Q_2}{4 \,k R}\,  \frac{\Delta^{2 k n} \sin^2\theta}{\Sigma }  \, \big(  b_1 b_2 \, e^{- 2 i n \zeta} +  \bar b_1 \bar b_2 \,  e^{2 i n \zeta} \big) \nonumber \\ 
& + \frac{ Q_1 Q_2}{4 \,k R}\,  \frac{\Delta^{k n} \sin^2\theta}{\Sigma }  \, \big(  (b_1 + b_2) \, e^{-  i n \zeta} +  (\bar b_1 + \bar b_2)\,  e^{ i n \zeta} \big) \nonumber \\   
& + \frac{Q_1 Q_2}{2 \,k  R }( b_1 \bar b_2  +   b_2 \bar b_1 ) \,\frac{(r^2 + a^2)}{\Sigma} \,  \bigg( \frac{\big( \Delta^{2 k n}-1\big)\sin^2 \theta}{(r^2 + a^2 \sin^2 \theta)} +  \frac{1}{a^2} \bigg)   ~+~ \frac{J}{a^2  }\,\frac{(r^2 + a^2)}{\Sigma}   +  c_1  \,,\label{om1sol1}  
\end{align}
\begin{align}
\omega_2 ~=~ & \frac{ Q_1 Q_2}{4 \,k R}\,  \frac{\Delta^{2 k n} \cos^2\theta}{\Sigma }  \, \big(  b_1 b_2 \, e^{- 2 i n \zeta} +  \bar b_1 \bar b_2 \,  e^{2 i n \zeta} \big) \nonumber \\ 
& + \frac{ Q_1 Q_2}{4 \,k R}\,  \frac{\Delta^{k n} \cos^2\theta}{\Sigma }  \, \big(  (b_1 + b_2) \, e^{-  i n \zeta} +  (\bar b_1 + \bar b_2)\,  e^{ i n \zeta} \big) \nonumber \\  
& + \frac{ \Delta^{k n} }{4 \,k R}\,  \big(  (b_1 Q_1+ b_2 Q_2) \, e^{-  i n \zeta} +  (\bar b_1 Q_1 + \bar b_2 Q_2)\,  e^{ i n \zeta} \big) \nonumber \\     
& + \frac{Q_1 Q_2}{2 \,k  R }( b_1 \bar b_2  +   b_2 \bar b_1 ) \,\frac{r^2 }{\Sigma} \,  \bigg( \frac{\big( \Delta^{2 k n}-1\big)\cos^2 \theta}{(r^2 + a^2 \sin^2 \theta)} -  \frac{1}{a^2} \bigg) ~-~ \frac{J}{a^2  }\,\frac{r^2 }{\Sigma} + c_2  \,,\label{om2sol1}  \\
\mathcal{F}~=~ &  \frac{Q_1 Q_2}{4 \,k^2 R^2}\,( b_1 \bar b_2  +   b_2 \bar b_1 ) \, \frac{1}{(r^2 + a^2 \sin^2 \theta)} \, \big(1 -  \Delta^{2 k n}\big)\,, \label{omZ3sol1}
\end{align}
where $J$, $c_1$ and $c_2$ are constants to be determined.  Note that  $\mathcal{F}$  vanishes at infinity.

The solutions for $(Z_4,\Theta_4)$ will be allowed to have different moding from $(Z_A,\Theta_B)$, where $\{A,B\} = \{1,2\}$.  Using  (\ref{Z4osc1})  and (\ref{T4osc1}) and the solutions for ``Source 3,'' we find
\begin{align}
 \omega_r ~=~ &   \frac{i }{4 \,k R}\,  \frac{\Delta^{2 k p}}{r (r^2 + a^2) }  \, \big(  b_4^2  \, e^{- 2 i p \zeta} -  \bar  b_4^2 \,  e^{2 i p \zeta} \big) \,, \qquad \omega_\theta  ~=~  0 \,, \label{omrthsol2} \\
\omega_1 ~=~ &- \frac{1}{4 \,k R}\,  \frac{\Delta^{2 k p} \sin^2\theta}{\Sigma }  \, \big(  b_4^2 \, e^{- 2 i n \zeta} +  \bar b_4^2   \,  e^{2 i n \zeta} \big) - \frac{|b_4|^2 }{k  R }\,\frac{(r^2 + a^2)}{\Sigma} \,  \bigg( \frac{\big( \Delta^{2 k n}-1\big)\sin^2 \theta}{(r^2 + a^2 \sin^2 \theta)} +  \frac{1}{a^2} \bigg)    \,,\label{om1sol2}  \\
\omega_2 ~=~ & - \frac{ 1}{4 \,k R}\,  \frac{\Delta^{2 k p} \cos^2\theta}{\Sigma }  \, \big(  b_4^2  \, e^{- 2 i p \zeta} +  \bar  b_4^2 \,  e^{2 i p \zeta} \big)  - \frac{|b_4|^2}{k  R } \,\frac{r^2 }{\Sigma} \,  \bigg( \frac{\big( \Delta^{2 k p}-1\big)\cos^2 \theta}{(r^2 + a^2 \sin^2 \theta)} -  \frac{1}{a^2} \bigg)  \,,\label{om2sol2}  \\
\mathcal{F}~=~ & {}- \frac{|b_4|^2}{2 \,k^2 R^2}\, \frac{1}{(r^2 + a^2 \sin^2 \theta)} \, \big(1 -  \Delta^{2 k p}\big)\,. \label{omZ3sol2}
\end{align}

One should note that these solutions are singular: $\omega_r$ diverges at $r=0$.  We therefore need to smooth these solutions out by adjusting the Fourier coefficients appropriately. 

\subsection{Coiffuring and regularity}
\label{ss:STReg}

As we have discussed in Section \ref{ss:SIflip}, the core  of this solution can be obtained  via spectral interchange \cite{Niehoff:2013kia}.   Moreover, supertube regularity requires that the charge density functions  satisfy (\ref{regcondd}) and (\ref{regconde}).    The important point here this the (\ref{regconde}) may be viewed as the continuum analog of (\ref{STreg0}) and, as such, determines $\rhohat$ in terms of $\rhoone$ and $\rhotwo$ This can easily be implemented explicitly in a finite Fourier expansion.   On the other hand   (\ref{regcondd}) and (\ref{regconde}) together mean that  one cannot have a regular solution that involves {\it finite} Fourier series for both $\rhoone$ and $\rhotwo$:  Regularity with only two fluctuating charge densities means (at least) one of the two Fourier series must be infinite.  Since the solution in this paper is the spectral interchange of such a charge density fluctuation, the conclusion will be exactly the same.

In \cite{Bena:2015bea}, regularity was achieved in a different manner: If one introduces one more charge species one cancels the singular terms between the species in a process that is known as ``coiffuring'' \cite{Mathur:2013nja,Bena:2014rea,Bena:2013ora}.  In the discussion above, the addition of the additional species is represented by a new charge density, $\rhofour$, and replaces the $\rhoone \rhotwo$ terms by $\rhoone \rhotwo-  \rhofour^2$; one can cancel the problematic quadratic terms and achieve regularity by simple linear constraints on Fourier coefficients and this can be implemented in a finite Fourier series.   Thus coiffuring simply represents a very convenient way to solve the standard supertube regularity conditions using finite Fourier expansions.

At a practical level, our problem is simply to cancel all the $\frac{1}{r}$ singularities in $\omega_r$ and there are two natural ways to achieve this.


\refstepcounter{subsubsection}
\subsubsection*{\thesubsubsection ~~ Coiffuring:  Style 1}
\label{ss:Coiffure1}

The first is to give $(Z_4,\Theta_4)$ the same mode-dependence as the $(Z_A,\Theta_B)$.  That is, to take $p=n$ in Eqs.\;(\ref{Z12osc1})--(\ref{T4osc1}) and then combine the corresponding contributions to $\omega$ and $\mathcal{F}$. 

The singular terms that depend on $e^{\pm 2 i n \zeta}$ are then cancelled by setting
\begin{equation}
b_4^2 ~=~  Q_1 Q_2 \, b_1 b_2   \,.
\label{breln1}
\end{equation}
There are still singular terms that depend upon $e^{\pm  i n \zeta}$ and these can be cancelled (at $r=0$) by setting
\begin{equation}
(b_1 + b_2) \, Q_1 Q_2 ~+~ a^2 \, (b_1 Q_1+ b_2 Q_2)  ~=~  0    \,.
\label{breln2}
\end{equation}
Eliminating $b_1$ and $b_2$ in terms of $b_4$ gives
\begin{equation}
b_1 ~=~  \frac{i\,b_4}{Q_1} \sqrt{ \frac{Q_1 + a^2}{Q_2 + a^2}} \,, \qquad   b_2 ~=~ - \frac{i\,b_4}{Q_2} \sqrt{ \frac{Q_2 + a^2}{Q_1 + a^2}}  \,.
\label{breln0}
\end{equation}
The solution for $\mathcal{F}$ and $\omega$ then reduces to:
\begin{align}
 \omega_r ~=~ &   - \frac{i}{4 \,k R}\,  \frac{r \, \Delta^{ k n}}{(r^2 + a^2) }  \, \big[(b_1 Q_1+ b_2 Q_2) e^{-  i n \zeta}  -   (\bar b_1 Q_1+ \bar b_2 Q_2)e^{  i n \zeta}   \big]  \qquad \quad \label{omrcoiff1}  \\
 \omega_\theta  ~=~ &  \frac{i \, \Delta^{ k n} }{4 \,k R}\, \tan \theta \,   \big[ (b_1 Q_1+  b_2 Q_2)   e^{-  i n \zeta}  -  (\bar b_1 Q_1+ \bar b_2 Q_2)   e^{  i n \zeta} \big] \,, \label{omthcoiff1} \\
\omega_1 ~=~  & - \frac{a^2}{4 \,k R}\,  \frac{\Delta^{k n} \sin^2\theta}{\Sigma }  \, \big[ (b_1 Q_1+ b_2 Q_2) \, e^{-  i n \zeta} + (\bar b_1 Q_1+ \bar b_2 Q_2)\,  e^{ i n \zeta} \big]\nonumber \\   
& - \frac{2\,|b_4|^2}{k  R } \,\frac{(r^2 + a^2)}{\Sigma} \,  \bigg( \frac{\big( \Delta^{2 k n}-1\big)\sin^2 \theta}{(r^2 + a^2 \sin^2 \theta)} +  \frac{1}{a^2} \bigg)   ~+~ \frac{J}{a^2}\,\frac{(r^2 + a^2)}{\Sigma}   +  c_1  \,,\label{om1coiff1}  \\
\omega_2 ~=~&   \frac{r^2}{4 \,k R}\,  { \frac{\Delta^{k n}}{\Sigma }}  \, \big(   (b_1 Q_1+ b_2 Q_2)  \, e^{-  i n \zeta} +  (\bar b_1 Q_1+ \bar b_2 Q_2) \,  e^{ i n \zeta} \big)  \nonumber \\     
& - \frac{2\,|b_4|^2}{ k  R }\,\frac{r^2 }{\Sigma} \,  \bigg( \frac{\big( \Delta^{2 k n}-1\big)\cos^2 \theta}{(r^2 + a^2 \sin^2 \theta)} -  \frac{1}{a^2} \bigg)  ~-~ \frac{J}{a^2  }\,\frac{r^2 }{\Sigma}  + c_2  \,,\label{om2coiff1}  \\ 
\mathcal{F}~=~ &  - \frac{|b_4|^2}{k^2 R^2}\, \frac{1}{(r^2 + a^2 \sin^2 \theta)} \, \big(1 -  \Delta^{2 k n}\big)\,, \label{omZ3sol1a}
\end{align}
where $J$, $c_1$ and $c_2$ are constants to be determined.

One can, of course, do gauge transformations of the form (\ref{gaugeh}) and set $\omega_r$ or  $\omega_\theta$ to zero.

It is amusing to note that if we choose $Q_1 =Q_2$ then (\ref{breln0}) implies $b_1 = -b_2$ and every oscillating term cancels from our expression for $\omega$:
\begin{align}
 \omega_r ~=~ &   \omega_\theta  ~=~ 0 \,, \label{omsimple0} \\
\omega_1 ~=~  &    - \frac{2\,|b_4|^2}{k  R } \,\frac{(r^2 + a^2)}{\Sigma} \,  \bigg( \frac{\big( \Delta^{2 k n}-1\big)\sin^2 \theta}{(r^2 + a^2 \sin^2 \theta)} +  \frac{1}{a^2} \bigg)   ~+~ \frac{J}{a^2}\,\frac{(r^2 + a^2)}{\Sigma}   +  c_1  \,,\label{omsimple1}  \\
\omega_2 ~=~&   - \frac{2\,|b_4|^2}{ k  R }\,\frac{r^2 }{\Sigma} \,  \bigg( \frac{\big( \Delta^{2 k n}-1\big)\cos^2 \theta}{(r^2 + a^2 \sin^2 \theta)} -  \frac{1}{a^2} \bigg)  ~-~ \frac{J}{a^2  }\,\frac{r^2 }{\Sigma}  + c_2  \,,\label{omsimple2} 
\end{align}
This is analogous to the completely coiffured black rings and microstate geometries discussed in \cite{Bena:2013ora,Bena:2014rea}.

Returning to the more general Style 1 coiffuring (with independent $Q_1$ and $Q_2$), we wish to examine the necessary conditions  to avoid  closed time-like curves\footnote{Following standard practice, we show that there are no CTC's near the supertube and no CTC's at infinity.  This is usually sufficient to guarantee the absence of CTC's globally.}.
As in Section \ref{ss:rndstube} we  require that $\omega$ vanish at infinity and, as $r \to \infty$, one finds that $\beta$ vanishes and 
\begin{equation}
\omega + \beta ~\to~   \bigg(\frac{J}{a^2} - \frac{2\,|b_4|^2}{k  R \,a^2}  +  c_1 \bigg)  d\varphi_1  -  \bigg(\frac{J}{a^2} -  \frac{2\,|b_4|^2}{k  R \,a^2}  -  c_2 \bigg) d\varphi_2 \,.
\label{ominfty2}
\end{equation}
At the center of space, $r=0, \theta =0$,   the $\varphi_1$ and $\varphi_2$ circles pinch off in the base metric $ds_4^2(\cB)$. At $r=0, \theta =0$ one finds: 
\begin{equation} 
\omega +\beta    ~\to~  \bigg(\frac{J}{a^2} - \frac{2\,|b_4|^2}{k  R \,a^2}  +  c_1 \bigg)  d\varphi_1 +  \big( c_2 -2 k R   \big) d\varphi_2\,
\label{cspace2a}
\end{equation} 
which for the absence of CTC's must vanish. Thus we require that:
\begin{equation} 
 c_1= - 2 \,k R  \,, \qquad  c_2 ~=~  2\, k R\,,  \qquad J~=~ \frac{2\,|b_4|^2}{k  R } +2\,k R \, a^2 \,.
\label{cspace2b}
\end{equation} 

As noted earlier, regularity of the metric near the supertube means that as one approaches $\Sigma =0$, or  $r=0, \theta =\frac{\pi}{2}$, the metric must remain smooth.   The only potentially singular terms are proportional to $d\varphi_1^2$ but compared to simple supertube of  Section \ref{ss:rndstube}, $\mathcal{F}$ is now finite as one approaches the supertube and so  (\ref{JSTdiv1}) generalizes to:
\begin{equation} 6
- \frac{2}{\sqrt{\cP}}\,\beta_1 \, \big( \omega_1 + \coeff{1}{2}\, \mathcal{F}\, \, \beta_1 \big)  ~+~  \sqrt{\cP} \,a^2  \, d\varphi_1^2 \, 
\label{nearST3}
\end{equation} 
 Collecting the singular terms terms and requiring that they vanish leads to a simple generalization of (\ref{STreg0}):
\begin{equation} 
  J  ~=~   \frac{1}{4 kR} \, \big[ Q_1 Q_2 +  4 \, |b_4|^2  \big]    \,,
\label{STreg3}
\end{equation} 
and combined with  (\ref{cspace2b}) we obtain 
\begin{equation} 
a^2  ~=~     \frac{1}{8\, k^2 R^2} \, \big[ Q_1 Q_2 - 4\, |b_4|^2  \big]  \,,
\label{STreg4}
\end{equation} 
which determines the radius of the supertube in terms of its electric charges.

Of particular significance is that, at infinity, one has
\begin{equation} 
- \frac{\mathcal{F}}{2} ~\sim~   \frac{|b_4|^2}{2k^2 R^2}\, \frac{1}{r^2} ~\sim~  \frac{Q_P}{r^2} \,,
\label{Z3infty1}
\end{equation} 
which implies that the supertube now carries a momentum charge of 
\begin{equation} 
Q_P~=~  \frac{|b_4|^2}{2k^2 R^2}\,.
\label{QP1}
\end{equation} 
Note also that (\ref{STreg4}) implies the following bounds:
\begin{equation} 
 |b_4|^2 ~<~  \frac{Q_1 Q_2}{4}   \qquad \Rightarrow \qquad  Q_P  ~<~  \frac{Q_1 Q_2}{8\,k^2 R^2} ~=~ \frac{Q_1 Q_2}{k^2 R_y^2}  \,.
\label{bounds1}
\end{equation} 
More generally, it is instructive to rewrite (\ref{STreg3}) and  (\ref{STreg4}) in terms of the momentum charge:
\begin{equation} 
  J  ~=~   2\, kR \,(a^2 + 2 Q_P )\,, \qquad   a^2  ~=~     \frac{Q_1 Q_2 }{8\, k^2 R^2}   -  Q_P  \,.
\label{STreg-c1}
\end{equation} 

For future reference, it is convenient to extract the components of $\omega_1$ and $\omega_2$ that do not contain powers of $\Delta$:
\begin{align}
\hat \omega_1 ~\equiv~  &  - \frac{2\,|b_4|^2}{k  R } \,\frac{(r^2 + a^2)}{\Sigma} \,  \bigg(- \frac{ \sin^2 \theta}{(r^2 + a^2 \sin^2 \theta)} +  \frac{1}{a^2} \bigg)   ~+~ \frac{J}{a^2}\,\frac{(r^2 + a^2)}{\Sigma}   +  c_1  \,,\label{om1coiff1a}  \\
\hat  \omega_2 ~\equiv~&  - \frac{2\,|b_4|^2}{ k  R }\,\frac{r^2 }{\Sigma} \,  \bigg(- \frac{ \cos^2 \theta}{(r^2 + a^2 \sin^2 \theta)} -  \frac{1}{a^2} \bigg)  ~-~ \frac{J}{a^2  }\,\frac{r^2 }{\Sigma}  + c_2  \,.\label{om2coiff1a} \end{align}
The terms involving powers of $\Delta$ represent higher multipoles arising from the oscillations and, when $k n$ is sufficiently large, these are highly suppressed in the regions  $r >> a$.    Thus the $\hat \omega_i$ are the `higher-multipole-free' components of the angular momentum.
Substituting (\ref{cspace2b})   into (\ref{om1coiff1a}) and (\ref{om2coiff1a}) yields 
\begin{equation}
\hat \omega_1 ~\equiv~     \frac{2\,|b_4|^2}{k  R } \,\frac{a^2 \,\sin^2 \theta \cos^2 \theta }{\Sigma \,(r^2 + a^2 \sin^2 \theta) }    ~+~ \frac{J}{\Sigma}\, \sin^2 \theta \,, \qquad 
\hat  \omega_2 ~\equiv~ - \frac{2\,|b_4|^2}{k  R } \,\frac{a^2 \,\sin^2 \theta \cos^2 \theta }{\Sigma \,(r^2 + a^2 \sin^2 \theta) }    ~+~ \frac{J}{\Sigma}\, \cos^2 \theta   \,.\label{omcoiff1b} 
\end{equation}
Note that the first terms in these expressions vanish as $r^{-4}$ when $r \to \infty$ and hence 
for sufficiently large $kn$, the asymptotic structure of $\omega$ is determined entirely by $J$:
\begin{equation}
\hat \omega    ~\sim~     \frac{J}{r^2}\, (\sin^2 \theta  d\varphi_1 + \cos^2 \theta  d\varphi_2)  \,.
\label{omasymp1}
\end{equation}
Recall from (\ref{betaform3}) that one  has:
\begin{equation}
\beta    ~\sim~     \frac{2\, kR a^2}{r^2}\, (\sin^2 \theta  d\varphi_1 - \cos^2 \theta  d\varphi_2)  \,.
\label{betaasymp1}
\end{equation}
It therefore follows that this configuration has angular momenta (here we switch back to the physical $y$ radius $R_y=2\sqrt{2}R$ for later use)
\begin{equation}
J_1 ~=~ \frac{1}{\sqrt{2}}\, (J +   2\, kR a^2) ~=~ \frac{Q_1 Q_2 }{k R_y} \,, 
\qquad J_2 ~=~ \frac{1}{\sqrt{2}}\,  (J -   2\, kR a^2) ~=~ k R_y Q_P \,.
\label{angmom1a}
\end{equation}
and $J$ should be identified with 
\begin{equation}
J_L ~\equiv~  \frac{1}{2}\, (J_1 +  J_2) ~=~ \frac{J}{\sqrt{2}} 
~=~ \frac{1}{2}  \frac{Q_1 Q_2 }{k R_y}  +  \frac{1}{2} k R_y Q_P .
\label{angmomL1}
\end{equation}
Also note that 
\begin{equation}
J_R ~\equiv~  \frac{1}{2}\,\, (J_1- J_2) ~=~ \sqrt{2}\, kR a^2 
~=~ \frac{1}{2}  \frac{Q_1 Q_2 }{k R_y}  -  \frac{1}{2} k R_y Q_P .
\label{angmomR1}
\end{equation}
For later use, we record that in terms of $b_4$, the angular momenta are
\begin{equation}
J_L ~=~ \frac{1}{2}  \frac{Q_1 Q_2 }{k R_y}  +  \frac{2|b_4|^2}{k R_y} \,,
\qquad \quad
J_R ~=~ \frac{1}{2}  \frac{Q_1 Q_2 }{k R_y}  -  \frac{2|b_4|^2}{k R_y} \,.
\label{angmomsty1-2}
\end{equation}
We observe that, compared to the angular momenta of the original supertube solution in Section \ref{ss:rndstube}, $J_1$ is unchanged, $J_L$ has increased and $J_R$ has decreased. We will interpret this in the CFT shortly.

Supersymmetric BMPV black holes~\cite{Breckenridge:1996is} with macroscopic horizons exist in the regime of parameters
\begin{equation} 
  Q_1 Q_2 Q_P - J_L^2 > 0 \,, \qquad J_R = 0 \,.
\label{overspin}
\end{equation}
Indeed, see Appendix \ref{app:BMPV} and, specifically (\ref{BMPVbound}), where we have given the metric of the BMPV black hole in our conventions.

It is useful to parameterize the momentum charge via:
\be
Q_P ~=~ c_p \frac{Q_1 Q_2 }{k^2 R_y^2} \,, \qquad 0 \le c_p<1 \,,
\ee
where the upper bound on $c_p$ is a rewriting of \eq{bounds1}.
Then using (\ref{angmomL1}) we find  
\begin{equation} 
  Q_1 Q_2 Q_P - J_L^2  ~=~   - \frac{(1-c_p)^2}{4} \left(\frac{Q_1 Q_2}{k R_y} \right)^2
\label{overspin-2}
\end{equation} 
and so, in terms of the quantum numbers of a BMPV black hole, this geometry is ``overspinning''  and becomes extremal in the scaling limit:
\begin{equation} 
Q_P ~\to~ \frac{Q_1 Q_2 }{k^2 R_y^2}   \quad \Rightarrow \quad  a^2  ~\to~    0  \,.
\label{scaling1}
\end{equation} 
To understand why our solutions are overspinning, note that the original supertube of Section \ref{ss:rndstube} is overspinning ($c_P=0$) and as we add momentum $Q_P$,  (\ref{STreg-c1}) shows that we must also add a corresponding amount of angular momentum, and that $a$ is adjusted according to (\ref{STreg-c1})  such that we obtain (\ref{overspin}).

\refstepcounter{subsubsection}
\subsubsection*{\thesubsubsection ~~ Coiffuring:  Style 2}
\label{ss:Coiffure2}
  
For our second style of coiffuring, we employ the coiffuring technique used in~\cite{Bena:2015bea}. We will see in due course that the holographic dictionary is somewhat simpler for these solutions.
The first step is to set $b_2 =0$ and take $n = 2p$ in Eqs.\;(\ref{Z12osc1})--(\ref{T4osc1}).  The leading $r^{-1}$ singularities are cancelled by taking:
\begin{equation}
Q_1 (Q_2 + a^2) b_1 ~=~    b_4^2   \,,
\label{breln3}
\end{equation}
which fixes the Fourier coefficient $b_1$ in terms of $b_4$.

This leads to the solution:
\begin{align}
 \omega_r ~=~  & - \frac{i \,Q_1}{4 \,k R}\,  \frac{\Delta^{ 2 k p} \, r  }{(r^2 + a^2) }  \, \big[  b_1  e^{-  2 i p \zeta}   -  \bar b_1 e^{   2 i p \zeta}   \big] \,,   \qquad  
 \omega_\theta  ~=~  \frac{i \,Q_1\, \Delta^{ 2 k p } }{4 \,k R}\,\tan \theta \, \big( b_1  e^{-2 i p  \zeta}  - \bar b_1 e^{ 2 i p  \zeta} \big) \,, \label{omthsol3} \\
\omega_1 ~=~ & - \frac{ Q_1 \, a^2}{4 \,k R}\,  \frac{\Delta^{2 k p} \sin^2\theta}{\Sigma }  \, \big(  b_1  \, e^{- 2 i p \zeta} +  \bar b_1 \,  e^{2 i p \zeta} \big) 
~-~  \frac{|b_4|^2}{k  R } \,\frac{(r^2 + a^2)}{\Sigma} \,  \bigg( \frac{\big( \Delta^{2 k p}-1\big)\sin^2 \theta}{(r^2 + a^2 \sin^2 \theta)} +  \frac{1}{a^2} \bigg) \nonumber \\
& ~+~ \frac{J}{a^2  }\,\frac{(r^2 + a^2)}{\Sigma}     +  c_1  \,,\label{om1sol3}  \\
\omega_2 ~=~ & \frac{ Q_1 }{4 \,k R}\,  \frac{\Delta^{2 k p} \, r^2\cos^2\theta}{\Sigma }  \, \big(  b_1  \, e^{- 2 i p \zeta} +  \bar b_1 \,  e^{2 i p \zeta} \big)~-~\frac{|b_4|^2}{k R } \, \frac{r^2 }{\Sigma} \,  \bigg( \frac{\big( \Delta^{2 k p}-1\big)\cos^2 \theta}{(r^2 + a^2 \sin^2 \theta)} -  \frac{1}{a^2} \bigg) \nonumber \\ &~-~ \frac{J}{a^2  }\,\frac{r^2 }{\Sigma}  + c_2  \,,\label{om2sol3}  \\
\mathcal{F}~=~ & - \frac{|b_4|^2}{2 \,k^2 R^2}\, \frac{1}{(r^2 + a^2 \sin^2 \theta)} \, \big(1 -  \Delta^{2 k p}\big)\,, \label{omZ3sol3}
\end{align}

The analysis of the absence of CTC's proceeds exactly as before, giving:
\begin{equation} 
 c_1= -2\,k R  \,, \qquad  c_2 ~=~ 2\, k R\,,  \qquad J~=~ \frac{|b_4|^2}{k  R } + 2\,k R \, a^2 \,.
\label{cspace3}
\end{equation} 
Regularity at the supertube once again requires (\ref{nearST3}) to be finite at $\Sigma =0$.  This yields
\begin{equation} 
  J  ~=~   \frac{1}{4kR} \, \big[ Q_1 Q_2 +  2\, |b_4|^2  \big]    \,,
\label{STreg5}
\end{equation} 
and combined with  (\ref{cspace3}) we obtain 
\begin{equation} 
a^2  ~=~      \frac{1}{8\, k^2 R^2} \, \big[ Q_1 Q_2 -  2\, |b_4|^2  \big]    \,,
\label{STreg6}
\end{equation} 
which, again, determines the radius of the supertube in terms of its electric charges.

At infinity we now have
\begin{equation} 
-\frac{\mathcal{F}}{2}~\sim~    \frac{|b_4|^2}{4\,k^2 R^2}\, \frac{1}{r^2} \,,
\label{Z3infty2}
\end{equation} 
which implies that the supertube now carries a momentum charge of 
\begin{equation} 
Q_P~=~  \frac{|b_4|^2}{4\,k^2 R^2}\,.
\label{QP2}
\end{equation} 

Since we have set $b_2 =0$ we have, in a sense, half as many oscillations and this leads to halving of various quantities in this style of coiffuring.

As before, the positivity of (\ref{STreg6}) places a bound on $|b_4|$ which, in turn, results in the same bound on the momentum charge:
\begin{equation} 
 |b_4|^2 ~\le~  \frac{Q_1 Q_2}{2}   \qquad \Rightarrow \qquad  Q_P  ~\le~  \frac{Q_1 Q_2}{8\,k^2 R^2} 
~=~ \frac{Q_1 Q_2}{k^2 R_y^2}  \,.
\label{bounds2}
\end{equation} 
More generally, when    (\ref{STreg5}) and  (\ref{STreg6}) are rewritten in terms of the momentum charge we obtain exactly the same conditions as in (\ref{STreg-c1}):
\begin{equation} 
  J  ~=~   2\, kR \,(a^2 + 2Q_P )\,, \qquad   a^2  ~=~     \frac{Q_1 Q_2 }{8\, k^2 R^2}   -   Q_P  \,.
\label{STreg-c2}
\end{equation} 
Furthermore, in terms of $Q_P$ and $J$,  the `higher-multipole-free' components of the angular momentum are identical to those of (\ref{omcoiff1b}):
\begin{equation}
\hat \omega_1 ~\equiv~    \frac{ 4\, k R \, a^2 \, Q_P  \sin^2 \theta \cos^2 \theta }{\Sigma \,(r^2 + a^2 \sin^2 \theta) }    ~+~ \frac{J}{\Sigma}\, \sin^2 \theta \,, \qquad 
\hat  \omega_2 ~\equiv~ - \frac{4\, k R \,  a^2 \, Q_P  \sin^2 \theta \cos^2 \theta }{\Sigma \,(r^2 + a^2 \sin^2 \theta) }    ~+~ \frac{J}{\Sigma}\, \cos^2 \theta   \,.\label{omcoiff2b} 
\end{equation}
Thus the discussion of the asymptotic angular momenta is the same as in Section \ref{ss:Coiffure1}, and we again have
\begin{equation}
J_L ~=~ \frac{1}{2}  \frac{Q_1 Q_2 }{k R_y}  +  \frac{1}{2} k R_y Q_P \,,
\qquad \quad
J_R ~=~ \frac{1}{2}  \frac{Q_1 Q_2 }{k R_y}  -  \frac{1}{2} k R_y Q_P \,.
\label{angmomsty2}
\end{equation}
The difference between Style 1 and Style 2 comes when we express the angular momenta in terms of the respective coefficients of the oscillating terms. For Style 2, we obtain
\begin{equation}
J_L ~=~ \frac{1}{2}  \frac{Q_1 Q_2 }{k R_y}  +  \frac{|b_4|^2}{k R_y} \,,
\qquad \quad
J_R ~=~ \frac{1}{2}  \frac{Q_1 Q_2 }{k R_y}  -  \frac{|b_4|^2}{k R_y} \,.
\label{angmomsty2-2}
\end{equation}
In the limit of $k=1$, and taking $b_4$ to be real, our Style 2 solution is the extension to asymptotically flat space of a particular subset%
\footnote{This subset of the solutions in~\cite{Bena:2015bea} is given by taking a single mode of that construction and setting $m = k$ in the notation of that paper.} 
of the solutions constructed in~\cite{Bena:2015bea}.

\refstepcounter{subsubsection}
\subsubsection*{\thesubsubsection ~~ The lowest harmonics}
\label{ss:lowharms}

Recalling the form of $\Delta$ in (\ref{Deltadefn}),
\begin{equation} 
\Delta ~\equiv~  \frac{a \, \cos  \theta}{(r^2 + a^2)^\frac{1}{2}} \,,
\end{equation} 
we see that for  low values  of $k,n$ and $p$, the powers of $\Delta$ do not fall off strongly at infinity and do not vanish very strongly at the ring ($r=0$, $\theta =\frac{\pi}{2}$).  This can potentially lead to apparently singular behavior at the ring and unusual asymptotics at infinity.  We now examine this more carefully. 

First, note that for $kn =1$ there is an additional singularity at $r=0, \theta =\frac{\pi}{2}$ in the first term of $\omega_1$ in (\ref{om1coiff1}), and $\omega_\theta$ and $\omega_2$ both contain terms that oscillate and fall off as as $r^{-1}$. However these terms are absent when $Q_1=Q_2$, and also in the decoupling limit, since they arise from the solution to Source 2 as described in Section \ref{ss:Layer2}. Therefore there is a good asymptotically AdS solution for $kn=1$.

 Restricting attention now to $kn \ge 2$, in ``Style 1'' one sees that $\omega_\theta$ and $\omega_2$ both contain terms that oscillate and fall off as as $r^{-k n}$, while $\omega_r$ falls off as  $r^{-(k n+1)} dr$.  Similarly, in  ``Style 2'', one sees that $\omega_\theta$ and $\omega_2$ both contain terms that oscillate and fall off as as $r^{-2 k p}$, while $\omega_r$ falls off as  $r^{-(2k p+1)} dr$.   Since we normally expect the angular momentum to appear as the leading term and fall off as $r^{-2}$ at infinity, the $r^{-2}$ terms may, at first, seem  anomalous.

However, these  oscillating terms do not  present a problem.  The most direct way to see this is to observe that they oscillate around the compactified $y$-circle and so average to zero in any measurement of asymptotic charge at infinity in the non-compact space.    Such terms have also been encountered in other holographic solutions.  Indeed, oscillating terms that fall off as $r^{-2}$ were encountered in \cite{Mathur:2011gz,Mathur:2012tj,Lunin:2012gp} and \cite{Giusto:2013bda} (see Eq.\;(5.21g))
 where they arose through the action of the underlying {\it global} chiral algebra.     Consequently, angular momentum modes that oscillate along $y$  and fall off as  $r^{-2}$ in flat space represent physical solutions, and upon taking the decoupling limit, the corresponding asymptotically-AdS solutions are dual to well-defined CFT states.

\subsection{Regularity bounds and CTC's} 
\label{ss:Pbound}

As we have seen, there is a bound on the Fourier coefficients, $|b_4|$, that resulted in a bound on the momentum charge that was independent of the coiffuring style:
\begin{equation} 
 Q_P  ~\le~  \frac{Q_1 Q_2}{4\,k^2 R^2}  \,.
\label{QPbound}
\end{equation} 
In addition, the  coiffuring conditions relate $|b_4|$ to the $|b_A|$ via (\ref{breln1}) or (\ref{breln3}) so that we have:
\begin{equation} 
|b_1 b_2 |  ~=~\frac{|b_4|^2}{Q_1 Q_2 }  ~\le~ \frac{1}{4}   \qquad {\rm or} \qquad |b_1|  ~=~\frac{|b_4|^2}{Q_1 (Q_2 +a^2) }  ~<~ \frac{1}{2}\,, 
\label{bbound}
\end{equation} 
depending upon the coiffuring style.  These conditions are completely consistent with the bounds that we obtained earlier, (\ref{babound}), based upon the regularity of the $Z_A$.

One can also examine the possibility of CTC's in the `intermediate region'  where $ a^2 \ll r^2 \ll Q_X$ for all charges, $Q_1, Q_2$ and $Q_P$.  We also assume $k n$ or $k p$ is sufficiently large that we can drop such powers of $\Delta$ everywhere and, in particular,  work with the `higher-multipole-free' components, $\hat \omega_i$, of the angular momentum.  In this intermediate region we have $Z_I \sim \frac{Q_I}{r^2}$ and the configuration looks like a BMPV black hole.  Moreover,  this region also contains the scaling limit (\ref{scaling1}).

Since $|\mathcal{F}| \gg 1$ in the intermediate region,  it is more natural to complete the squares in the metric (\ref{sixmet}) by writing
\begin{equation}
ds_6^2 ~=~    \frac{1}{\mathcal{F} \, \sqrt{\cP}} \, \big(du +  \omega)^2 ~-~\frac{\mathcal{F}}{\sqrt{\cP} } \,   \big(dv+\beta +  \mathcal{F}^{-1}(du +  \omega)  \big)^2  
~+~  \sqrt{\cP} \, ds_4^2(\cB)\,,  \label{sixmet2}
\end{equation}
If one considers displacements only in the $(v,\varphi_1,\varphi_2)$ directions and chooses $dv$ so that the middle term in (\ref{sixmet2}) vanishes then the absence of closed timelike curves (CTC's) requires:
\begin{equation}
   - \omega ^2 ~-~ \mathcal{F} \, \cP \, \big( (r^2 + a^2) \sin^2 \theta \, d\varphi_1^2 +  r^2  \cos^2 \theta \, d\varphi_2^2 \big) ~\ge~ 0\,. \label{CTC1}
\end{equation}
Dropping all powers of $\Delta^k$ we can replace $\omega$ by $\hat \omega$.  We will retain the $1$'s in the $Z_A$'s, relabelling them by $\varepsilon_0$ so as to keep track of them.  
The absence of a negative eigenvalue in this two dimensional metric results in an inequality on the determinant that may be simplified to:
\begin{equation}
 (Q_1 + \varepsilon_0 \,\Sigma) (Q_2 + \varepsilon_0\, \Sigma) Q_P ~-~  J^2  ~\ge~ \frac{4\, k^2 R^2 a^8 \, \sin^2 \theta \cos^2 \theta}{r^2 (r^2 +a^2)} \,, \label{CTC2}
\end{equation}
where $\varepsilon_0 =1$. Using (\ref{overspin}) this identity simplifies to 
\begin{equation}
\varepsilon_0 \,  (Q_1 +  Q_2 + \varepsilon_0\, \Sigma) \, Q_P  ~\ge~ \frac{a^4 \,  (r^2 +a^2\sin^2 \theta) }{r^2 (r^2 +a^2)} \,, \label{CTC-3}
\end{equation}
which is generically  satisfied in the intermediate region.  Thus we have  solutions without CTC's\footnote{Strictly speaking we have only shown  that there are no CTC's near the supertube, in the intermediate region and at  infinity.  Again, this should be sufficient to guarantee the absence of CTC's globally.}  but that have the charges of overspinning BMPV black holes.

\section{Dual CFT states}
\label{Sect:CFT}

The spacetime CFT dual to gravity on AdS$_3\times\bbS^3\times \cM$ is a non-linear sigma model on the moduli space of instantons on $\cM=\bbT^4$ or $K3$~\cite{Vafa:1995bm,Douglas:1995bn,Maldacena:1997re}.  As is usual in AdS/CFT duality, the CFT is strongly coupled where gravity is weakly coupled, and vice versa.  There is a locus in the moduli space where the target space of the CFT is the symmetric orbifold $\cM^N/S_N$~\cite{Vafa:1995zh,Bershadsky:1995qy} (see also the review~\cite{David:2002wn}), and since the BPS spectrum does not change in the passage from weak to strong coupling, one can hope to identify the CFT states in the orbifold theory which, when transported across moduli space to the regime where supergravity is weakly coupled, are dual to our geometries.

In the orbifold theory, the duals to black-hole states are the twisted sectors of the orbifold containing long cycles that permute many copies of $\cM$.  Most of the entropy comes from oscillator excitations with fractional moding, and it has proven challenging to construct solutions that map to CFT states involving such fractional oscillators (for some previous examples, see~\cite{Giusto:2012yz,Chakrabarty:2015foa}). A major motivation for our construction is that it provides a large class of supergravity solutions whose CFT duals involve fractionally-moded oscillators.

In this section we begin with a review of the structure of the symmetric product orbifold CFT~-- covering both the structure of its supersymmetric ground states (in Section~\ref{ss:symprod twistops}), and the relation between BPS operators in the CFT and linearized mode operators in supergravity (in Section~\ref{ss:sugra modes}).  

Previous studies have considered {\it spectral flow} as a means of introducing momentum charge to the system starting from a two-charge seed solution~\cite{Giusto:2004id,Giusto:2004ip,Lunin:2004uu}.  In CFT states where all strands have windings which have a common divisor greater than one, there is the possibility to perform {\it fractional spectral flow}~\cite{Martinec:2001cf,*Martinec:2002xq} which can be used to generate three-charge solutions~\cite{Giusto:2012yz,Chakrabarty:2015foa}. After a brief review of spectral flow in Section~\ref{ss:CFT specflow}, a proposal is made in Section~\ref{ss:CFT dual states} for the CFT states dual to our geometries, built from fractional spectral flow on a {\it subset of strands} of a suitable two-charge BPS seed state. 
These candidate dual states are shown to carry the appropriate conserved quantum numbers, and reproduce at leading order the selection rules on the vevs of CFT operators dual to supergravity modes.  The precise content of the dual CFT states is then specified at the fully non-linear level by finding the CFT states dual to a two-charge supertube profile that yield the CFT states we construct by fractional spectral flow.

For the purpose of comparison, it will be somewhat more convenient to work in the F1-NS5 duality frame, where the background fields are all from the NS sector. NS-R parity is then manifest (it is simply fermion parity in the CFT), and is an additional tool which can be used to characterize states and operators.

\subsection{\nBPS{4} states: Twisted sector ground states of the CFT}
\label{ss:symprod twistops}

The vast majority of the \nBPS{4} states of the symmetric orbifold $(\cM)^N/S_N$ CFT are the twisted sector ground states under the symmetric group.  There is an independent twisted sector for each conjugacy class in the orbifold group.  In the symmetric group, one may write elements of the group as {\it words} consisting of   products of (non-overlapping) cyclic permutations of the copies of $\cM$.  The conjugacy class of a word is characterized simply by the lengths of all the cycles in the word.  Thus the conjugacy class is specified by the number $n_\kappa$ of cycles of length $\kappa$, $\kappa\in\{1,\dots N\}$, and the total length (including cycles of length one) is $\sum_i n_i = N$.

We will mostly focus on $\bbT^4$, and comment on the modifications that result when $K3$ is realized as $\bbT^4/\bbZ_2$, though, of course, the Ramond ground state structure is the same anywhere on the $K3$ moduli space.  The sigma model on the $\ell^{\rm th}$ copy of $\cM$ has bosonic fields $X^{(\ell)}_{A\Abar}$ and fermions $\chi^{(\ell)}_{A\alpha},\chibar^{(\ell)}_{A\alphabar}$.  These carry labels under a variety of $SU(2)$ symmetries:
\begin{itemize}
\item
The doublets $\alpha,\alphabar$ of the left and right $(SU(2) \times SU(2))_\cR$ $\cR$-symmetry.  
\item
The doublet $\Abar$ under a custodial $SU(2)_\cC$ which is a global symmetry of the $\cN=(4,4)$ superalgebra.  The supercurrents carry spin one-half under $SU(2)_\cC$ as well as under the $\cR$-symmetry.
\item
The doublet $A$ under an auxiliary $SU(2)_\cA$.  This $SU(2)_\cA$ is a symmetry for $\cM=\bbT^4$, but is broken by the holonomy of the connection for $\cM=K3$.
\end{itemize}
In a given twisted sector cycle, the bosons $X^{(\ell)}_{A\Abar}$ and fermions $\chi^{(\ell)}_{A\alpha},\chibar^{(\ell)}_{A\alphabar}$ of the individual $\bbT^4$ CFTs are cyclically permuted:
\be
X^{(\ell)}(e^{2\pi i} z) = X^{(\ell+1)}(z)
\quad,\qquad \ell=0,\dots,\kappa-1~,
\ee
where $X^{(\kappa)}\equiv X^{(0)}$; similarly for the fermions 
$\chi^{(\ell)},\chibar^{(\ell)}$.  
The twist operator for such a cyclic orbifold is most conveniently expressed in terms of fields that diagonalize the twist action.  Define the ``clock'' fields that are discrete Fourier transforms of these ``shift'' fields
\be
\label{Xclock monodromy}
\Xclock^{(\nu)} = \sum_{\ell=0}^{\kappa-1} \exp\Big[ 2\pi i \frac{\nu\ell}{\kappa} \Big] \, X^{(\ell)}
\quad,\qquad \nu = 0,\dots,\kappa-1~,
\ee
and similarly for the fermions 
$\chiclock^{(\nu)},\chiclockbar^{(\nu)}$.
The clock fields diagonalize the cyclic permutation
\be
\Xclock^{(\nu)}(e^{2\pi i} z) = e^{2\pi i \nu/\kappa} \Xclock^{(\nu)}(z)~.
\ee
A twist operator that implements these boundary conditions is the tensor product of standard $\bbZ_\kappa$ orbifold twist operators $\sigma_{(\nu/\kappa)}$ for each clock sector.%
\footnote{The full orbifold is of course non-abelian, but for the purpose of describing the spectrum, one can use abelian orbifold terminology.}  
These have dimension $h_{\nu,b} = \nu(\kappa-\nu)/\kappa^2$ for the bosonic twist operators, and $h_{\nu,NS} = (\nu/\kappa)^2$ for NS sector fermion twist operators, or $h_{\nu,R} = (\nu/\kappa - 1/2)^2$ for the R sector fermion twists.  Taking the product over all the clock sectors yields the full twist operator for the cycle
\be
\Sigma^{(\kappa)} = \prod_{\nu=0}^{\kappa-1} \sigma_{(\nu/\kappa)}
\quad,\qquad 
h_\Sigma = 
\begin{cases} (\kappa-1)/2 ~~ , & {\rm NS}~; \\  \kappa/4 ~~ , & {\rm R}~. \end{cases}
\ee
These NS sector twist ground state operators are spin $(\kappa-1)/2$ under both left- and right-moving $SU(2)_\cR$ $\cR$-symmetries, as may be seen by bosonizing the clock fermions and building the fermion twist operators as exponentials.  Thus, these operators, and the states that they create from the NS sector vacuum, are \nBPS{4}, breaking half the supersymmetries of each chirality.  Additional BPS operators are obtained by combining the lowest-dimension twist operator with the center-of-mass ($\nu=0$) fermion field.
Similarly, the $\kappa$-cycle R sector ground state operators preserve one quarter of the Ramond supersymmetries.

The monodromy~\eqref{Xclock monodromy} results in fractional mode expansions for the $\Xclock^{(\nu)}$
\begin{align}
\partial_z\Xclock^{(\nu)}(z) &= \sum_{m\in\bbZ} \xmode^{(\nu)}_{m+\nu/\kappa} \,z^{-m-\nu/\kappa-1}
\nonumber\\
\chiclock^{(\nu)}(z) &= \sum_{m\in\bbZ} \chimode^{(\nu)}_{m+\nu/\kappa} \,z^{-m-\nu/\kappa} ~,
\end{align}
together with the oscillator commutation relations
\begin{align}
[\xmode^{(\nu)}_{m+\nu/\kappa} , \xmode^{(\kappa-\nu)}_{-m'-\nu/\kappa} ] &= \alpha'  (m+\nu/\kappa)\delta_{mm'}
\nn\\
[\chimode^{(\nu)}_{m+\nu/\kappa} , \chimode^{(\kappa-\nu)}_{-m'-\nu/\kappa} ] &= \alpha' (m+\nu/\kappa)\delta_{mm'}
~~~,
\end{align}
where to reduce clutter the tangent space indices on the modes and fields have been suppressed in these expressions.

At this point, one can assemble all the different clock sector modes into a single set of ``untwisted'' (integer moded) $\bbT^4$ scalar fields $\xcover_{A\Abar}$ and fermions $\chicover_{A\alpha}$, $\chibarcover_{A\alphabar}$ living on the $\kappa$-fold cover of the cylinder.  In order that the oscillator commutation relations remain canonical, one must rescale the effective string tension $\alpha'$ by a factor of $\kappa$, to ${\widehat\alpha}'=\alpha'/\kappa$; the fractionated oscillator mode energies are also $\kappa$ times smaller than the energies of the the untwisted oscillator modes.

The covering-space picture makes it clear that the R ground states carry spinor quantum numbers in the target space, since the structure is the same as the worldsheet theory of free perturbative strings.  We can label the Ramond ground states for $\bbT^4$ as
\be
\label{R gd states}
\ket{\alpha\alphabar}~,\quad
\ket{A B}~,\quad
\ket{\alpha B}~,\quad
\ket{A\alphabar}~;
\ee
for $\bbT^4/\bbZ_2$ the fixed points provide sixteen more.
One moves around in the space of ground states by the action of the zero modes of the fermions 
$\chicover_{A\alpha}$, $\chibarcover_{B\alphabar}$, which act as gamma matrices.
We will focus on two ground states in particular -- the highest weight state $\ket{++}$ of the spin-1/2 multiplet, and the ``singlet'' combination of the auxiliary $SU(2)_\cA$ bispinor
\be
\ket{00} \equiv \epsilon^{AB}\ket{AB}~.
\ee

The covering space picture also leads to a somewhat more geometrical picture of the \nBPS{4} states in the Ramond sector.
The conformal dimension of the ground state in the twisted sector is determined by the covering space transformation $z\to z^\kappa$, for which the Schwarzian derivative contribution to the stress tensor leads to
\be
\label{twist base dim}
h_0^{(\kappa)} = \frac{\kappa}{4} - \frac{1}{4\kappa} ~.
\ee
One can then apply any operator $\cO$ of the $\cM=\bbT^4$ (or $K3$) SCFT to this ground state; recalling that energies of the covering space theory are reduced by a factor of $\kappa$ leads to the spectrum
\be
\label{twistdims}
h_\cO^{(\kappa)} = h_0^{(\kappa)}  + \frac{h_\cO}{\kappa} ~.
\ee

The Ramond ground states of $\cM$ are in one-to-one correspondence with the cohomology of $\cM$;%
\footnote{Since topologically twisting the supersymmetry of the sigma model relates the cohomology of the supersymmetry charges to the cohomology of a Dolbeault-type operator on the target space.}
for instance, for $\cM=\bbT^4$ the special spin-1/2 multiplet $\ket{\alpha\alphabar}$ is associated to the $(0,0)$, $(0,2)$, $(2,0)$ and $(2,2)$ cohomology, while the $j=0$ states $\ket{AB}$ are associated to the $(1,1)$ cohomology.  
Thus, in the $\kappa$-cycle twisted Ramond sector for $\bbT^4$, the supersymmetric ground states with $h=\kappa/4$ consist of one $(j,\bar j)=(1/2,1/2)$ multiplet and a quartet of singlets.  In addition there are representations $(1/2,0)$ and $(0,1/2)$ which correspond to the odd cohomology.  The action of the fermion zero modes of $\chi,\chibar$ moves one among these various representations.  A similar story holds for $\cM=K3$.
If we realize $K3$ as a $\bbT^4/\bbZ_2$ orbifold, we obtain 16 additional singlets from the 2-cohomology associated to the 16 fixed points of the orbifold, however there is no odd cohomology and so no $(1/2,0)$ or $(0,1/2)$ representations. 

Under spectral flow, the operators that create these Ramond ground states from the vacuum transform into BPS short multiplet operators in the NS sector.  We will discuss spectral flow in more detail below; here we simply wish to note that spectral flow generates from any operator with quantum numbers $(L_0,J_3)=(h,j)$ a related set of flowed operators with quantum numbers
\be
\label{int spec flow}
L_0 = h + 2 \,j\, s + \frac{c}{6} \,s^2 ~~,\quad J_3 = j + \frac{c}{6} \, s ~~,\quad s\in \frac12 \bbZ~.
\ee
If the initial operator is in the R (NS) sector, then spectral flow by integer amounts leads to another R (NS) operator, while spectral flow by odd half-integer amounts leads to NS (R) operators.

Finally, the twist operator for the full word conjugacy class in the symmetric group is given by the product of the twist operators for the $n$-component cycles in the word, for instance
\be
\label{twistops}
\Sigma = \prod_{i=1}^n\Sigma_i^{(\kappa_i)}
\quad,\qquad 
h_\Sigma = 
\begin{cases} (N-n)/2 ~~ , & {\rm NS}~; \\  N/4 ~~ , & {\rm R}~, \end{cases}
\ee
where we have used the fact that the sum of all the $\kappa_i$ is $N$.
In the NS sector, only if all the polarizations $\alpha_i, \alphabar_i$ are aligned is the state BPS, since only then does the $\cR$-charge equal (plus or minus) the scaling dimension.  In the R sector, any choice of polarizations will do, and one obtains a large degeneracy of BPS ground states carrying any $\bbS^3$ angular momentum in the tensor product $(\frac12)^{\otimes n}$.  Note also that all the Ramond ground states are at zero energy once we include the Casimir energy $E_0=-c/24=-N/4$ for the CFT on a cylindrical geometry.  The geometries constructed in  Sections~\ref{Sect:sixD}--\ref{Sect:MomST} are dual to CFT states in the Ramond sector, so henceforth we specialize to this sector.  On the other hand, linearized excitations are NS sector operators, and so we will be interested in these NS operators when probing a CFT state to see what vevs of the supergravity fields are turned on in the supergravity background dual to this CFT state.

\subsection{\nBPS{4} operators: Linearized supergravity modes}
\label{ss:sugra modes}

The spectrum of linearized supergravity on AdS$_3\times \bbS^3$ and its relation to the symmetric product was worked out in~\cite{deBoer:1998ip,Larsen:1998xm} (see also~\cite{Deger:1998nm,Kutasov:1998zh}).  
The bosonic spectrum consists of 
\begin{itemize}
\item
$\bbT^4$: the graviton, 5 self-dual (SD) plus 5 ASD tensors, 16 vectors, and 25 scalars; 
\item
$K3$: the graviton, 5 SD plus 21 ASD tensors, and 105 scalars.
\end{itemize}
All these fields lie in short multiplets of the $\cN\!=\!(4,4)$ superconformal algebra.  The $\cR$-charge of these multiplets is a combination of the spatial momentum on $\bbS^3$ and the tensor structure of the fields.  
The left-moving $\cR$-charge content of an NS sector short multiplet consists of
\be
\begin{matrix}
\rm{state} & j & j' & h \\
\midrule  
\ket{\Psi} & n/2 & 0 & n/2 \\
G_{-\half}\ket{\Psi} & (n-1)/2 & 1/2 & (n+1)/2 \\
(G_{-\half})^2 \ket{\Psi} & (n-2)/2 & 0 & (n+2)/2 \\
\end{matrix}
\ee
\vskip .3cm
\noindent
where $j$ is the spin under the $SU(2)_\cR$ $\cR$-symmetry, and $j'$ is the spin under the global (custodial) $SU(2)_\cC$ of the $N=4$ algebra; similarly for the right-moving structure.
{Short multiplets may also carry an additional auxiliary $SU(2)_\cA$ quantum number $A,B$ associated to the fermions $\chi_{\alpha A}, \chibar_{\alphabar A}$ for $\bbT^4$.  For $K3$, this is the $SU(2)$ for which the connection has holonomy, and so is not generically a good quantum number, however it is an ``accidental'' symmetry for untwisted states of the $\bbT^4/Z_2$ orbifold locus and we can continue to use this labelling.}

Consider the highest weight component of a short multiplet operator
\be
\cO^{(\kappa)}_{{\bf m},{\bf \bar m}} = 
\cO_{(\alpha_1...\alpha_{\bf m}),(\alphabar_1...\alphabar_{\bf  \bar m}) } 
\ee
of $\cR$-charge spins $(2j+1,2\bar j+1)=({\bf m},{\bf\bar m})$.  Its single descendants are thus
\be
\cO_{(\alpha_2...\alpha_{\bf m}),(\alphabar_2...\alphabar_{\bf  \bar m}) }^{\Abar\Bbar}
=
G_{-\half}^{\alpha_1 \Abar} \bar G_{-\half}^{\alphabar_1 \Bbar}
\cO_{(\alpha_1...\alpha_{\bf  m}),(\alphabar_1...\alphabar_{\bf  \bar m}) } ~,
\ee
where $\Abar,\Bbar$ are custodial $SU(2)_\cC$ indices (not to be confused with the auxiliary $SU(2)_\cA$ labels $A,B$ for the ground states in equation~\eqref{R gd states}); the double descendants are
\be
\cO_{(\alpha_3...\alpha_{\bf m}),(\alphabar_3...\alphabar_{\bf  \bar m}) }
=
\big(\epsilon_{\Abar\Bbar}G_{-\half}^{\alpha_1 \Abar}G_{-\half}^{\alpha_2 \Bbar} \big)
\big(\epsilon_{\Abar'\Bbar'}\bar G_{-\half}^{\alphabar_1 \Abar'}\bar G_{-\half}^{\alphabar_2 \Bbar'}\big)
\cO_{(\alpha_1...\alpha_{\bf m}),(\alphabar_1...\alphabar_{\bf  \bar m}) } \,.
\ee
One also has the helicity $({\bf m}-{\bf\bar m} \pm 2)/2$ fields one gets by taking the double-descendant only on one side.\footnote{Note that the action of $G_{-\half}$ lowers the $SU(2)$ spin while raising the $SL(2)$ spin, so that the six-dimensional helicity stays constant; similarly for $\bar G_{-\half}$.  Thus the short multiplets with ${\bf m}-{\bf \bar m}=\pm 2$, whose highest weight has spin one in both $SU(2)$ and $SL(2)$, contain the six-dimensional spin-two graviton polarizations.}
These comprise the bosonic content of the supermultiplet.\footnote{The lowest BPS operator in the short multiplet $(k,k)_S$ has special properties at low $k$.
For $k=1$ this operator is the identity operator, and the higher components of the superfield are absent.  For $k=2$, the lowest component has dimension $h=\bar h=1/2$, and the double-descendant is null.  Not until $k=3$ is there a non-trivial double-descendant operator.}

According to~\cite{deBoer:1998ip,Larsen:1998xm}, the spectrum of $\cN\!=\!(4,4)$ short multiplets for type IIB supergravity on AdS$_3\times\bbS^3\times K3$ is
\begin{align}
\label{K3 short}
&\oplus_{m\ge1}\big[ ({\bf m},{\bf m+2})_S+({\bf m+2},{\bf m})_S
+  ({\bf m+2},{\bf m+2})_S\big] 
+ n_T\big[\oplus_{m\ge2} ({\bf m},{\bf m})_S \big] ~,
\end{align}
where $\bf m$ is the dimension of the $SU(2)_\cR$ representation.
These supermultiplets expand into a set of $\bbS^3$ harmonics (ignoring special restrictions at low angular momentum)
\begin{align}
\oplus_{m} \big( & ({\bf m},{\bf m\pm 4}) + 4({\bf m},{\bf m\pm 3}) + (n_T+7) ({\bf m},{\bf m\pm 2})\nn\\
&+(4n_T+8) ({\bf m},{\bf m\pm 1}) +  (6n_T+8) ({\bf m},{\bf m}) \big) ~.
\end{align}
The number of ASD tensors $n_T=21$ is dictated by anomaly cancellation.
These quantum numbers result from the product of spherical harmonics on the $\bbS^3$ with the representations of the $SO(4)_{\cL}$ little group
\be
({\bf 3},{\bf 3}) + 4({\bf 2},{\bf 3}) + 5({\bf 1},{\bf 3}) + 
n_T({\bf 3},{\bf 1}) + 4 n_T ({\bf 2},{\bf 1}) + 5 n_T ({\bf 1},{\bf 1})  ~.
\ee
Similarly, the spectrum of short multiplets for AdS$_3\times\bbS^3\times\bbT^4$ is~\cite{deBoer:1998ip,David:2002wn}
\begin{align}
\label{T4 short}
&\oplus_{m\ge1}\big[ ({\bf m},{\bf m+2})_S+({\bf m+2},{\bf m})_S
+  ({\bf m+2},{\bf m+2})_S\big] \nn\\
&\qquad\qquad
+ 5\big[\oplus_{m\ge2} ({\bf m},{\bf m})_S \big] 
+ 4\oplus_{m\ge2}\big[ ({\bf m},{\bf m+1})_S+({\bf m+1},{\bf m})_S\big] ~.
\end{align}

There is a one-to-one correspondence between single-particle supergravity modes and $\kappa$-cycle Ramond ground states, by starting with the operators associated to the latter and performing a single unit of spectral flow to the NS sector.  The operators associated to a single cycle of the symmetric group correspond to the single-particle modes in supergravity.  The cycle winds together $\kappa$ copies of $\cM$, and thus has central charge $c=6\kappa$.  Under the spectral flow operation~\eqref{int spec flow}, 
the $j=1/2$ Ramond operators flow to one $h=j=(\kappa-1)/2$ NS operator (from the $j_3=-1/2$ polarization), and one $h=j=(\kappa+1)/2$ NS operator (from the $j_3=+1/2$ polarization).%
\footnote{Note that the latter operator can also be obtained from the former by tensoring with the center-of-mass current $J^+$ of the $\kappa$-cycle.  By $\kappa$-cycle currents, or center-of-mass currents, we mean the total $SU(2)$ currents built of the copies of $\bbT^4$ being sewn together in a particular cycle of length $\kappa$ in a symmetric group word, rather than the total $\cR$-currents of the entire theory.}  
Similarly, the $j=0$ operators flow to $h=j=\kappa/2$ operators.

The special spin-1/2 multiplet thus yields $SU(2)_\cR$ representations ${\bf m}=\kappa,\kappa+2$ after spectral flow, while the spin-0 multiplets yield representation ${\bf m}=\kappa+1$ after spectral flow. Combining left- and right-movers yields the spectra~\eqref{K3 short},~\eqref{T4 short}.  Note that for $K3$, the Ramond operators have the same fermion parity on left and right, while for $\bbT^4$, the left and right fermion parity can be chosen independently since one can act with fermion zero modes on left and right independently; this is the origin of the short multiplets $({\bf m}, {\bf m\pm 1})_S$ which comprise the harmonic expansion of the vector supermultiplets.  

The special spin $(1/2,1/2)$ Ramond ground states are universal, and upon spectral flow to the NS sector are associated to the six-dimensional graviton, dilaton and NS B-field.  Their harmonics on the spatial $\bbS^3$ organize themselves into $\cN\!=\!(4,4)$ short multiplets
 \be
 \label{graviton sector}
 ({ \kappa},{ \kappa})_S + ({ \kappa},{ \kappa\!+\!2})_S+({ \kappa\!+\!2},{ \kappa})_S + ({\kappa\!+\!2},{\kappa\!+\!2})_S 
\ee
comprising the lowest spin chiral primary $\cO^{(\kappa)}_{\bf m,m}$ of the $\kappa$-cycle, which has ${\bf m} = \kappa$, together with the three additional chiral primaries built by tensoring with the $\kappa$-cycle currents $J^+$ and/or $\bar J^+$ on 
$\cO^{(\kappa)}_{\kappa,\kappa}$.
The bosonic content of these multiplets consists of two six-dimensional supergravity supermultiplets.  The first, the graviton supermultiplet, contains the graviton plus the self-dual part of the B-field, as well as four more self-dual, six-dimensional, two-form tensor fields from the RR sector.  The second, a six-dimensional tensor multiplet, contains the ASD six-dimensional polarizations of the NS B-field, as well as the dilaton and four six-dimensional scalars made from the triplet of RR fields $C_2^+$ (the RR tensor which is self-dual on $\bbT^4$) together with the self-dual combination $v_4 C_0+C_4$ of the RR scalar and $\bbT^4$ four-form.  The zero modes of these latter four scalars are moduli in the F1-NS5 duality frame.  The ${\bf m=\bar m}$ CFT primaries map to linear combinations of supergravity field modes that diagonalize the linearized field equations.%
\footnote{See~\cite{Taylor:2007hs} for a discussion of subtleties in this map.}
In general, there are also non-linear corrections to the map between CFT operators and supergravity field modes; in a typical correlator, these corrections are suppressed by powers of the gravitational coupling, but in so-called extremal correlators (where the conformal dimension of one operator is the sum of all the others) these non-linearities can contribute at leading order.

The highest weight, together with the double descendants of the quartet of superfields~\eqref{graviton sector}, yield the harmonic expansion of the six-dimensional graviton, 
the six-dimensional NS B-field, and the six-dimensional dilaton. The quantum numbers $(h,\bar h, j,\bar j)$ are the resolution of the product of the spatial harmonic and the tensor structure onto states of definite total spin in both $SL(2)$ and $SU(2)$.  The graviton $g_{MN}$ and B-field $B_{MN}$ can have their tensor polarizations either along AdS$_3$, $M,N=\mu,\nu$, or along $\bbS^3$, $M,N=a,b$.  Of the two indices, one transforms under the left $SL(2)\times SU(2)$, and the other transforms under the right $SL(2)\times SU(2)$.
An analysis of~\cite{Kutasov:1998zh} 
shows that the physical combinations of tensor polarization and $SL(2)\times SU(2)$ spatial harmonic $(h,j)=(\lambda,\ell)$ are those whose total $SL(2)\times SU(2)$ quantum numbers are $(h,j)=(\lambda\pm1,\ell)$ or $(\lambda,\ell\pm 1)$.  One can thus trace the six-dimensional polarizations through the field transformation
and resolution onto components of definite total spin, in order to match supergravity fields with CFT operators at the linearized level.
As discussed above, beyond the leading order in the small field expansion, the map between CFT operators and supergravity modes is non-linear.

The single descendants have the opposite NSR parity, and comprise
a quartet of tensor harmonics;
this custodial $SU(2)_\cC$ bi-doublet can be decomposed into a scalar and self-dual tensor on $\bbT^4$/$K3$, and thus one obtains the self-dual six-dimensional polarizations of $C_2$, as well as $C_4$ having two legs in six-dimensional and two legs along $\bbT^4$/$K3$.
Note that for $\kappa=2$, one finds the four RR moduli of the background; a null vector truncates the representation from above, so that these components are in fact the highest components of the superfield -- a multiplet with $h=j=1/2$ is an {\it ultrashort} multiplet, and thus perturbing by the single-descendant operators preserves $\cN\!=\!(4,4)$ supersymmetry.

The remaining quartet $\cO^{(\kappa)}_{AB}$ of spin $j=\bar j=\kappa/2$ superfields (which have $\mm=\mmb=\kappa+1$), comprise four additional tensor multiplets containing 4 ASD tensors and 20 scalars.  The four tensors appear in the lowest and highest components plus the helicity $\pm 1$ one-sided double-descendants, and are the ASD parts of the RR tensors whose opposite chiralities are in the gravity supermultiplets~\eqref{graviton sector}.  These components also include the harmonics of four RR fixed scalars (the ASD combinations of $C_0,C_4$ and $C_2$ with polarization entirely on $\bbT^4/K3$).  The single-descendants comprise 16 NS sector scalars -- the polarizations of the graviton and B-field along $\bbT^4/K3$.  For $\kappa=1$, one has the 16 NS sector moduli of $\bbT^4$ (again these are ultrashort multiplets, so the single-descendant is the highest component).
The spectrum is then completed either with the $ ({\bf m},{\bf m\pm1})_S$ vector multiplets for $\bbT^4$; or 16 more $ (\kappa,\kappa)_S$ tensor multiplets for $K3=\bbT^4/\bbZ_2$, with similar content.  

We summarize the short multiplet content of the $\kappa$-cycle sector of the gravity sector supermultiplets in the following table, where $(\mm,\mmb) \in \{(\kappa,\kappa),(\kappa,\kappa\!+\!2),(\kappa\!+\!2,\kappa),(\kappa+2,\kappa+2)\}$:
\be
\label{graviton table}
\begin{array}{|c|c|c|c|c|}
\hline
\rm{multiplet} & (2j+1,2\bar j+1) & SU(2)_\cA & {\rm sugra~field} \\
\hline\hline
\cO^{(\kappa)}_{\mm,\mmb} & ({\mm,\mmb}) & {\mathbf 1} & G,B,\Phi  \\
\hline
(G_{-\half})^2  \cO^{(\kappa)}_{\mm,\mmb} & ({\mm-2,\mmb}) & {\mathbf 1} & G,B,\Phi  \\
\hline
 (\bar G_{-\half})^2 \cO^{(\kappa)}_{\mm,\mmb} & ({\mm,\mmb-2}) & {\mathbf 1} & G,B,\Phi  \\
\hline
(G_{-\half})^2 (\bar G_{-\half})^2 \cO^{(\kappa)}_{\mm,\mmb} & ({\mm-2,\mmb-2}) & {\mathbf 1} & G,B,\Phi  \\
\hline
G_{-\half}^{\Abar}{ \bar G}_{-\half}^{\Bbar} \cO^{(\kappa)}_{\mm,\mmb} & ({\mm-1,\mmb-1}) & {\mathbf 1\oplus \mathbf 3} & C_2^+,C_4^+,C_0  \\
\hline
\end{array}
\ee  
The RR six-dimensional tensor fields together with the six-dimensional tensor $B$ field comprise the five self-dual tensors in the six-dimensional $N=(2,0)$ graviton supermultiplet.

The remaining six-dimensional tensor supermultiplets contain the torus moduli fields, and consist of:
\be
\label{torus moduli table}
\begin{array}{|c|c|c|c|c|}
\hline
\rm{multiplet} & (2j+1,2\bar j+1) & SU(2)_\cA & {\rm sugra~field}  \\
\hline\hline
\cO^{(\kappa)AB}_{\kappa\!+\!1,\kappa\!+\!1} & (\kappa+1,\kappa+1) & {\mathbf 1} & C_0, C_2^-, C_4^- ~{\it tensors/scalars}  \\
\hline
(G_{-\half})^2  \cO^{(\kappa)AB}_{\kappa\!+\!1,\kappa\!+\!1} & (\kappa-1,\kappa+1) & {\mathbf 1} & C_2^-, C_4^- ~{\it tensors} \\
\hline
 (\bar G_{-\half})^2 \cO^{(\kappa)AB}_{\kappa\!+\!1,\kappa\!+\!1} & (\kappa+1,\kappa-1) & {\mathbf 1} & C_2^-, C_4^- ~{\it tensors} \\
\hline
(G_{-\half})^2 (\bar G_{-\half})^2 \cO^{(\kappa)AB}_{\kappa\!+\!1,\kappa\!+\!1} & (\kappa-1,\kappa-1) & {\mathbf 1} &   C_0^{\strut},C_2^-, C_4^- ~{\it tensors/scalars} \\
\hline
G_{-\half}^{\Abar}{ \bar G}_{-\half}^{\Bbar} \cO^{(\kappa)AB}_{\kappa\!+\!1,\kappa\!+\!1} & (\kappa,\kappa) & {\mathbf 1\oplus \mathbf 3} & G,B ~{\it moduli} \\
\hline
\end{array}
\ee
These four multiplets contain six-dimensional ASD RR tensors.
In all of the tables, the plus/minus superscript on tensors indicates their six-dimensional chirality.  The additional 16 ASD tensor supermultiplets of the $K3$ theory arising from the fixed points of $\bbT^4/\bbZ_2$ are similar in content to the above table.  It is straightforward to work out the vector multiplets, which only contain transverse vector polarizations and their fermionic superpartners.

Given the foregoing collection of $SL(2,R)\times SU(2)$ highest weights organized into $\cN\!=\!(4,4)$ multiplets, the action of the lowering operators $J^-$, $\bar J^-$ of $SU(2)_L\times SU(2)_R$ and raising operators $L_{-1}$, $\bar L_{-1}$ of $SL(2,R)_L\times SL(2,R)_R$ fills out a complete basis of six-dimensional spatial harmonics of the supergravity fields. 

\subsection{CFT spectral flow to \nBPS{8} states}
\label{ss:CFT specflow}

We need one more ingredient to specify the class of CFT states dual to the supergravity geometries above.  Spectral flow is a coherent deformation of the charge in a CFT with a $U(1)$ current.  Any primary field $\mathcal O$ in such a theory can be written
\be
\label{charged op}
{\mathcal O} = e^{i\sqrt2\alpha H} \Phi
\ee
where the $U(1)$ current is bosonized as $J=i \partial H$, and $\Phi$ is a $U(1)$ singlet operator.  Spectral flow is then the deformation along $\alpha$, which leads to a family of operators/states of dimension and charge
\be
\label{specflow}
h = h_\Phi + {\kappa\,\alpha^2 }
\quad,\qquad
q = {\sqrt2} \, { \kappa\,\alpha}
\ee
where the normalization of the current is 
\be
J(z) J(w) \sim \frac{\kappa}{(z-w)^2}  ~,
\ee
For an $\cN\!=\!(4,4)$ SCFT, the normalization of the $SU(2)$ $\cR$-current $J_3$ is set by the algebra, $\kappa=c/6$, and the $SU(2)$ spin of operators~\eqref{charged op} is $j_3=\alpha \kappa$.  

One can decompose the $1/4$-BPS twist operators under spectral flow as follows.  Consider the NS sector twist field for a cyclic permutation of order $\kappa$, with quantum numbers
\be
c=6\kappa
\quad,\qquad
h=j_3=\frac{ (\kappa-1)}2\quad.
\ee  
One can determine the dimension of the operator $\Phi$ via a spectral flow by an amount $\alpha=-(\kappa-1)/2\kappa$ that strips off the $j_3$ charge; in this way one finds
\be
\label{gd state dim}
h_\Phi = \frac{\kappa}{4} - \frac1{4\kappa} ~,
\ee
the dimension~\eqref{twist base dim} of the operator that implements the covering transformation $z\to z^\kappa$.
Spectral flow to the R sector shifts the $j_3$ charge by $\kappa/2$, from $j_3=(\kappa-1)/2$ to $j_3=-1/2$.  The $U(1)$ charge exponential now carries dimension $1/(4\kappa)$, resulting in the Ramond-sector twist operator dimension $\kappa/4$, equation~\eqref{twistops}.  Similarly, on $\bbT^4$ one may regard the $\cR$-symmetry singlet states $\ket{AB}$ as the result of acting on the state $\ket{\Phi,\alpha\!=\!0}$ by a spectral flow to spin $1/2$ in the {\it auxiliary}~$SU(2)_\cA$ (for $\bbT^4/\bbZ_2$, there are 16 additional states coming from the fixed points).  

One can now obtain new R-sector states from the cyclic twist ground states $\ket{\alpha\alphabar}$ and $\ket{AB}$ via spectral flow by an amount $s/\kappa$, $s\in \bbZ$.  This operation is {\it fractional spectral flow} on the $\kappa$-cycle, but {\it integer} spectral flow on the covering space; it results in a series of \nBPS{8} states, for example
\begin{align}
\label{specflow states}
\ket{++}_{\kappa,s}
\quad,\qquad  &h_{\kappa,s} = \frac{\kappa^2-1}{4\kappa} + \frac{(s+1/2)^2}{\kappa}
\quad, \hskip -2cm &&j^3_{\kappa,s} = s+\frac12
\nn\\
\ket{00}_{\kappa,s}
\quad, \qquad &h_{\kappa,s} = \frac{\kappa}{4} + \frac{s^2}{\kappa}
\quad, \hskip -2cm &&j^3_{\kappa,s} = s \quad ,
\end{align}
and corresponding operators.  For a general  conjugacy class in the symmetric group, one has the choice of independent spectral flow on each component cycle.

States that survive the symmetric group quotient have $h-\bar h\in \bbZ$ for each cycle.  In the twisted sectors, a cycle of length $\kappa$ has a $\bbZ_\kappa$ projection on its Hilbert space%
\footnote{In the $g$-twisted sector, one has a projection by the action of all group elements $h$ that commute with $g$, $hgh^{-1}=g$.  This includes $g$ itself, whose action imparts a phase to the fractional modes.}
that assigns charge $\nu/\kappa$ to the $\nu^{\rm th}$ clock sector, and neutrality under this projection guarantees that states have integer momentum, cycle by cycle.  To satisfy this requirement, one must have $s^2/\kappa\in\bbZ$ for the $\ket{00}$ state, or $s(s+1)/\kappa\in\bbZ$ for the $\ket{++}$ state.

The generic state in this construction is obtained by taking tensor products of $\ket{\alpha\alphabar}_{\kappa,s}$ and $\ket{AB}_{\kappa,s}$ chosen independently for each cycle; these states are built from fractional spectral flow under the $J_3$ pertaining to that cycle only, and are subject to the integer momentum constraint on each cycle (and the $\bbZ_2$ quotient for $K3=\bbT^4/\bbZ_2$).  
Note that the $U(1)$ currents that we are using to spectral flow are not present in the CFT away from the orbifold locus, apart from the overall $U(1)$.  Nevertheless, at the orbifold locus they serve to generate states for us that are protected by the BPS property as we move away from the orbifold locus in moduli space, and so we can continue to characterize them through the use of this special property of the orbifold theory.

The states spectrally flowed under $J_3$ have an equivalent description in terms of descendant states in the $SU(2)$ current algebra; for example
\be
\label{00 spec flow}
\ket{00}_{\kappa,s} = (J^+_{-s/\kappa})^{s}\, \ket{00}_\kappa ~.
\ee
This relation is straightforward to see in the covering space description, where this state can be thought of in terms of the raising operator $(J^+_{-s})^s$ acting on the current algebra vacuum (recall the moding is rescaled by a factor $\kappa$ on the covering space).  The covering space $SU(2)$ current algebra has level one, and is entirely accounted for through bosonization.  The operator $(J^+_{-s})^s$ is a Virasoro highest weight operator of spin $s$ and dimension $s^2$, and therefore must be an exponential $\exp[i\sqrt2 \,s\widehat H]$ of the boson $J_3=i\partial \widehat H$ on the covering space, hence~\eqref{00 spec flow} indeed implements a spectral flow transformation.

Finally, in the $\bbT^4$ SCFT, spectral flow has a third interpretation in terms of shifting the Fermi sea of the $\chicover_{A\alpha}$, by populating all the modes in the Hilbert space up to and including level $s/
\kappa$.  It is straightforward to check that this leads to the shifts~\eqref{specflow} in energy and charge.  

\subsection{CFT duals of our superstrata}
\label{ss:CFT dual states}

We now combine the ingredients discussed above to develop our proposal for the CFT duals of our superstrata. 
We observed below Eq.\;\eq{angmomsty2-2} that our Style 2 solution, in the limit of $k=1$, is the extension to asymptotically flat space of a particular subset of solutions constructed in~\cite{Bena:2015bea}. The proposed dual CFT states~\cite{Bena:2015bea,Giusto:2015dfa} can be written in terms of spectral flow on chiral primary states.  
This suggests we look to similar states for candidate CFT duals of our solutions.
The fact that the wavenumber in $v$ is a fraction $1/k$ of the wavenumber in $\psi$ suggests that we consider ``fractional spectral flow'' states of the sort described above.

The orbifold projection on cycles of length $\kappa$ enforces integer momenta on each strand.  Consider spectral flow on Ramond sector $\ket{00}_\kappa$ cycles.  Integer $h-\bar h$ means that $\alpha^2\kappa\in \bbZ$ in Eq.\;\eqref{specflow states}.  One also wants $j_3 = k (h-\bar h)$ so that the state corresponds to a supergravity solution whose fields have a phase dependence which is a multiple of $\zeta=\frac{v}{2R}-k\varphi_2$ (Eq.\;\eqref{zetadefn}).  Equation~\eqref{specflow} then requires $\alpha\kappa = k \alpha^2 \kappa$; thus $\alpha = 1/k$.  
Therefore $\kappa$ should be a multiple of $k^2$ in order for $h-\bar h\in\bbZ$, i.e. $\kappa=k^2 \hat{p}$ for some integer $\hat{p}$.  
Then $\alpha=s/\kappa$ gives $s= k \hat{p}$.  Thus, one component of the candidate CFT dual for Style 2 coiffuring is fractional spectral flow by an amount $\alpha = 1/k$ on cycles $\ket{00}_{k^2 \hat{p}}$ of length $k^2\hat{p}$.  

A second candidate component of the CFT dual arises from spectral flow on Ramond sector cycles $\ket{++}_{\kappa'}$.  Applying the same logic as above, one finds the criteria of integer $h-\bar h$ and spectral flow yielding $j_3-\bar j_3 = k (h-\bar h)$ result in a spectral flow by amount $s'=k \hat{n}$, 
and cycle length $\kappa'=k(k\hat{n}+1)$, for some $\hat{n}\in\bbZ$.

Finally, a third candidate component of the CFT dual uses spectral flow on Ramond sector cycles $\ket{--}_{\kappa''}$.  Once again the spectral flow amount is 
$s''=k\hat{m}$, and the cycle length is $\kappa''=k(k\hat{m}-1)$ for some $\hat{m}\in\bbZ$.

The cycles excited by fractional spectral flow can also be expressed in terms of the action of $J_{-1/k}^+$, and we will find it convenient to do this.
For $\ket{00}$ strands this follows from equation~\eqref{00 spec flow}, and similarly for $\ket{\pm\pm}$ strands using the strand lengths and amounts of spectral flow above.
In addition, the supergravity solution is built on a ``ground state'' which is a supertube of radius $a$ and angular momentum $Q_1Q_2/4kR$, whose CFT dual consists of length $k$ cycles $\ket{++}_k$ whose number is proportional to $a^2$.  

A state which combines these supertube strands with the above longer cycles excited via fractional spectral flow has the form 
\be
\label{CFT dual state}
\Big(\ket{++}_k^{\strut}\Big)^{n_1}
\!\prod_{\hm,\hn,\hp} 
\Big( (J_{-1/k}^+)^{k\hat{p}}\ket{00}^{\strut}_{k^2\hat{p}}\Big)^{\ntwop}
\Big( (J_{-1/k}^+)^{k\hat{n}}\ket{++}^{\strut}_{k(k\hat{n}+1)}\Big)^{\nthreen}
\Big( (J_{-1/k}^+)^{k\hat{m}}\ket{--}^{\strut}_{k(k\hat{m}-1)}\Big)^{\nfourm}
\ee 
with appropriate conditions on the strand numbers so that the state carries the same quantum numbers as the supergravity solution, Eqs.\;\eqref{Z12osc1}--\eqref{T4osc1}.  
Of course, one can write the above tensor product with a single index $\hp$, but we will temporarily carry along $\hm$ and $\hn$ to emphasize the differences between the strands. 
Ultimately, our proposed dual CFT states will be coherent states built from superpositions of the states \eq{CFT dual state}, as discussed in~\cite{Skenderis:2006ah,*Kanitscheider:2006zf,Giusto:2015dfa}. 
The first check that we have the right class of states is to show that the appropriate supergravity field modes are turned on under a small deformation away from the parent supertube solution (i.e.~when the total number of copies taken up by the excited strands is small compared to the total number of copies taken up by the unexcited base supertube strands).

\refstepcounter{subsubsection}
\subsubsection*{\thesubsubsection ~~ Expectation values of supergravity mode operators}

One can regard the cyclic twist components $\ket{00}_{k^2\hp}$ or $\ket{++}_{k(\knpo)}$ or $\ket{--}_{k(\kmmo)}$ as excitations above the supertube ``ground state''.
Supergravity field modes turned on by the coiffuring procedure are associated to operators in the CFT having expectation values in the coherent states built from the states \eq{CFT dual state}. These operators will include those that have a matrix element that annihilates one of the long cycles, and converts it into multiple copies of $\ket{++}_k$.  Indeed, the order $k\hp$ anti-cyclic permutation
\be
\label{annihilator}
 \big(k\hp k,(k\hp\!-\!1)k,(k\hp\!-\!2)k,\dots,k\big) 
\ee
acting on the cycle
\be
\big(1,2,3,4,\dots,k^2\hp\big) 
\ee
results in the tensor product of $k\hp$ cycles of length $k$
\be
 \big(1,\dots,k\big)\, \big(k\!+\!1,\dots,2k\big)\, \big(2k\!+\!1,\dots,3k\big) \cdots \big((k\hp\!-\!1)k\!+\!1,\dots,k\hp k\big) ~,
\ee
and similarly an anticyclic permutation of length $\knpo$ can convert a cycle of length $k(\knpo)$ into $\knpo$ cycles of length $k$, and an anticyclic permutation of length $\kmmo$ can convert a cycle of length $k(\kmmo)$ into $\kmmo$ cycles of length $k$.

The operators that mediate the appropriate matrix elements for the state~\eqref{CFT dual state} must also soak up the currents that spectrally flow the state from the $1/4$-BPS ground state in this twist sector.  Each $k^2\hp$-cycle in this state carries $(J_3,\bar J_3)$ charges $(k\hp,0)$, while the final state has $k\hp$ extra cycles $\ket{++}_{k}$, with charges $(k\hp/2,k\hp/2)$, and so the operator that effects the transition must have charge $(-k\hp/2,k\hp/2)$.  Similarly, each $k(\knpo)$-cycle has $(J_3,\bar J_3)=(k\hn+\half,\half)$, while the final state has charge $(\half(\knpo),\half(\knpo))$, and so the operator that mediates the transition must have charge $(-k\hn/2,k\hn/2)$; and each $k(\kmmo)$-cycle has $(J_3,\bar J_3)=(k\hm-\half,-\half)$, while the final state has charge $(\half(\kmmo),\half(\kmmo))$, and so the operator that mediates the transition must have charge $(-k\hm/2,k\hm/2)$.

The BPS twist operator whose conjugacy class contains the $k\hat{p}$-cycle~\eqref{annihilator} and which also has these $SU(2)_\cR$ quantum numbers, and obeys selection rules of the auxiliary $SU(2)_\cA$, is the NS sector operator
\be
\label{specflow vev 1}
\epsilon_{AB}\,(J^-_0)^{k\hp}\, \cO^{(k\hp)AB}_{k\hp\!+\!1,k\hp\!+\!1} ~.
\ee
This component operator has $(J_3,\bar J_3)=(-k\hp/2,k\hp/2)$ and so carries the appropriate $\cR$-charges to implement the matrix element that sends $(n_1,n_2,n_3,n_4)$ to $(n_1\!+\!k\hp,n_2\!-\!1,n_3,n_4)$.  Similarly, the NS sector operator
\be
\label{specflow vev 2}
(J^-_0)^{k\hn}\, \cO^{(\knpo)}_{\knpo,\knpo} 
\ee
has $(J_3,\bar J_3)=(-k\hn/2,k\hn/2)$ and so carries the appropriate quantum numbers to mediate the transition $(n_1,n_2,n_3)\to (n_1\!+\!\knpo,n_2,n_3\!-\!1,n_4)$; and the operator
\be
\label{specflow vev 3}
(J^-_0)^{k\hm}\, \cO^{(\kmmo)}_{\kmpo,\kmpo} 
\ee
has the $SU(2)_\cR$ quantum numbers $(J_3,\bar J_3)=(-k\hm/2,k\hm/2)$ and mediates the transition
$(n_1,n_2,n_3,n_4)\to (n_1\!+\!\kmmo,n_2,n_3,n_4\!-\!1)$.%
\footnote{The operator~\eqref{specflow vev 1} is the spectral flow to the NS sector of the Ramond operator corresponding to the state $\ket{00}_{k\hp}$ (flowed in opposite directions on left and right), while the operator~\eqref{specflow vev 2} is the spectral flow to the NS sector of the Ramond operator corresponding to the state $\ket{--}_{\kppo}$, and the operator~\eqref{specflow vev 3} corresponds to the NS to R flow of the state $\ket{++}_{\kmmo}$.}

Ward identities for the conformal group $SL(2,R)\times SL(2,R)$ guarantee a dependence $\exp[i\hp v/2R]$ for the matrix elements mediated by the operator~\eqref{specflow vev 1}.  The $SU(2)\times SU(2)$ $\cR$-symmetry quantum numbers of this operator ensures that the matrix elements have the angular dependence
\be
\label{ang dep}
\exp\!\big[{ik\hp}\,\varphi_2/2\big] \cos^{k\hp}\theta
\ee
on $\bbS^3$.  This angular dependence equates to the fact that the operator~\eqref{specflow vev 1} is a twisted chiral operator ($SU(2)_{\cR}$ lowest weight on the left and highest weight on the right).  Similarly, the operators~\eqref{specflow vev 2} and~\eqref{specflow vev 3} are twisted chiral operators that mediate analogous matrix elements, whose coordinate dependences are the same apart from the substitution $\hp\to \hn$ and $\hp\to \hm$, respectively.  The leading asymptotic power of $r$ in the matrix element is also dictated by the scale dimension of the operator, and a matching power of $a^{k\hp}$ in the quantity $\Delta^{k\hp}$ comes from the number of supertube $k$-cycles created when a cycle of length $k^2\hp$ is annihilated.

The general solution of type IIB supergravity compactified on $\bbT^4\times \bbS^1$
that preserves the same supercharges as the F1-NS5-P system and is
invariant under rotations of $\bbT^4$ has the form 
\begin{subequations}\label{ansatzSummary}
\allowdisplaybreaks
 \begin{align}
d s_{10}^2 &=-\frac{2Z_2}{\cP}\,\big(d v+\beta\big)\,\Big[d u+\omega + \frac{\mathcal{F}}{2}\big(d v+\beta\big)\Big]
+Z_2\,d s^2_4  +  d \hat{s}^2_{4}  ~,  \label{10dmetric}\\
e^{2\Phi}&=\frac{Z_2^2}{\cP}\, ,\\
B_2 &= -\frac{Z_2}{\cP}\,(d u+\omega) \wedge(d v+\beta) ~,\\
B_6 &=\widehat{\mathrm{vol}}_{4} \wedge \left[ -\frac{Z_1}{\cP}\,(d u+\omega) \wedge(d v+\beta)\right] +\dots\notag\\
C_0&=\frac{Z_4}{Z_2}\, ,\\
C_2&= \frac{Z_4}{\cP}\,(d u+\omega) \wedge(d v+\beta)+ \dots ~, \\
C_4 &= \frac{Z_4}{Z_2}\, \widehat{\mathrm{vol}}_{4} +\dots ~, \\
\end{align}
\end{subequations}
with
\begin{equation}
\cP   \equiv     Z_1 \, Z_2  -  Z_4^2 ~.
\label{Psimp-2}
\end{equation}
Here $ds^2_{10}$ is the ten-dimensional string-frame metric, $ds_4^2$ is the metric on the space transverse to the branes, $\Phi$ is the dilaton, $B_p$ and $C_p$ are the NS-NS and RR gauge forms. (It is useful to also list $B_6$, the 6-form dual to $B_2$, to make explicit the appearance of $Z_1$ and $Z_2$ as the magnetic and electric components of the NS B-field.) The flat metric on $\bbT^4$ is denoted by $d \hat{s}^2_4$ and the corresponding volume form by $\widehat{\mathrm{vol}}_{4}$.  For further discussion, see~\cite{Skenderis:2006ah,*Kanitscheider:2006zf,Giusto:2013rxa,Bena:2015bea,Giusto:2015dfa}. 

In the supergravity solution~\eqref{ansatzSummary}, the harmonic function $Z_4$ appears in the RR scalars $C_0$ and $C_4$ as well as in the six-dimensional $C_2$ tensor field in the F1-NS5 frame, and carries the quantum numbers leading to the angular dependence~\eqref{ang dep}.   
The operator~\eqref{specflow vev 1} corresponds (in the F1-NS5 duality frame) to the scalar $C_0 v_4 - C_4$ and the six-dimensional tensor $C_2^-$, according to~\eqref{torus moduli table}, and its matrix elements carry the appropriate angular dependence.
Thus for both coiffuring styles, we expect that there should be a vev of this operator proportional to $b_4$.  When one builds coherent states out of the building blocks~\eqref{CFT dual state}, one determines the average number of of strands $\bar n_2$ such that it reproduces this vev~\cite{Skenderis:2006ah,*Kanitscheider:2006zf,Giusto:2015dfa}.

The harmonic functions $Z_{1,2}$ appear in the electric and magnetic components of the six-dimensional NS B-field and the dilaton in this duality frame, with angular dependence of the form~\eqref{ang dep}, where $\hp=n=p$ for Style 1 and $\hp=n=2p$ for Style 2.  The operators~\eqref{specflow vev 2}, \eqref{specflow vev 3} correspond to the supermultiplet~\eqref{graviton table} containing the six-dimensional NS B-field and the dilaton.  Their matrix elements also have angular dependence of the form~\eqref{ang dep}, with the same replacements for the two coiffuring styles, and imply the corresponding vevs for the dual CFT coherent states.

Thus we have all the ingredients to reproduce the coiffured supergravity solutions of Section~\ref{Sect:MomST} from the CFT, and it is natural to anticipate that the average numbers of excited $\ket{00}_{k^2\hat{p}}$ strands $\bar{n}_{2,\hat{p}}$ will be related to the coefficient $b_4$, and average number of excited $\ket{++}_{k^2\hat{n}+k}$ and $\ket{--}_{k^2\hat{m}-k}$ strands $\bar{n}_{3,\hat{n}}$, $\bar{n}_{4,\hat{m}}$ will be related to the coefficients $b_{1}$, $b_{2}$. This is indeed what we will find.

The coiffuring construction imposes relations on the mode amplitudes and frequencies in order that the supergravity solution is regular at $r=0$.  These restrictions are not apparent in the CFT states in the  linearized analysis, which, {\it a priori}, allows independent values for the cycle length quantum numbers $(\hp,\hn,\hm)$ and the corresponding amplitudes $(n_{2,\hat{p}},n_{3,\hat{n}},n_{4,\hat{m}})$.  

For coiffuring Style 1 in supergravity, one has $n=p$ and the amplitude relations~\eqref{breln1}-\eqref{breln2}; for Style 2, one has $n=2p$ together with the amplitude restrictions $b_2=0$ and~\eqref{breln3}.  For Style 2, at leading order there is no amplitude for $b_1$ (since $b_1\sim b_4^2$) and $b_2=0$, hence there is no linearized vev for the NS B-field; this suggests that Style 2 corresponds to a state with $n_{3,\hat{n}}=n_{4,\hat{m}}=0$ in the CFT.  The vev of the B-field at second order in $b_4$ could be accounted for by the non-linearities in the CFT-supergravity mode map, and indeed we will show this to be true in the next subsection.  

For Style 1, an amplitude at leading order in $b_4$ is present for all of the $m=n=p$ modes of the NS B-field, but in the CFT it seems that the amplitude of the corresponding vevs can be independently varied~-- at this point there appears to be no restriction on the relative numbers $(n_{2,\hp},n_{3,\hn},n_{4,\hm})$ of the different kinds of strands at leading order.  Again, to understand these restrictions, it is necessary to understand the  relation between supergravity modes and CFT fields at the non-linear level.  

Instead of trying to carry through the somewhat daunting task of determining the non-linear corrections to the supergravity-CFT map, we will instead proceed somewhat differently, and determine what strands are present in the CFT (and in what amounts) by an analysis of the two-charge solutions on which the coiffured solutions are based.

\refstepcounter{subsubsection}
\subsubsection*{\thesubsubsection ~~ Information from two-charge solutions}

Our proposed dual CFT states involve fractional spectral flow on a two-charge \nBPS{4} state, and spectral flow does not change the strand content of a BPS state.  Therefore, we can determine the amounts of the various types of strands present (at the fully non-linear level) in both styles of coiffuring, by studying the known map between CFT and supergravity for the two-charge system%
~\cite{Lunin:2001jy,Lunin:2002bj,Skenderis:2006ah,*Kanitscheider:2006zf,Kanitscheider:2007wq}.

The harmonic functions determining the geometry of a circular F1-NS5 supertube in the decoupling limit are
\begin{subequations}\label{generaltwocharge}
\begin{align}
\label{Z2profile}
& Z_2 = \frac{Q_2}{L} \int_0^{L} \frac{1}{|x_i -g_i(\v)|^2}\, d\v~, \qquad
  Z_4 = - \frac{Q_2}{L} \int_0^{L} \frac{\dot{g}_5(\v)}{|x_i -g_i(\v)|^2} \, d\v \,,\\ 
\label{Z1profile}
& Z_1 = \frac{Q_2}{L} \int_0^{L} \frac{|\dot{g}_i(\v)|^2+|\dot{g}_5(\v)|^2}{|x_i -g_i(\v)|^2} \, d\v ~, 
\\
& A = - \frac{Q_2}{L} \int_0^{L} \frac{\dot{g}_j(\v)\,dx^j}{|x_i -g_i(\v)|^2} \, d\v ~, \quad~~
 dB = - *_4 dA~, \quad~~ ds^2_4 = dx^i dx^i~, \\
& \beta = \frac{-A+B}{\sqrt{2}}~, \qquad \omega = \frac{-A-B}{\sqrt{2}}~, \qquad {\mathcal{F}}=0~,
\end{align}
\end{subequations}
where the dot on the profile functions indicates a derivative with respect to $\v$ and $*_4$ is the dual with respect to the flat transverse $\mathbb{R}^4$ parametrized by $x_i$.

The onebrane charge is given by
\begin{equation}
\label{Q1int}
 Q_1={Q_2\over L}\int_0^L \bigl(|\dot{g}_i(\v)|^2+|\dot{g}_5(\v)|^2\bigr)d\v ~.
\end{equation}
The quantities $Q_1$, $Q_2$ are related to quantized onebrane and fivebrane
numbers $n_1$, $n_5$ by
\begin{equation}
Q_1 = \frac{(2\pi)^4\,n_1\,g_s^2\,\alpha'^3}{V_4}\,,\qquad Q_2 = n_5\,\alpha' ~,
\label{Q1Q5_n1n5}
\end{equation}
where $V_4$ is the coordinate volume of $\bbT^4$. 

The circular supertube profile is is given by
\be
\label{circular}
g_1+ig_2 = a \exp[2\pi i k \v/L] ~.
\ee
It will prove convenient to denote $x=x_1+ix_2$, $y=x_3+ix_4$, and parametrize the profile by $\xi\equiv2\pi k\v/L$. Since the supertubes of interest run around the same profile $k$ times, the integral is simply $k$ times the integral over the range $\xi\in(0,2\pi)$.
The further change of variables $z=e^{i\xi}$, and the use of $\bar z =1/z$ for an integral along the unit circle in $z$, converts the integrals into contour integrals for which we can use the method of residues, for example
\be
\label{Hint}
Z_2=\frac{Q_2}{2\pi i}\oint \frac{dz}{z} \frac{1}{(x-a z)(\bar x - a/z)+y\bar y}  ~.
\ee 
The poles in the integrand are located at
\be
z_\pm = \frac{\tilde{w} \pm\sqrt{\tilde{w}^2-4x\bar x a^2}}{2\bar x a} ~,
\ee
where $\tilde{w}=x\bar x+y\bar y+a^2$, and so
\be
Z_2=\frac{Q_2}{\sqrt{\tilde{w}^2-4x\bar x a^2}} ~.
\ee
Converting from Cartesian coordinates to spherical bipolar ones
\begin{align}
x = \tilde r \sin\tilde\theta e^{i\varphi_1}
~~&,\quad
y = \tilde r \cos\tilde\theta e^{i\varphi_2} \nn\\
\tilde r =\sqrt{r^2 +a^2\sin^2\theta}
~~&,\quad
\cos\tilde\theta = \frac{r\cos\theta}{\sqrt{r^2+a^2\sin^2\theta}}
\end{align}
leads to the correct form of $Z_2$ in the decoupling limit,
\be
\label{Hanswer}
Z_2 = \frac{Q_2}{r^2+a^2\cos^2\theta} =\frac{Q_2}\Sigma ~.
\ee
Next, we introduce  $\nu = kp$ for convenience and we add a $g_5$ term to the profile function,
\be
g_5(\v) ~=~ - \frac{2b_4}{\nu k R_y} \sin \left( \frac{2\pi k}{L}  \nu \, \v \right) ~=~ \frac{-b_4}{i\nu k\R} \, (z^\nu-z^{-\nu})  \,,
\label{g5}
\ee
where $b_4$ is real, and corresponds to the magnitude of the quantity $b_4$ in the supergravity~\eq{Z4osc1}. The quantity that corresponds to the phase of the supergravity $b_4$ is a shift in $\v$ in \eq{g5}. 
In what follows, for both Style 1 and Style 2, we will take $b_4$ to be real, both for convenience and for ease of comparison to~\cite{Bena:2015bea}.
This $g_5$ term in the profile function gives rise to the following contour integral expression for the harmonic function $Z_4$:
\be
\label{Aint}
Z_4 = b_4\frac{1}{2\pi i}\oint \frac{dz}{z} \frac{z^\nu+z^{-\nu}}{(x-a z)(\bar x-a/z)+y\bar y} \,.
\ee
The $z^\nu$ term yields the result
\be
b_4\,\frac{ z_-^\nu}{(-a\bar{x})(z_- - z_+)} ~.
\ee
The denominator gives again the factor of $\Sigma$;
furthermore, one has
\be
\frac{z_-}{z_+} ~=~ \frac{a^2\sin^2\theta}{r^2+a^2} \,, \qquad\quad z_+ z_-  ~=~ e^{2i\varphi_1} \,.
\ee
One can then rewrite $z_-^\nu$ as
\be
\label{2-charge Delta}
z_-^\nu ~=~ \left( \frac{z_-}{z_+} \right)^{\nu/2} \! (z_+ z_-)^{\nu/2} ~=~ \Big(\frac{a^2 \sin^2\theta}{r^2+a^2}\Big)^{\nu/2} 
e^{i\nu\varphi_1} ~.
\ee
The harmonic function depends on the combination $\,\sin\theta \,e^{i\varphi_1}\,$ rather than $\,\cos\theta\, e^{i\varphi_2}\,$ because the linearized supergravity modes getting a vev correspond to chiral rather than twisted-chiral operators in the CFT.  The fractional spectral flow operation converts one to the other.

One is looking to match the structure in~\cite{Bena:2015bea} equation (3.11c) which is
\be
\left(\frac{a^2 \sin^2\theta}{r^2+a^2}\right)^{\nu/2} \frac{\cos \nu\varphi_1}{\Sigma}
~=~ \frac{(z_-/z_+)^{\nu/2} \big(  (z_+ z_-)^{\nu/2} +  (z_+ z_-)^{-\nu/2}\big)}{2(-a\bar{x})(z_- -z_+)} ~;
\ee
this is exactly what is found once the contribution from the $z^{-\nu}$ term in~\eqref{Aint} is added. So the seed $Z_4$ is
\be
Z_4 ~=~ 2 b_4 \left(\frac{a^2 \sin^2\theta}{r^2+a^2}\right)^{\nu/2} \frac{\cos \nu\varphi_1}{\Sigma} \,.
\ee

\refstepcounter{subsubsection}
\subsubsection*{\thesubsubsection ~~ Style 2}

For coiffuring Style 2, the harmonic function $Z_1$ exhibited in~\cite{Bena:2015bea} equation (3.11a) for the corresponding two-charge seed solution, translated into our conventions is
\be
\label{Z1 style2}
Z_1 ~=~ \frac{Q_1}{\Sigma} + \frac{2b_4^2}{Q_2} \left(\frac{a^2 \sin^2\theta}{r^2+a^2}\right)^{\nu} \frac{\cos 2\nu\varphi_1}{\Sigma}
\ee
which follows from equation~(\ref{generaltwocharge}b).  
Thus we see that Style 2 coiffuring is the result solely of exciting $\ket{00}$ strands of length $\kappa=k\nu=k^2p$; the strength $b_1\propto b_4^2$ of the vev is entirely accounted for by non-linear effects of the $\ket{00}$ strands, and so no additional contribution corresponding to nonzero $n_{3,\hat{n}},n_{4,\hat{m}}$ is necessary.  The corresponding two-charge solution is precisely as in~\cite{Bena:2015bea}, and the spectral flow that adds the third charge simply turns vevs from chiral to twisted-chiral~-- under fractional spectral flow, the factor $\,\sin^\nu\theta\,e^{i\nu\varphi_1}\,$ turns into $\,\cos^\nu\theta\,e^{i\nu\varphi_2}\,$, which is what we see in the coiffured harmonic functions of Section~\ref{Sect:MomST} above.

\refstepcounter{subsubsection}
\subsubsection*{\thesubsubsection ~~ Style 1}
\label{sec:sty1cft}

It remains to match Style 1 to a set of supertube strands in a two-charge solution prior to spectral flow and coiffuring.  We expect to at least have strands of length $k(kp+c)$ for $c=\{-1,0,+1\}$, since vevs for the operators~\eqref{specflow vev 1}-\eqref{specflow vev 3} must appear at linear order in $b_4$.  The $c=\pm1$ strands correspond to $\ket{\pm\pm}$ cycles and so in general will affect the location of the supertube profile in the transverse~$\bbR^4$.  
Introducing $\ket{++}_{k(kp+1)}$ and $\ket{--}_{k(kp-1)}$ strands, the deformation profile becomes
\be
\label{style 1 profile}
g_1+ ig_2 = a z + b_- z^{-(kp-1)} + b_+ z^{kp+1} = z \big( a+b_-z^{-\nu} +b_+z^{\nu}\big) ~,
\ee
where $\nu=kp$ and $z=\exp(2\pi i k\v/L)$.

For small amplitude deformation $b_\pm\ll a$, there are no new poles inside the contour of integration, and the pole in the integrand will still be close to $z_{-}$.  
We can map the profile back to a unit-velocity circular profile (i.e. $g_{1}+ig_2 = a e^{i\u}$) via a single-valued conformal map, at the cost of a Jacobian for the transformation.  In general this leads to an infinite series in the expressions for $Z_{1}$ and $Z_2$ if there are only one or two lengths of strand in the profile $g_{1}+ig_2$; since we wish to engineer a finite Fourier series for $Z_{1}$ and $Z_2$, the dual CFT state will have a series of strand lengths involving all possible multiples of $p$.  

Working firstly to leading order in $b_\pm$, consider the profile
\be
(g_1+ig_2)(\v) = a \exp\big[ i \u\big(\xi(\v)\big)\big] \,,\qquad \xi(\v) =\frac{2\pi k\v}{L} \,, \qquad 
\u(\xi) = \xi - \frac{b}{\nu} \cos\nu \xi + \dots
\ee
where we set $b_+=b_-\equiv - i ab/2\nu$ in order that the map is a proper element of ${\it Diff}(\bbS^1)$. Here again $\xi$ serves as a rescaled periodic coordinate which ranges over $[0,2\pi k)$. The motivation for considering such a profile comes from coiffuring -- the idea is that coordinate transformations on the supertube worldvolume apply a density perturbation to the round supertube without perturbing its location in space.  The fivebrane and onebrane charge densities will no longer be constant along the supertube.
Expanding this profile out to leading order in $b$ reproduces~\eqref{style 1 profile}.  Such a change of variable has no effect on $Z_4$ (which is reparametrization invariant) but it will change $Z_{1}$ and $Z_2$. The integration measure picks up a factor
\be
d\v \to d\u \Big(\frac{d\u}{d\v}\Big)^{-1} , \qquad  
\frac{d\u}{d\v} = \frac{d\u}{d\xi}\frac{d\xi}{d\v}
\ee
where we have expanded the Jacobian factor to clarify that $\u$ means 
$\u\big(\xi(\v)\big)$ as above.
Similarly the ``energy densities'' in the numerator of $Z_1$ in~\eqref{generaltwocharge} pick up a factor
\be
| (\dot{g}_1+i\dot{g}_2)(\v)|^2 + |\dot g_5(\v)|^2 \to  \Big(\frac{d\u}{d\v}\Big)^{2}\big( | (g'_{1}+ig'_{2})(\u)|^2 + | g'_5(\u)|^2 \big) 
\ee
where primes denote derivatives with respect to the argument.
Again evaluating the factors of $(d\u/d\v)$ to leading order in $b$ one finds that in the new integration variable $\u$, the integrand of $Z_2$ is modified by a factor of $1-b\sin \nu \u$ in~\eqref{Z2profile}, while the integrand of $Z_1$ gets a factor of $1+b\sin\nu \u$ (in addition to the corresponding factors of $2\pi k/L$).  We then find the same sort of integral we encountered in~\eqref{Z1profile}, with the same result.  Our primitive approximations give $b_1=-b_2$, which is appropriate for $a^2\ll Q_1,Q_2$; the latter is a consequence of the decoupling limit.

In principle one can proceed order-by-order in a series expansion in $b_I$, $(I=1,2,4)$, working out the non-linear map between the $\v$ coordinate frame in which the strand content is specified, and the $\u$ coordinate frame in which the supertube profile is a constant velocity parametrization of a circle.  However, ultimately we are interested in the harmonic functions $Z_{I}$ having a single non-trivial Fourier coefficient.  In Style 1, the perturbation to $Z_2$ looks like~\eqref{Aint} with $\nu=kp$ and a coefficient $b_2$; and $Z_4$ is the same but with a coefficient $b_4$.  These simple forms suggest that the more straightforward route is to work directly in the $\u$ coordinate frame and only implicitly specify the coordinate map via its inverse,
\be
\label{implicit map}
\v(\xi) = \frac{L}{2\pi k} \xi \,, \qquad\quad \xi(\u) = \u + \frac{b}{\nu } \cos\nu \u \,.
\ee
Plugging this into~\eqref{Z2profile} gives exactly the right result for $Z_2$ and $Z_4$ in Style 1, using
\bea
(g_1+ig_2)(\v) ~=~ a \exp\big[ i \u\big(\xi(\v)\big)\big] \,, \qquad 
g_5(\v) ~=~ - \frac{2 b_4}{\nu k\R} \, \sin \big[ \nu \, \u\big(\xi(\v)\big) \big].
\eea
We have now used up almost all our freedom to specify the state; all that remains are the amplitudes $b$, $b_4$.  The integral for $Z_1$ is 
\begin{align}
\label{Kint}
Z_1 &= \frac{k^2\R^2}{2\pi Q_2} \int_0^{2\pi} \!d\u \Big(\frac{d\u}{d\xi}\Big)^{-1}\,
\frac{\big(| (g'_{1}+ig'_{2})(\u)|^2 + | g'_5(\u)|^2\big)(d\u/d\xi)^2}{|x-(g'_{1}+ig'_{2})(\u)|^2+|y|^2}
\nn\\
&= \frac{k^2\R^2}{2\pi Q_2}\int_{0}^{2\pi}\!\!d\u\; 
\frac{a^2+(2b_4/k\R)^2\cos^2\nu \u}{1-b \sin\nu \u}\,\frac{1}{|x-a e^{i\u}|^2+|y|^2}
\end{align}
where we have used the relation
\be
L=\frac{2\pi Q_2}{\R} ~.
\ee
Let us choose 
\be
b^2 ~=~ \frac{(2b_4/k\R)^2}{(2b_4/k\R)^2+a^2} ~=~ \frac{4b_4^2}{Q_1Q_2}  ~,
\ee
where the second equality is the analog of the gravity regularity condition \eq{STreg4}. Then the factor in the integrand becomes
\be
\frac{a^2+(2b_4/k\R)^2\cos^2\nu \u}{1-b \sin\nu \u} 
= (a^2+(2b_4/k\R)^2)\big(1+b\sin\nu \u \big)  
= \frac{Q_1Q_2}{k^2\R^2} \big(1+b\sin\nu \u \big) ~,
\ee
where the last equality comes from evaluating the expression~\eqref{Q1int}.
All harmonic functions have only terms that are constant or a single $\bbS^3$ harmonic of the form~\eqref{2-charge Delta}, with $m=n=p$ and $b_1=-b_2= i b_4/\sqrt{Q_1Q_2}$.  These results agree precisely with the decoupling limit $a^2\ll Q_1,Q_2$ of the amplitude relations~\eqref{breln0} of Style 1 coiffuring in supergravity.

For these two-charge solutions, the mode amplitude restrictions do not come from requiring regularity of the supergravity solution -- all the two-charge solutions are non-singular.  Rather, the restriction comes from the somewhat arbitrary requirement that the harmonic functions contain only a single Fourier mode rather than a combination of modes of different wavenumbers.

It is worth reiterating that the map between supergravity and CFT takes place in the $\v$ coordinate frame, which is only implicitly specified above through the relation~\eqref{implicit map}.  In the $\v$ coordinates the solution is very complicated and has in principle all values of $\hm,\hn,\hp$ turned on.  The non-zero values of $\hm$, $\hn$ are given by the non-zero Fourier coefficients of $(g_1+ig_2)(\v) = a \exp[ i \u(\xi(\v))]$,
\be
\label{Fcoeff}
c_n = \int_{0}^{2\pi} \frac{d\xi}{2\pi}\, e^{-in \xi} \, e^{i\u(\xi)} 
= \int_{0}^{2\pi} \frac{d\u}{2\pi} \frac{d\xi}{d\u} e^{-in\xi(\u)+i\u}
\ee
and are predominantly concentrated on the lowest modes.  Expanding in $b$,
\be
c_n = \int_0^{2\pi}\! \frac{dw}{2\pi}\, \big(1+b\cos(\nu\u)\big)\,e^{i(1-n)w}
\sum_{\ell=0}^\infty \frac1{\ell!}\Big(\frac{-inb\cos\nu\u}{\nu}\Big)^\ell 
\label{eq:Fourier-sty1}
\ee
one sees that the only nonzero Fourier coefficients occur for $n=1+q\nu$, $q\in\bbZ$, generalizing~\eqref{style 1 profile}. Of these non-zero Fourier coefficients, the positive values of $n$ give the non-zero values of $\hm$, and the negative values of $n$ give the non-zero values of $\hn$. Thus we see that $\hm$ and $\hn$ must be multiples of $p$, the mode number of the supergravity solution. Similarly, the non-zero values of $\hp$ are all multiples of $p$.

In addition, reality of the conformal map implies $c_{1+q\nu}=c_{1-q\nu}^*$, which in turn means that for each $q$, the average numbers of $\ket{++}_{k(qkp+1)}$ and $\ket{--}_{k(qkp-1)}$ strands are equal.
Finally, note that in the quantum theory, there is a maximum mode number $N=N_1N_5$ and so one cannot precisely generate Style 1 because the Fourier expansion is necessarily finite; the result will differ at the $1/N$ level.

For $kp=1$, this family of states has a somewhat degenerate limit, since the length of the $\ket{--}_{k(kp-1)}$ strands is zero. This simply means that this particular strand type is absent for $kp=1$, while the other strands remain as described above. In particular, the average numbers of $\ket{++}_{k(qkp+1)}$ and $\ket{--}_{k(qkp-1)}$ strands are equal for $q\ge2$.

\refstepcounter{subsubsection}
\subsubsection*{\thesubsubsection ~~ Summary of proposed dual CFT states}

In both Style 1 and Style 2, we start with a two-charge seed solution, determined by a profile function.
The general dictionary for two-charge states is discussed in \cite{Skenderis:2006ah,*Kanitscheider:2006zf,Kanitscheider:2007wq,Giusto:2015dfa}.  We now describe how it applies to our two-charge seed states.
Given a profile function, the non-zero Fourier coefficients specify the types of strands involved in the dual CFT state, and the values of the Fourier coefficients control the coefficients of the individual terms in the coherent state superposition. 

\vspace{2mm}
\noindent
{\bf Style 2}

As shown in the previous subsection, the Style 2 seed solution is determined by the profile function \eq{circular}, \eq{g5}:
\bea
(g_1+ig_2)(\v) ~=~ a \exp\left(i \frac{2\pi k}{L} \v \right) \,, \qquad 
g_5(\v) ~=~ - \frac{2b_4}{\nu k R_y} \sin \left( \frac{2\pi k}{L}  \nu \, \v \right).
\label{eq:sty2profile}
\eea
Since both $g_1+ig_2$ and $g_5$ have only a single Fourier mode, the dual CFT state contains just two types of strands,
\be
\label{CFT dual state-sty2seed}
\ket{++}_k^{\strut}  \,, \qquad\quad
\ket{00}^{\strut}_{k^2 p} \,,
\ee 
where the excited strands are only of one type, given by $\hat{p}=p$.

To form the coherent state, one considers all partitions of the $N_1 N_5$ copies of the CFT into strands of the two above types. Then one forms a sum in which the coefficients of these different partitions are controlled in a specific way by the two non-zero Fourier coefficients of the profile function \eq{eq:sty2profile} (for more details, see in particular the discussion in~\cite{Giusto:2015dfa}). 

Given this seed two-charge state, we excite all strands except for the $\ket{++}_k$ strands in the way described in Section \ref{ss:CFT dual states}, so that the resulting three-charge state is composed only of strands of type
\be
\label{CFT dual state-sty2}
\ket{++}_k^{\strut}  \,, \quad~~
(J_{-1/k}^+)^{k p}\ket{00}^{\strut}_{k^2 p} \,.
\ee 
The coefficients in the coherent state sum remain as in the two-charge seed solution. 

\vspace{2mm}
\noindent
{\bf Style 1}

For Style 1, the seed solution is given by the profile function
\bea
(g_1+ig_2)(\v) ~=~ a \exp\big[ i \u\big(\xi(\v)\big)\big] \,, \qquad 
g_5(\v) ~=~ - \frac{2b_4}{\nu k\R} \, \sin \big[ \nu \, \u\big(\xi(\v)\big) \big]
\label{eq:sty12chseed}
\eea
where from \eq{implicit map} we specify the map implicitly through its inverse,
\be
\label{implicit map-1}
\v(\xi) = \frac{L}{2\pi k} \xi \,, \qquad\quad \xi(\u) = \u + \frac{b}{\nu } \cos\nu \u \,.
\ee
Since both $g_1+ig_2$ and $g_5$ have an infinite Fourier series, 
the types of CFT strands present are those of type
\be
\ket{++}_k^{\strut} \,, \qquad
 \ket{00}^{\strut}_{k^2\hat{p}}\,, \qquad
\ket{++}^{\strut}_{k(k\hat{n}+1)}\,, \qquad
\ket{--}^{\strut}_{k(k\hat{m}-1)} \,,
\ee
where $\hat{m}$, $\hat{n}$, $\hat{p}$ can be any independent multiples of $p$, compatible with the total number of strands being $N_1 N_5 $. The  coherent state has many more ingredients, however the coefficients in the superposition are again fully specified by the Fourier coefficients of the profile function \eq{eq:sty12chseed}.
Therefore all the coefficients are determined by the parameter $b_4$ (since $b_4$ fixes $b$ and $a$).

Given this seed two-charge state, we again excite all strands except for the $\ket{++}_k$ strands in the way described in Section \ref{ss:CFT dual states}, so that the resulting three-charge state is composed only of strands of type
\be
\label{CFT dual state-gen}
\ket{++}_k^{\strut}  \,, \quad~~
(J_{-1/k}^+)^{k\hat{p}}\ket{00}^{\strut}_{k^2\hat{p}}  \,, \quad~~
(J_{-1/k}^+)^{k\hat{n}}\ket{++}^{\strut}_{k(k\hat{n}+1)}  \,, \quad~~
(J_{-1/k}^+)^{k\hat{m}}\ket{--}^{\strut}_{k(k\hat{m}-1)}  \,,
\ee 
where again the values of $\hat{m}$, $\hat{n}$, $\hat{p}$ are independent multiples of $p$, compatible with the total number of strands  being $N_1 N_5 $, and the coefficients in the coherent state sum remain as in the two-charge seed solution.

Finally,  note that, at the level of counting free parameters in the solutions, we expect there to be good agreement more generally between coiffured deformations of circular supertubes on the supergravity side, and fractional spectral flows of circular two-charge seed solutions on the CFT side.  On the CFT side, one has two functional degrees of freedom~-- the specification of the profile of the $\ket{00}$ strands embodied in the function $g_5$, and the diffeomorphism $\u(\xi)$ that changes the parametrization of the round supertube.  On the supergravity side, the diffeomorphism $\u(\xi)$ corresponds to the charge densities $\rhoone$ and $\rhofour$ discussed in Section~\ref{ss:SIexample}, and the profile of the $\ket{00}$ strands corresponds to the function $Z_4$. In Section~\ref{ss:SIexample} we saw that in the absence of $Z_4$ there are three functions and two functional constraints, leaving one functional degree of freedom; adding in $Z_4$ gives two functional degrees of freedom, which agrees with the CFT.

There are interesting parallels between the supergravity construction of Section \ref{Sect:SpecInter} and the appearance of density fluctuations in the CFT.
However the relationship is not direct. In the CFT, the density profile appears in the two-charge seed solutions before applying fractional spectral flow; on the gravity side, the density perturbations were introduced in a spectrally inverted frame, and then  a second spectral inversion was applied to transform back to the original frame.   The density fluctuations were thus applied to a supertube that does not have a simple, direct relationship to the original D1-D5 CFT.   There is also the technical distinction in that the construction of Section \ref{Sect:SpecInter} initially involves three apparently independent charge density functions that must then satisfy the constraints of supertube regularity (\ref{regcondb}) and (\ref{regcondc}), leaving only one independent density function.  In this section, the density fluctuation is introduced via a combination of the $g_5$ profile and (in Style 1) a conformal map of the round supertube profile, which, via the Lunin-Mathur map, automatically maintains the supertube regularity conditions.  It would be very interesting to investigate this relationship in more detail.

\subsection{Comparison of conserved charges}

We now compare the angular momenta $J^3$, $\bar{J}^3$ and the momentum charge $Q_P$, and demonstrate the agreement between our supergravity solutions and our proposed dual CFT states.  For ease of comparison to the supergravity discussion in Section~\ref{ss:STReg}, we revert to the D1-D5 duality frame.

The discussion that follows requires a certain amount of notation to write the charges explicitly, however the reasons that underlie the agreement can be stated simply. 

Firstly, all our momentum excitations can be expressed in terms of the action of powers of $J^+_{-\frac{1}{k}}$. Secondly, for each $\hat{p}$ the average numbers of the strands of length $k^2\hat{p}+k$ and $k^2\hat{p}-k$ are equal, because of their origin as the (real-valued) two-charge density profile. Therefore adding momentum $\hat{p}$ requires, on average, $k^2\hat{p}$ strands of the CFT. This fact leads to the relation between the angular momenta $J^3$, $\bar{J}^3$ and the momentum charge $Q_P$ observed in the supergravity, as we now show explicitly.

\refstepcounter{subsubsection}
\subsubsection*{\thesubsubsection ~~ Style 2}

For Style 2, we have a coherent state which is a sum of terms of the form
\be
\label{CFT dual state-style2}
\Big(\ket{++}_k^{\strut}\Big)^{n_1} 
\;\Big( (J_{-1/k}^+)^{kp}\ket{00}^{\strut}_{k^2p}\Big)^{n_2} \,,
\ee
where the sum runs over all $n_2$ such that 
\bea
k n_1 + (k^2 p) n_2 &=& N_1 N_5\,,
\eea
weighted with coefficients as described in the previous subsection.

For Style 2 coiffuring, from \eq{QP2} and \eq{angmomsty2} we have on the gravity side
\bea
Q_P ~=~   \frac{2|b_4|^2}{k^2 R_y^2} \,, \qquad 
J_R ~=~ \frac{1}{2}  \frac{Q_1 Q_2 }{k R_y}  -  \frac{1}{2} k R_y Q_P \,, \qquad 
J_L  ~=~ J_R + k R_y Q_P \,.
\eea
In the CFT, the expectation value of the momentum $L_0 - \bar{L}_0$ in the Style 2 state is
\bea
N_p &=& p \, \bar{n}_2\,.
\eea
The total number of strands is $N_1 N_5$; this determines $\bar{n}_1$ in terms of $\bar{n}_2$ (or $N_p$) as
\bea
\bar{n}_1 &=& \frac{N_1 N_5}{k} - k N_p \,.
\eea
Then the CFT ${\bar \jmath}^3$ is 
\bea
{\bar \jmath}^3 ~=~ \frac{\bar{n}_1}{2} ~=~ \frac{1}{2}\frac{N_1 N_5}{k} - \frac{k}{2} N_p \,.
\eea
We convert the supergravity charges to quantized charges using 
\bea
Q_1 ~=~ \frac{g_s N_1 \alpha'^3}{V} \,, \quad~~ Q_5 ~=~ g_s N_5 \alpha' \,, \quad~~ Q_P ~=~ \frac{g_s^2 N_P \alpha'^4}{R_y^2 V} \,, \quad~~ \frac{\pi}{4 G^{(5)}} ~=~ \frac{V R_y}{g_s^2 \alpha'^4}
\eea
which lead to the useful relations
\bea
\frac{\pi}{4 G^{(5)}} \frac{Q_1 Q_5}{R_y} ~=~ N_1 N_5 \,, \qquad 
\frac{\pi}{4 G^{(5)}} R_y Q_P ~=~ N_P \,.
\eea
Thus we obtain
\bea
{\bar \jmath}^3_{\rm grav} ~=~ \frac{\pi}{4 G^{(5)}} J_R ~=~ \frac{1}{2}\frac{N_1 N_5}{k} - \frac{k}{2} N_p \,
\eea
which agrees with the CFT. Next, the CFT $j^3$ is 
\bea
j^3 ~=~ \frac{\bar{n}_1}{2} + (kp) \bar{n}_2 ~=~ {\bar \jmath}^3 + k N_p \,.
\eea
Comparing to the gravity solution we have
\bea
{j}^3_{\rm grav} ~=~ \frac{\pi}{4 G^{(5)}} J_L ~=~ {\bar \jmath}^3_{\rm grav} + k  N_p \,
\eea
which is also in agreement. Then by comparing the momentum charge we obtain the map between $|b_4|^2$ and $\bar{n}_2$:
\bea
N_P &=& \frac{\pi}{4 G^{(5)}} R_y Q_P  \qquad \Rightarrow \qquad p \, \bar{n}_2 ~=~ \left(\frac{\pi}{4 G^{(5)}} \right)\frac{2|b_4|^2}{k^2 R_y} \,.
\eea

\refstepcounter{subsubsection}
\subsubsection*{\thesubsubsection ~~ Style 1}
\label{sec:sty1charges}

For Style 1, we first consider $kp > 1$. As described above, the ingredients in the coherent state sum are
\be
\label{CFT dual state-sty1}
\Big(\ket{++}_k^{\strut}\Big)^{n_1}
\prod_{\hat{p}\in p\bbZ} 
\;\Big( (J_{-1/k}^+)^{k\hat{p}}\ket{00}^{\strut}_{k^2\hat{p}}\Big)^{\ntwop}
\;\Big( (J_{-1/k}^+)^{k\hat{p}}\ket{++}^{\strut}_{k(k\hat{p}+1)}\Big)^{\nthreep}
\;\Big( (J_{-1/k}^+)^{k\hat{p}}\ket{--}^{\strut}_{k(k\hat{p}-1)}\Big)^{\nfourp}
\ee 
For Style 1 coiffuring, from \eq{QP1}, \eq{angmomL1} and \eq{angmomR1} we have on the gravity side
\bea
Q_P ~=~   \frac{4|b_4|^2}{k^2 R_y^2} \,, \qquad 
J_R ~=~ \frac{1}{2}  \frac{Q_1 Q_2 }{k R_y}  -  \frac{1}{2} k R_y Q_P \,, \qquad 
J_L  ~=~ J_R + k R_y Q_P \,.
\eea
In the CFT, each individual element in the coherent state sum has momentum eigenvalue
\bea
\sum\limits_{\hat{p}\in p\bbZ} \, \hat{p} \left( \ntwop + \nthreep + \nfourp\right)
\eea
and so the expectation value of $L_0-\bar{L}_0$ again involves the average numbers of strands,
\bea
N_p &=& \sum\limits_{\hat{p}\in p\bbZ} \, \hat{p} \left( \barntwop + \barnthreep + \barnfourp\right).
\eea
Since the total number of strands is $N_1 N_5$, we have
\bea
k \bar{n}_1 + \sum\limits_{\hat{p}\in p\bbZ}
\left[
(k^2 \hat{p}) \barntwop 
+ (k^2 \hat{p}+k) \barnthreep 
+ (k^2 \hat{p}-k) \barnfourp \right]
&=& N_1 N_5 \,.
\eea
Because the density profile function $w(\xi)$ is real, we have the relation on the average numbers 
\bea
\barnthreep &=& \barnfourp  \qquad \mathrm{for~all} ~~ \hat{p} \,.
\eea
Therefore we have 
\bea
k \bar{n}_1 + \sum\limits_{\hat{p}\in p\bbZ} (k^2 \hat{p}) \left(\barntwop+\barnthreep+\barnfourp\right)  &=& N_1 N_5
\eea
and so, as for the Style 2 states, $\bar{n}_1$ is given by
\bea
\bar{n}_1 &=& \frac{N_1 N_5}{k} - k N_p   \,.
\eea
Next, the CFT ${\bar \jmath}^3$ is 
\bea
{\bar \jmath}^3 ~=~ \frac12 \left[ \bar{n}_1 + \sum\limits_{\hat{p}\in p\bbZ} \left( \barnthreep-\barnfourp \right)\right]
 ~=~ \, \frac{\bar{n}_1}{2} \,~=~\, \frac{1}{2}\frac{N_1 N_5}{k} - \frac{k}{2} N_p \,.
\eea
The gravity ${\bar \jmath}^3$ is 
\bea
{\bar \jmath}^3_{\rm grav} ~=~ \frac{1}{2}\frac{N_1 N_5}{k} - \frac{k}{2} N_p \,,
\label{eq:sty1jbarcft}
\eea
so we find perfect agreement.

The CFT $j^3$ is 
\bea
{j}^3 ~=~ \frac12 \left[ \bar{n}_1 + \sum\limits_{\hat{p}\in p\bbZ} \left( \barnthreep-\barnfourp \right)\right] 
+ \sum\limits_{\hat{p}\in p\bbZ}(k\hat{p}) \left(\barntwop+\barnthreep+\barnfourp\right) ~=~ {\bar \jmath}^3 + k N_p \,.
\eea
Comparing to the gravity we have
\bea
{j}^3_{\rm grav} ~=~ \frac{\pi}{4 G^{(5)}} J_L ~=~ {\bar \jmath}^3_{\rm grav} + k  N_p \,
\label{eq:sty1j3cft}
\eea
and so we again find perfect agreement.

Finally, by comparing the momentum charge we obtain the map between $|b_4|^2$ and the average numbers of excited CFT strands,
\bea
N_P &=& \frac{\pi}{4 G^{(5)}} R_y Q_P  \qquad
\Rightarrow \qquad \sum\limits_{\hat{p}\in p\bbZ} \, \hat{p} \left( \barntwop + \barnthreep + \barnfourp\right) ~=~ \left(\frac{\pi}{4 G^{(5)}} \right)\frac{4|b_4|^2}{k^2 R_y} \,. \qquad~~
\eea

For  $kp=1$, the analysis contains minor differences, however the expressions for $j^3$ and $\bar{\jmath}^3$ in terms of $N_1$, $N_5$, $N_P$ are the same as those given in \eq{eq:sty1jbarcft} and \eq{eq:sty1j3cft}, as we show in Appendix \ref{sec:sty1chargeskp1}. Thus the conserved charges agree for all values of $k$ and $p$.

Therefore we find exact agreement of conserved charges between gravity and CFT, providing supporting evidence for our proposal.
It would be interesting to scrutinize our proposal further with the tools of precision holography~\cite{Skenderis:2006ah,*Kanitscheider:2006zf,Kanitscheider:2007wq,Taylor:2007hs,Giusto:2015dfa}.

\subsection{Comments on momentum fractionation}
\label{ss:fractionation}

The fractional spectral flow that we perform results in filled Fermi seas on the excited strands. One way to see this is to observe that the $SU(2)_\cR$ current algebra has the identity 
\bea
\label{specflow strand}
 (J_{-1/k}^+)^{k\hat{p}}\,\ket{00}^{\strut}_{k^2\hat{p}} &=&
J_{-\frac{2k\hat{p}-1}{k^2\hat{p}}}^+ \cdots J_{-\frac{5}{k^2\hat{p}}}^+ J_{-\frac{3}{k^2\hat{p}}}^+ J_{-\frac{1}{k^2\hat{p}}}^+   
\ket{00}^{\strut}_{k^2\hat{p}} \,.
\eea
Similar expressions apply for  the $\ket{++}$ and $\ket{--}$ strands.

So the CFT state can be written in different ways, and in one way of looking at our states, we excite modes with the lowest possible energy compatible with the constraint of integer momentum per strand. 
Saying this another way, spectral flow creates a state with the lowest possible energy for a given angular momentum, or equivalently maximal angular momentum for a given energy, so that there is no available free energy for thermal excitations of the state.  

As one backs away from maximal angular momentum, one has the freedom to excite different modes, and the entropy increases.  For instance, if we change one of the current raising operators on the right-hand side of~\eqref{specflow strand} from a $J^+$ to a $J^3$, the angular momentum is decreased by one unit but the energy and momentum remain the same; and there are $kp$ distinct ways to do this.  Decrease the angular momentum by one more unit, and we can either have one $J^-$ or two $J_3$ with the rest remaining $J^+$, and there are of order $(kp)^2$ choices; and so on.

Such a deformation away from maximal spin preserves the BPS property of the CFT state.  
It is interesting to ask what the gravitational description of such excitations will be, and whether they will match those of the CFT. 
If we change the lowest modes with energy/momentum of order $1/k^2p$, we would expect to have made a change in the geometry in the places with the deepest red-shift.  Note that such BPS deformations are not available in the two-charge seed on which the three-charge coiffured solution is based. 

Since the CFT state has strands of length of order $k^2p$, there are also non-BPS excitations that have zero momentum and angular momentum, and energy of order $1/k^2p$. Such excitations are also present in the  two-charge seed states.  
In the supergravity, the non-BPS excitations are described at the linearized level by solving wave equations in the superstratum geometry.

The supergravity solutions do not appear to have excitations at the scale $1/k^2p$ suggested by the CFT, however; in general, there seems to be a mismatch between the gap in supergravity and in the CFT.  The two-charge seed for Style 2 coiffuring is quite similar to a class of two-charge solutions studied in~\cite{Lunin:2002iz}, for which the gap was estimated to be $a/b$ (with $a$ related to the number of $\ket{++}$ strands, $b$ the number of $\bbT^4$ strands including $\ket{00}$ strands).  
In the CFT, the gap depends only on the length $\kappa$ of the strands and is independent of the relative amounts $a$ and $b$ of the different kinds of strands.  A preliminary study of the foregoing three-charge geometries indicates that, similar to the examples of~\cite{Lunin:2002iz}, the red-shift depends on the amplitudes $a$ and $b$, and that the deepest red-shifts are not $kp$ times deeper than those of the parent $k$-wound supertube. 

In general, one can arrange that the throat in supergravity is deeper and results in a smaller gap than in the orbifold CFT ({\it e.g.} supergravity duals to CFT states discussed in~\cite{Lunin:2002iz} having only short cycles but low total angular momentum), and in yet other examples the throat in supergravity is shallower and results in a larger gap than in the orbifold CFT ({\it e.g.} the coiffured geometries discussed in this paper when $b$ is finite but much less than~$a$).
It would be useful to understand better the cause of this discrepancy. 

The two-charge seed geometries of Section~\ref{ss:CFT dual states} offer a qualitative explanation of the gap in supergravity.  The dual of the F1-P source in the Lunin-Mathur construction of two-charge geometries~\cite{Lunin:2001fv}  is a D1-D5 supertube smeared over the compact directions -- the circle parametrized by $y$ and the compactification manifold $\cM$~\cite{Lunin:2001jy,Lunin:2002bj,Lunin:2002iz,Kanitscheider:2007wq}.  When segments of the unsmeared source approach one another, a throat opens and deepens in the geometry. This property explains why the profile~\eqref{circular} results in a red-shift of order $k$~-- the supertube source traces the same profile in the transverse space $k$ times in the course of the supertube winding the $y$ circle, and is $k$ times more compact (in ${\mathbb R}^4$ coordinates); as a consequence, the harmonic functions are $k$ times bigger at their maximum, and the throat is $k$ times deeper.  

For a small perturbation of this profile, it may be that the oscillations of the profile are $kp$ times faster than the $k$-fold spiral of the supertube, but this is a small perturbative wiggle and does not make the profile $kp$ times more bunched together, and hence the deepest parts of the throat do not exhibit a red-shift $kp$ times deeper.  However, as one shifts more of the strands from $\ket{++}$ type to $\ket{00}$ type, the angular momentum is reduced, the source becomes more compact, and the throat deepens.

It remains a puzzle why there is such a mismatch between the behavior of supergravity and that of the CFT for such a coarse property of the geometry.  The gap to non-BPS excitations is of course not a robust property of the system, and could change dramatically as one passes from the regime where the CFT is weakly coupled to the regime where it is strongly coupled and gravity is a good approximation.  Nevertheless, there are examples (see for instance~\cite{Giusto:2012yz}) where the gap can be matched on both sides of the duality. 
The presence or absence of strands polarized in the $\bbT^4$ directions appears to be an ingredient which influences whether this quantity agrees between gravity and CFT; it would be useful to understand fully when this comparison does and does not work.

\section{Discussion}
\label{Sect:Discussion}

This work has expanded the construction of superstrata to include momentum-carrying modes in deep AdS$_3$ throats, in which the red-shift at the bottom of the throat is $k$ times that of 
a singly-wound supertube. Our construction started from a  $k$-wound circular supertube geometry. We performed spectral inversion on this solution, then altered its angular momentum by adding charge density fluctuations along the supertube with a wavenumber $kp$ for some integer $p$, without deforming the shape of the supertube. We then brought the solution back to the original frame, where these fluctuations became momentum-carrying excitations. 

Our construction also produced the first examples of asymptotically-flat superstrata. We built  two classes of solutions, corresponding to two different ways of arranging the Fourier coefficients in order to obtain smooth solutions (with the usual $Z_k$ orbifold singularities at the location of the supertube). 

Taking the decoupling limit to obtain the corresponding asymptotically-AdS solutions, we derived a proposal for the dual CFT states, for both classes of solutions.  
The starting supertube is built from a macroscopic ensemble of cycles of length $k$ in the twisted sector of the symmetric orbifold CFT.  The angular excitations in the CFT description are coherent fractional spectral flows on additional cycles of the twisted sector state, whose length is of order $k^2p$.  This fractional spectral flow can also be thought of either as acting of order $kp$ times with the fractionally-moded raising operator $J^+_{-1/k}$, or as raising the Fermi seas on these cycles by filling all the fermion modes with positive $\cR$-charge up to a level of order $1/k$.

In our states, the fractionally-moded quanta in the CFT correspond to perfectly regular, local excitations in the supergravity theory and not to non-geometric or multi-valued perturbations. The bulk reflection of the fractional momentum carriers is rather the red-shift of the perturbations down the supertube throat.

A small puzzle that remains is the apparent mismatch in the excitation gap of orbifold CFT states and supergravity geometries discussed in Section~\ref{ss:fractionation}.  
A very similar mismatch was previously noted for certain two-charge solutions~\cite{Lunin:2002iz}.  In the CFT, the gap is determined by the length of the longest cycles in the twisted sector ground state.  In the geometry, the depth of the throat depends on other quantities, such as the relative proportions of the different strands. The supergravity gap can be larger or smaller than the orbifold CFT gap. The gap to non-BPS excitations is not protected in general, so this is not a serious problem for the holographic duality. However there are examples (see for instance~\cite{Giusto:2012yz}) where the gap matches between gravity and CFT. It would be interesting to understand when the gap should agree, and when it should not.

Our solutions do not have all desired features of typical black-hole microstates: Their angular momenta are over-spinning and the throats are not as deep as those of typical states.  
The corresponding orbifold CFT states contain strands having length of order $k^2 p$, and so $k$ can at most be of order $\sqrt{N_1N_5}$, while the longer wavelength scale $k^2p$ is not apparent in the geometry. 
Thus we regard the supergravity solutions presented here as a ``proof of concept'' of a supergravity realization of momentum fractionation on superstrata, much like the solutions in  \cite{Bena:2015bea} are a proof of concept of the existence of superstrata solutions parameterized by arbitrary functions of two variables.

For the future, one would like to improve on both of these (related) features: To lower the angular momenta, and to deepen the throat further.
First, regarding the angular momenta, in Section \ref{ss:fractionation} we identified CFT excitations that move away from the maximally spinning/overspinning regime by reducing the angular momentum through a change in the polarization of the $\cR$-symmetry currents acting on the two-charge seed.  Using this freedom, one can make available some of the free energy to wiggle the throat while remaining BPS.  Where in the throat the excitation lies should correlate with the degree of fractionation of the modes whose polarizations are being adjusted in the CFT.  

One place to look for these more general solutions on the supergravity side is to consider more generic superstrata, described by arbitrary functions of two variables. In this work we have focused on a sub-class of solutions which are parameterized by functions of one variable. This has been a choice made for technical convenience, to focus on the physics of momentum fractionation in a tractable system. It would be interesting to generalize our solutions to superstrata which are parametrized by functions of two variables and which exhibit momentum fractionation.
Looking further ahead, the generic CFT state deformations discussed above, which stay BPS by deforming the polarizations of the spectral flow $\cR$-currents, will correspond to deformations of the supergravity solution that depend on all all three angular variables $(v,\varphi_1,\varphi_2)$.

The next essential step in the study of superstrata is to construct states with deeper throats, that are in a macroscopic scaling regime.   Our solutions have throats $k$ times deeper than the first superstrata constructed in \cite{Bena:2015bea}, and so represent progress in this direction. 
The standard way to obtain a macroscopic scaling solution is to use at least three Gibbons-Hawking centers, but it may also be possible to construct scaling solutions with two centers when the supertubes fluctuate.
As we noted above, for technical reasons we have focussed on some very particular modes and this choice of modes meant that whenever we added momentum to the supertube we also added a similar amount of angular momentum.  Thus our solutions remained over-spinning or extremal.  As a result, we could not access the scaling region that is usually associated with the microstates of a black hole with macroscopic horizon area.  In this paper we added charges to the supertube in a manner that precluded us from exploring such deep, scaling geometries.

In addition to the excitations discussed above that lower the angular momenta, more broadly one can consider excitations that either have no angular momentum, or have negative angular momentum. In principle, by using these excitations one can add momentum to the supertube in a way that takes the charges into the BMPV regime.  The corresponding black hole would then have a macroscopic horizon and the microstate geometry should then scale and exhibit larger red-shifts and lower holographic energy gaps.  This is presently under investigation.

More generally, one may desire to embed superstrata and the kind of twisted-sector structure elucidated here, in multi-centered deep, scaling geometries since this is (as yet) the only known way to access {\it typical} twisted-sector CFT states within the supergravity approximation. On a technical level this will be challenging, since it means going beyond two centers and yet our construction has made very heavy use of the flat $\IR^4$ base and the separability of various wave equations in bipolar coordinates. However, this does not mean that it is impossible: The scalar Green functions for charge density fluctuations in generic ambipolar backgrounds were discussed in \cite{Bena:2010gg}, and a three-centered Green function was constructed explicitly.  So while this may be very difficult, it is not completely out of reach.  Moreover, we hope to find physical arguments that illuminate what the geometries constructed in this paper will probe once they are combined  with generic superstrata and embedded  in deep, scaling geometries. 

Looking further to the future, it would be of great interest to study momentum fractionation in non-supersymmetric microstates, as done in~\cite{Chakrabarty:2015foa}. The recent construction of multi-bubble non-BPS black-hole microstate geometries~\cite{Bena:2015drs} offers the prospect of progress in this direction.

\section*{Acknowledgments}

\vspace{-2mm}
We would like to thank Stefano Giusto, Rodolfo Russo and Masaki Shigemori for helpful discussions.
The work of IB and DT was supported by John Templeton Foundation Grant 48222 and by a grant from the Foundational Questions Institute (FQXi) Fund, a donor advised fund of the Silicon Valley Community Foundation on the basis of proposal FQXi-RFP3-1321 (this grant was administered by Theiss Research). 
The work of EJM was supported in part by DOE grant DE-SC0009924. The work of DT was supported in part by a CEA Enhanced Eurotalents Fellowship. The work of
NPW was supported in part by DOE grant DE-SC0011687. For hospitality during the course of this work, EJM and NPW are very grateful to the IPhT, CEA-Saclay; IB, EJM, and DT thank the Centro de Ciencias de Benasque Pedro Pascual; and DT and NPW thank the Yukawa Institute for Theoretical Physics, Kyoto University.

\appendix

\section{The BMPV black hole}
\label{app:BMPV}

To help establish normalizations, it is useful to give the standard BMPV black-hole metric~\cite{Breckenridge:1996is} in terms of the Ansatz used in this paper. 
Everything is, of course, $v$-independent and the vector field, $\beta$, and the $\Theta_I$, are set to zero.  For a BMPV black hole located at the center of space ($r=0$, $\theta =0$) the $Z_I$ are appropriately-sourced harmonic functions:
\begin{equation}
Z_1 ~=~ 1 ~+~ \frac{Q_1}{\Lambda} \,, \qquad Z_2 ~=~ 1 ~+~ \frac{Q_2}{\Lambda} \,, \qquad \mathcal{F} ~=~ - \frac{2Q_3}{\Lambda}   \,, \qquad Z_4 ~=~ 0 \,. 
\label{BMPV-Zs}
\end{equation} 
 The angular momentum vector, $\omega$, is then simply the ``harmonic''  solution to the homogeneous equation (\ref{BPSlayer2a}) with source at the center of space:
\begin{equation}
\omega ~=~ \frac{J}{\Lambda^2} \,  \big( (r^2 + a^2)\sin^2 \theta \, d \varphi_1  - r^2 \cos^2 \theta\, d \varphi_2  \big)  \,. 
\label{BMPV-om}
\end{equation} 
Note that as $r \to \infty$ one has
\begin{equation}
Z_I ~\sim~  1+\frac{Q_I}{r^2} \,, \qquad I=1,2,3 \,, \quad\qquad \omega ~\sim~ \frac{J}{r^2} \,  \big(\sin^2 \theta \, d \varphi_1  -  \cos^2 \theta\, d \varphi_2  \big)  \,,
\label{BMPV-Zs-2}
\end{equation} 
which determine the charges and angular momenta of the black hole.

To make the asymptotic analysis of the metric in the vicinity of the center of space using more standard spherical coordinates in the infinitesimal neighborhood of  $r=0$, $\theta =0$, one can simply take:
\begin{equation}
r ~=~ \lambda\, \sin  \chi  \,, \qquad \theta ~=~ \frac{\lambda}{a}\,  \cos  \chi   \,. 
\label{infsph}
\end{equation} 
and expand to lowest order in $\lambda$.  One then finds that the leading part of the metric becomes:
\begin{align} 
{ds}_5^2 ~=~  \sqrt{Q_1Q_2} \,  \bigg[  \, & \frac{d\lambda^2}{\lambda^2}  ~+~   d \chi^2   ~+~\sin^2 \chi \cos^2 \chi  \, (d \varphi_1- d \varphi_2)^2 \nonumber \\ &  ~+~  \frac{2 Q_3}{Q_1 Q_2}  \, \Big(dv - \frac{J}{2Q_3}  \,  (\cos^2 \chi \, d \varphi_1+\sin^2 \chi\, d \varphi_2) \Big)^2  \nonumber \\ &  ~+~   \Big(1 -\frac{J^2}{2 Q_1 Q_2 Q_3} \Big)\, (\cos^2 \chi \, d \varphi_1+\sin^2 \chi\, d \varphi_2)^2\, \bigg] \,.
\label{BMPVnear}
\end{align} 
In particular, we see that with our normalizations one must impose the condition:
\begin{equation}
J^2 ~\le~ 2 Q_1 Q_2 Q_3  \qquad \Leftrightarrow \qquad   J_L^2 ~\le~ Q_1 Q_2 Q_P  \,,
\label{BMPVbound}
\end{equation} 
where $J_L = J/ \sqrt{2}$ and $Q_P =Q_3$.

\section{Reduction to five dimensions}
\label{app:5dlimit} 

There are two standard ways of reducing the six-dimensional solution, and the system of BPS equations~\cite{Gutowski:2003rg,Cariglia:2004kk,Bena:2011dd,Giusto:2013rxa}, to the standard, five-dimensional analogs found in may references (see, for example, \cite{Bena:2007kg,Giusto:2012gt}). These two  choices of reduction come from different embeddings of the five-dimensional fields in the six-dimensional  formulation; we summarize these two standard choices here.   The five-dimensional BPS equations are: 
\begin{eqnarray}
 \Theta^{(I)}  &~=~&  \star_4 \, \Theta^{(I)} \label{5dBPSeqn:1} \,, \\
 \nabla^2  Z_I &~=~&  \coeff{1}{2}  \, C_{IJK} \star_4 (\Theta^{(J)} \wedge
\Theta^{(K)})  \label{5dBPSeqn:2} \,, \\
 d\mathbf{k} ~+~  \star_4 d\mathbf{k} &~=~&  Z_I \,  \Theta^{(I)}\,.
\label{5dBPSeqn:3}
\end{eqnarray}
Our goal will be to take $v$-independent, six-dimensional solutions and compactify on an $S^1$ fiber so that the system equations 
(\ref{BPSlayer1a})--(\ref{BPSlayer1c}),  (\ref{BPSlayer2a}) and  (\ref{BPSlayer2b}) reduce to the five-dimensional system.

\subsection{Reduction 1}

This is the canonical choice if  $\mathcal{F}$ never vanishes and in particular, when $\mathcal{F} \to -1$ at infinity.   One can then write the metric \eq{sixmet} globally as 
\begin{equation}
ds_6^2 ~=  \frac{1}{\sqrt{\cP}\, \mathcal{F}} \,(du +  \omega)^2 ~-~ \frac{\mathcal{F}}{\sqrt{\cP}}\, \big(dv+\beta + \mathcal{F}^{-1}  (du +  \omega)  \big)^2 
~+~  \sqrt{\cP} \, ds_4^2(\cB)\,. \label{sixmet-sq}
\end{equation}
Upon making the identifications
\begin{equation}
\mathcal{F}  ~=~ - Z_3\,, \qquad u ~=~t \,, \qquad v~=~t+y \,, \qquad \mathbf{k} ~=~ \omega\,,\qquad \Theta_3 ~=~ d \beta \,, 
\label{identifications1}
\end{equation}
the six-dimensional metric is given by
\begin{equation}
ds_6^2 ~=  -\frac{1}{Z_3\, \sqrt{\cP}} \, (dt +  \mathbf{k})^2 ~+~ \frac{Z_3}{\sqrt{\cP}}\,
\Big[ dv+\beta - Z_3^{-1}  (dt+\mathbf{k})  \Big]^2 
~+~  \sqrt{\cP} \, ds_4^2(\cB)\,,  \label{Bsixmet-sq1}
\end{equation}
which can also be written as
\begin{equation}
ds_6^2 ~=  -\frac{1}{Z_3\, \sqrt{\cP}} \, (dt +  \mathbf{k})^2 ~+~ \frac{Z_3}{\sqrt{\cP}}\, 
\Big[dy + (1-Z_3^{-1})  (dt + \mathbf{k}) + (\beta-\mathbf{k})  \Big]^2 
~+~  \sqrt{\cP} \, ds_4^2(\cB)\,.  \label{Bsixmet-sq2}
\end{equation}
Compactifying on the $y$-circle yields an overall warp factor of $(\frac{Z_3}{\sqrt{\cP}})^{1/3}$ on the five-dimensional metric and leads to 
\begin{equation}
ds_5^2 ~=  -\big(Z_3\,\cP \big)^{-\frac{2}{3}} \, (dt +  \mathbf{k})^2   ~+~  \big(Z_3\,\cP \big)^{\frac{1}{3}} \, ds_4^2(\cB)\,,  \label{Bfivemet-sq}
\end{equation}
%

These identifications reduce the six-dimensional BPS system used in this paper directly to the canonical five-dimensional system; this is the origin of how we have chosen to normalize the flux fields like $\Theta_I$.  However we have chosen the $t$, $y$ coordinates (\ref{uvdefn}), meaning that $\mathcal{F} \to 0$ at infinity, leading to a canonical embedding more closely associated with supertubes.  We will now describe this in more detail. 

\subsection{Reduction 2}
In this reduction we use the coordinates (\ref{uvdefn}):
\begin{equation}
u ~\equiv~   \frac{1}{\sqrt{2}}\, (t-y) \,,  \qquad v ~\equiv~    \frac{1}{\sqrt{2}}\, (t+y)  \,.  \label{Buvdefn}
\end{equation}
Then as described in \eq{eq:Z3k}, we introduce
\bea \label{eq:Z3k-2}
Z_3 ~=~ 1- \frac{\cF}{2} \,, \qquad  \mathbf{k} ~=~ \frac{\omega+\beta}{\sqrt{2}} \,, 
\eea
and complete the squares in the metric as in \eq{sixmet-sqty} to obtain 
\begin{equation}
ds_6^2 ~=  -\frac{1}{Z_3\sqrt{\cP} } \, (dt +  \mathbf{k})^2 \,+\, 
\frac{Z_3}{\sqrt{\cP}}\, \left[dy  +\left(1- Z_3^{-1}\right)  (dt +  \mathbf{k}) +\frac{\beta-\omega}{\sqrt{2}} \right]^2   +  \sqrt{\cP} \, ds_4^2(\cB)\,.
\label{sixmet-sqty-2}
\end{equation}

With these identifications one must make the following replacements and re-definitions for the quantities defined in the body of this paper 
\begin{equation}
\Theta_I ~\to~ \sqrt{2}\, \Theta_I \,, \quad I=1,2,4 \,; \qquad \qquad \Theta_3 ~=~ \sqrt{2}  \, d \beta \,.
\label{rescales2}
\end{equation}
Doing this, the BPS equations (\ref{BPSlayer1a})--(\ref{BPSlayer1c}),  (\ref{BPSlayer2a}) and  (\ref{BPSlayer2b}) reduce to the five-dimensional system \eq{5dBPSeqn:1}--\eq{5dBPSeqn:3}.  In particular, the terms arising from the constant in $\mathcal{F}  = - 2(Z_3-1)$  cancel in (\ref{BPSlayer2a}) against the terms $D\beta +*_4 D\beta$ arising from the replacement $\omega =\sqrt{2}  \mathbf{k}-\beta$.

\section{The lowest Style 1 modes}
\label{sec:sty1chargeskp1}

In this appendix we demonstrate the agreement of conserved charges for the lowest possible modes in Style 1, those with $kp=1$, following the analysis for $kp>1$ done in Section \ref{sec:sty1charges}.

For $kp=1$, the dual CFT state is a particular superposition of states of the Style 1 type \eq{CFT dual state-sty1},
\be
\label{CFT dual state-sty1_app}
\Big(\ket{++}_1^{\strut}\Big)^{n_1}
\prod_{\hat{p}\in \bbZ} 
\;\Big( (J_{-1}^+)^{\hat{p}}\ket{00}^{\strut}_{\hat{p}}\Big)^{\ntwop}
\;\Big( (J_{-1}^+)^{\hat{p}}\ket{++}^{\strut}_{\hat{p}+1}\Big)^{\nthreep}
\;\Big( (J_{-1}^+)^{\hat{p}}\ket{--}^{\strut}_{\hat{p}-1}\Big)^{\nfourp} 
\ee 
 As explained at the end of Section \ref{sec:sty1cft}, the average numbers of $\ket{++}_{k(k\hat{p}+1)}$ and $\ket{--}_{k(k\hat{p}-1)}$ strands are equal for $\hat{p}\ge 2$,
\bea
\bar{n}_{3,\hat{p}} &=& \bar{n}_{4,\hat{p}}  \qquad \mathrm{for~all} ~~ \hat{p}\ge 2 \,,
\label{eq:kp1balance}
\eea
while the excited $\ket{--}$ strands that would be counted by $n_{4,1}$ would have length zero, which does not exist. Therefore we set 
\bea
n_{4,1} &=& 0 \,.
\eea
This means that the excited $\ket{++}_2$ strands that are counted by $n_{3,1}$ are not balanced out by corresponding $\ket{--}$ strands. Nevertheless, the conserved charges will work out properly, as we now show.

Since the total number of strands is $N_1 N_5$, using \eq{eq:kp1balance} we obtain
\bea
\bar{n}_1 + \sum\limits_{\hat{p}\in \bbZ}
\left[
\hat{p} \barntwop 
+ (\hat{p}+1) \barnthreep 
+ (\hat{p}-1) \barnfourp \right]
&=& N_1 N_5 \, \cr
\Rightarrow\qquad \bar{n}_1 + \sum\limits_{\hat{p}\in \mZ} \left[ \hat{p} (\bar{n}_{2,\hat{p}}+\bar{n}_{3,\hat{p}}+\bar{n}_{4,\hat{p}})\right] +  \bar{n}_{3,1}  &=& N_1 N_5\,
\eea
and so $\bar{n}_1$ is given by 
\bea
\bar{n}_1 &=& N_1 N_5  -   N_p  - \bar{n}_{3,1}  \,.
\eea
Next, the CFT ${\bar \jmath}^3$ is 
\bea
{\bar \jmath}^3 ~=~ \frac12 \left[ \bar{n}_1 + \sum\limits_{\hat{p}} \left( \bar{n}_{3,\hat{p}}-\bar{n}_{4,\hat{p}} \right)\right]
 ~=~ \frac{1}{2}\left( \bar{n}_1 + \bar{n}_{3,1} \right) ~=~ \frac{1}{2}N_1 N_5 - \frac{1}{2} N_p  \,,
\label{eq:jbarkp1-cft}
\eea
in perfect agreement with the value of ${\bar \jmath}^3$ computed from the gravity.

The CFT $j^3$ is 
\bea
{j}^3 &=& \frac12 \left[ \bar{n}_1 + \sum\limits_{\hat{p}} \left( \bar{n}_{3,\hat{p}}-\bar{n}_{4,\hat{p}} \right)\right] 
+ \sum\limits_{\hat{p}\in \mZ}\hat{p} \left(\bar{n}_{2,\hat{p}}+\bar{n}_{3,\hat{p}}+\bar{n}_{4,\hat{p}}\right) \cr
&=& \frac12 \left( \bar{n}_1 + \bar{n}_{3,1} \right)
+ \sum\limits_{\hat{p}\in \mZ}\hat{p} \left(\bar{n}_{2,\hat{p}}+\bar{n}_{3,\hat{p}}+\bar{n}_{4,\hat{p}}\right)
 ~=~ {\bar \jmath}^3 + k N_p \,.
\label{eq:j3kp1}
\eea
which again agrees exactly with the value of $j^3$ computed from the gravity.
The momentum charge determines $b_4$ just as for $kn>1$, and so all conserved charges agree.

This agreement shows that comparing conserved charges alone does not put any constraint on the value of $\bar{n}_{3,1}$.
Of course, our proposal of Section \ref{ss:CFT dual states} fixes $\bar{n}_{3,1}$ unambiguously, since we have specified in principle all coefficients in the coherent state. To scrutinize our proposal further, one would have to perform further holographic tests.

One can see how this agreement works in another way:
Relative to the unexcited base supertube $\ket{++}_1$ strands, the difference in conserved charges is as follows. For each excited $\ket{++}_2$ strand, the change in $\bar{\jmath}^3$ is $\Delta \bar{\jmath}^3 = - 1/2$; for $j^3$ we have 
$\Delta j^3=1-1/2 = 1/2$; and we have $\Delta P=1$. So regardless of the value of $\bar{n}_{3,1}$, the above expressions for $j^3$ and $\bar{\jmath}^3$ in terms of $N_1$, $N_5$, $N_P$ are the same.


\newpage

\begin{adjustwidth}{-3mm}{-3mm} 


\bibliographystyle{utphysM}      


\bibliography{microstates}       

\end{adjustwidth}

\end{document}